\documentclass[prd, twocolumn, amsfonts, amssymb, amsmath, showkeys, nofootinbib, superscriptaddress]{revtex4-2}
\usepackage[english]{babel}
\usepackage[utf8]{inputenc}
\usepackage{amsthm}
\usepackage{mathtools}
\usepackage{physics}
\usepackage{xcolor}
\usepackage{graphicx}
\usepackage{adjustbox}
\usepackage{placeins}
\usepackage[T1]{fontenc}
\usepackage{lipsum}
\usepackage{csquotes}
\newcommand{\msun}{\mathrm{\, M_\odot}}
\usepackage{placeins}
\usepackage{tabularx}
\usepackage{multirow}
\usepackage{hhline}
\usepackage[pdftex, pdftitle={Article}, pdfauthor={Author}]{hyperref} 
\usepackage{xcolor}
\usepackage{amsmath}

\begin{document}

\newcolumntype{P}[1]{>{\centering\arraybackslash}p{#1}}

\title{Neutron star-black hole mergers in next generation gravitational-wave observatories}

\author{Ish Gupta}
\email[Correspondence email address: ]{ishgupta@psu.edu}
\affiliation{Institute for Gravitation and the Cosmos, Department of Physics, Pennsylvania State University, University Park, PA 16802, USA}
    
\author{Ssohrab Borhanian}
\affiliation{Theoretisch-Physikalisches Institut, Friedrich-Schiller-Universit\"at Jena, 07743, Jena, Germany}

\author{Arnab Dhani}
\affiliation{Institute for Gravitation and the Cosmos, Department of Physics, Pennsylvania State University, University Park, PA 16802, USA}

\author{Debatri Chattopadhyay}
\affiliation{Gravity Exploration Institute, School of Physics and Astronomy, Cardiff University, Cardiff, CF24 3AA, UK}

\author{Rahul Kashyap}
\affiliation{Institute for Gravitation and the Cosmos, Department of Physics, Pennsylvania State University, University Park, PA 16802, USA}

\author{V.~Ashley Villar}
\affiliation{Department of Astronomy and Astrophysics, Pennsylvania State University, University Park, PA 16802, USA}
\affiliation{Institute for Gravitation and the Cosmos, Pennsylvania State University, University Park, PA 16802, USA}
\affiliation{Institute for Computational and Data Sciences, Pennsylvania State University, University Park, PA 16802, USA}

\author{B.S. Sathyaprakash}
\affiliation{Institute for Gravitation and the Cosmos, Department of Physics, Pennsylvania State University, University Park, PA 16802, USA}
\affiliation{Gravity Exploration Institute, School of Physics and Astronomy, Cardiff University, Cardiff, CF24 3AA, UK}
\affiliation{Department of Astronomy and Astrophysics, Pennsylvania State University, University Park, PA 16802, USA}

\date{\today} 

\begin{abstract}
Observations by the current generation of gravitational-wave detectors have been pivotal in expanding our understanding of the universe. Although tens of exciting compact binary mergers have been observed, neutron star-black hole (NSBH) mergers remained elusive until they were first confidently detected in 2020. The number of NSBH detections is expected to increase with sensitivity improvements of the current detectors and the proposed construction of new observatories over the next decade. In this work, we explore the NSBH detection and measurement capabilities of these upgraded detectors and new observatories using the following metrics: network detection efficiency and detection rate as a function of redshift, distributions of the signal-to-noise ratios, the measurement accuracy of intrinsic and extrinsic parameters, the accuracy of sky position measurement, and the number of early-warning alerts that can be sent to facilitate the electromagnetic follow-up. Additionally, we evaluate the prospects of performing multi-messenger observations of NSBH systems by reporting the number of expected kilonova detections with the Vera C. Rubin Observatory and the Nancy Grace Roman Space Telescope. We find that as many as $\mathcal{O}(10)$ kilonovae can be detected by these two telescopes every year, depending on the population of the NSBH systems and the equation of state of neutron stars.
\end{abstract}


\maketitle

\section{Introduction} \label{sec:intro}
By the end of 2019, the Advanced Laser Interferometer Gravitational-Wave Observatory (aLIGO) \cite{LIGOScientific:2014pky,aLIGO:2020wna,Tse:2019wcy} and the Advanced Virgo (AdV) \cite{VIRGO:2014yos,Virgo:2019juy} detector had made a multitude of gravitational-wave (GW) detections coming from binary black hole (BBH) and binary neutron star (BNS) mergers \cite{LIGOScientific:2020ibl,LIGOScientific:2021djp}. In January 2020, the network made its first detection of a binary comprising a neutron star and a black hole, marking the first discovery ever of neutron star-black hole (NSBH) binaries in astronomy \cite{LIGOScientific:2021qlt}. This discovery not only proved the existence of NSBH systems that merge within Hubble time, but it has also provided the first direct constraint on the local merger rate of these systems \cite{LIGOScientific:2021qlt,LIGOScientific:2021psn}. 

Detecting NSBH mergers is crucial for a diverse range of astrophysical pursuits. Multiple formation channels are proposed to explain the formation and merger of NSBH systems, such as the isolated binary formation channel \cite{1976ApJ...207..574S}, dynamical formation in globular \cite{PortegiesZwart:1999nm,Clausen:2012zu,Downing:2009ag} or young stellar clusters \cite{Santoliquido:2020bry,Rastello:2020sru}, population III stars \cite{Belczynski:2016ieo} and others. These channels have varying, and often distinct, predictions for the mass and spin distributions of black holes (BHs) and neutron stars (NSs). The detection of NSBH mergers will enhance our understanding of the population characteristics and also help identify the preferred scenarios for the formation of the NSBH binaries in the universe \cite{Vitale:2015tea}. An extensive catalog of NSBH events will provide the redshift distribution of such systems, giving information about the star-formation rate (SFR) and preferred time-delay models that best explain their evolution. Just like BNS systems, NSBH systems are also expected to be sources of short gamma-ray bursts and kilonova (KN) emissions \cite{Li:1998bw,Metzger:2019zeh,Goodman:1986az,Eichler:1989ve}, making them interesting candidates for multi-messenger astronomy (MMA). NSBH detections followed by short gamma-ray bursts can be used as GW standard sirens \cite{Nissanke:2009kt}. One can also measure the fraction of short gamma-ray bursts produced by BNS and NSBH systems \cite{Sarin:2022cmu}, giving information about the preferred production mechanism of short gamma-ray bursts. NSBH detections have also been shown to be potential candidates for the measurement of Hubble constant \cite{Vitale:2018wlg,Feeney:2020kxk,Gupta:2022fwd} and are capable of estimating it to larger distances than the BNS systems. GWs from NSBH mergers, with or without an electromagnetic (EM) counterpart, can also be used to constrain the NS equation of state (EOS) \cite{Lackey:2011vz,Ascenzi:2018mwp,Coughlin:2018fis,Kawaguchi:2020osi,Barbieri:2019sjc,Tiwari:2021gfl,Brege:2018kii,Darbha:2021rqj}.

\begin{figure}[htbp]
{\centering \includegraphics[scale=0.57]{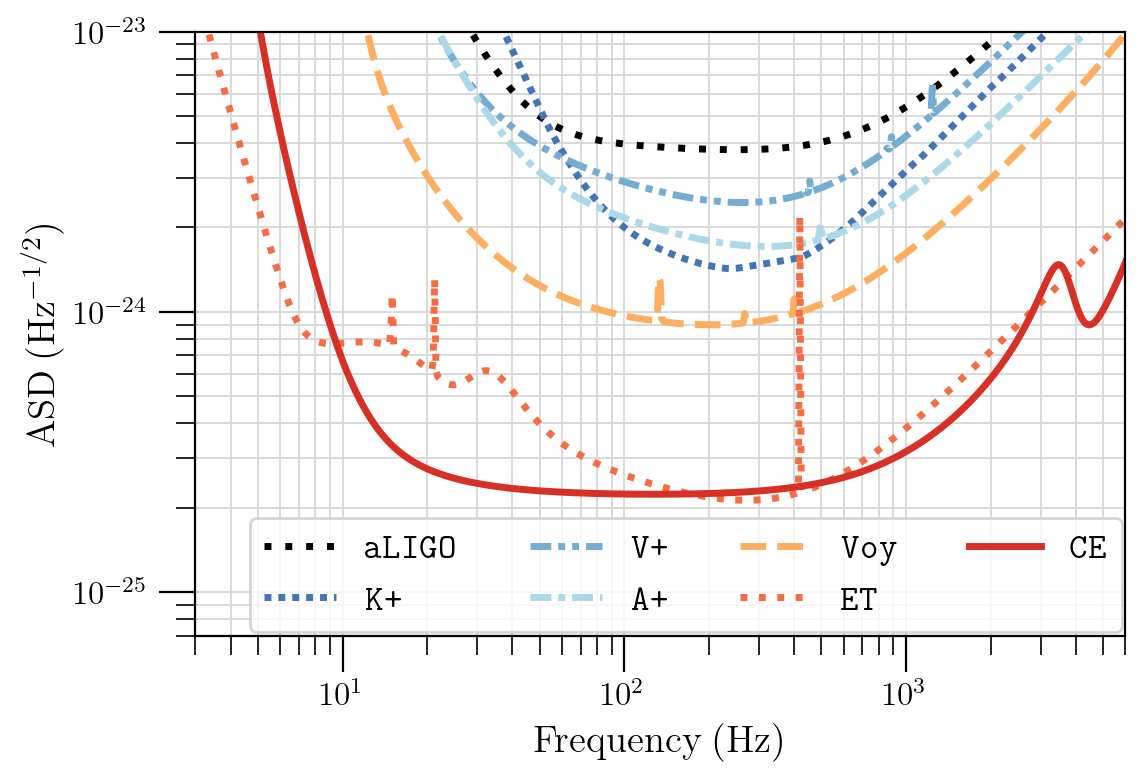}}
\caption{\label{fig:det_sens} The amplitude spectral densities (ADSs) for the proposed advancements to the current detectors as well as for the planned XG detectors. \texttt{V+}, \texttt{K+} and \texttt{A+} refer to the VIRGO detector, KAGRA detector, and the LIGO detectors at \texttt{A+} sensitivity, respectively. \texttt{Voy} refers to the LIGO detectors at Voyager sensitivity. \texttt{CE} and \texttt{ET} refer to the Cosmic Explorer and the Einstein Telescope, respectively. We also include the ASD for the aLIGO sensitivity for comparison.}
\end{figure}

Fortunately, with the proposed advancements to current GW detectors and plans in place to construct more sensitive detectors, we expect both the quantity and the quality of the NSBH detections to improve. Some of these improvements include:
\begin{itemize}
    \item \textbf{A}\texttt{+}\textbf{ sensitivity \cite{Miller:2014kma,KAGRA:2013rdx}-} The LIGO detectors at Hanford, Livingston, and the planned detector in Aundha, India \cite{LIGO-India} (referred to in this paper as LIGO-India or LIGO-I), the Virgo detector in Italy, and the KAGRA \cite{Somiya:2011np,KAGRA:2020agh,Aso:2013eba} detector in Japan are expected to upgrade to A\texttt{+} or similar sensitivities, with lower quantum noise and thermal coating noise, improving the aLIGO sensitivity by about $50\%$. The A\texttt{+} sensitivity increases the range for BNS detection to $1.9$ times and BBH sources to $1.6$ times that achieved by the aLIGO detectors. 
    \item \textbf{Voyager sensitivity \cite{LIGO:2020xsf}-} The Voyager upgrade intends to improve the aLIGO sensitivities by about 2 to 4 times. This is accomplished by reducing the quantum shot noise and using cryogenically cooled test masses with an amorphous silicon coating that reduces the thermal noise associated with the mirrors. 
    \item \textbf{Cosmic Explorer \cite{Evans:2021gyd,LIGOScientific:2016wof,Reitze:2019iox}-} The Cosmic Explorer (CE) project refers to the proposed next-generation (XG) L-shaped detector design with $40$ km arms, i.e., $10$ times the size of the current LIGO detectors. Due to the scaled-up length of the arms, CE detectors result in $\mathcal{O}(10)-\mathcal{O}(100)$ improvement in sensitivity, depending on the frequency, as compared to A\texttt{+}. Currently, there are several proposals for the configuration of the CE detectors, including the option of having just one of the two detectors, or having two detectors such that the second detector is smaller with $20$ km arms instead, which can be tuned to BNS post-merger signals. 
    \item \textbf{Einstein Telescope \cite{Punturo:2010zz,Hild:2010id}-} Einstein Telescope (ET) is the proposed XG underground detector in Europe with three detectors placed along the vertices of an equilateral triangle of side $10$ km. The detectors are planned to have a \textit{xylophone} design with each side containing two interferometers. With the longer arms, triangular-xylophone design, and measures to suppress fundamental noise sources, ET is expected to have sensitivities similar to CE.
\end{itemize}
The amplitude spectral densities (ASDs) characterizing the noise features corresponding to these enhancements have been plotted in Fig. \ref{fig:det_sens}. In this study, we will analyze the performance of the six ground-based GW detector networks listed in Table \ref{tab:net}. These networks are expected to be operational over timescales ranging from five to twenty years.

Several studies have looked at the possible improvements in the detection of GWs from compact binaries with the onset of XG detector networks \cite{Borhanian:2022czq,Ronchini:2022gwk,Iacovelli:2022bbs,Evans:2021gyd}. In this study, we assess the detection capability of the six GW detector networks for NSBH mergers and the science that can be extracted from these detections. This is carried out using \texttt{GWBENCH} \cite{Borhanian:2020ypi}, a software package that computes the signal-to-noise ratio (SNR) and the Fisher information matrix (FIM) for a given GW network and waveform model from which one can obtain the errors in intrinsic and extrinsic parameters as well as the localization area of the signal on the sky. This work is the successor of \cite{Borhanian:2022czq}, which performs a similar study for BBH and BNS systems. 
\begin{table}[htbp] 
  \centering
  \caption{\label{tab:net}The six next-generation ground-based GW detector networks that are included in the analysis, with the abbreviation used to refer to the network.}
  \renewcommand{\arraystretch}{1.5} 
  \resizebox{8.3cm}{!}{
    \begin{tabular}{ l  c }
    \hhline{==}
    Network & Detectors  \\
    \hhline{--}
    HLVKI\texttt{+} & LIGO (HL\texttt{+}), Virgo\texttt{+}, KAGRA\texttt{+}, LIGO-I\texttt{+}\\
    VK\texttt{+}HLIv & Virgo\texttt{+}, KAGRA\texttt{+}, LIGO (HLI-Voy)\\
    HLKI\texttt{+}E & LIGO (HL\texttt{+}), KAGRA\texttt{+}, LIGO-I\texttt{+}, ET\\
    VKI\texttt{+}C & Virgo\texttt{+}, KAGRA\texttt{+}, LIGO-I\texttt{+}, CE-North\\
    KI\texttt{+}EC & KAGRA\texttt{+}, LIGO-I\texttt{+}, ET, CE-North\\
    ECS & ET, CE-North, CE-South\\
    \hhline{==}
    \end{tabular}
    }
\end{table}

The FIM is applicable under the high-SNR or the linear signal approximation limit \cite{Vallisneri:2007ev}. This brings into question the applicability of our results for signals with low $(\sim 10)$ SNR. While, in general, the measurement errors from FIM are expected to be an underestimation compared to the errors obtained using Bayesian analysis even for systems detected with an SNR $\sim 20$ \cite{Balasubramanian:1995bm}, studies have found that this is not always true \cite{Rodriguez:2013mla} (also see Ref. \cite{Mandel:2014tca} which attributes some of the results in Ref. \cite{Rodriguez:2013mla} to truncation effects). Studies have also shown that the $90\%-$credible sky-area is underestimated by the FIM approach for a threshold SNR of $12$, but there is a broad agreement in the results between the estimated from FIM and Bayesian approach for a threshold SNR of $25$ \cite{Magee:2022kkc}. In light of these results, we are currently working towards a comprehensive comparison between the FIM and the Bayesian estimates of measurement errors for various binary parameters as a function of SNR. 

We begin by generating populations of NSBH binaries based on our assumptions of their properties. The parameters that characterize these populations are described in Sec. \ref{sec:pop_method}. The section also explains the methodology used to assess the measurement abilities of the networks. Next, we compare the detection capabilities of the six GW detector networks. In Sec. \ref{sec:eff_rate}, we calculate the efficiency of the detector networks and list the \textit{reach} for each network. Using the efficiency and the estimated `event-based' local merger rate density for NSBH systems, we calculate the yearly detection rate for each detector network.  In Sec. \ref{sec:measure}, we present the quality of measurement of the NSBH detections and the accuracy with which several intrinsic and extrinsic parameters can be measured. In particular, we estimate how well events can be localized in the sky, to assess the possibility of an EM follow-up of GW signals from NSBH mergers. The MMA prospects concerning NSBH detections are discussed in Sec. \ref{sec:mma}, where we give the number of systems for which \textit{early-warning} alerts can be sent to facilitate the EM follow-up, as well as the number of KN detections we can expect based on the population model, the GW detector network, the NS EOS, and the EM telescopes used. In Sec. \ref{sec:concl}, we summarize our results and present our conclusions regarding the science that can be extracted from the NSBH detections using GW detector networks.


\section{Population and Methodology} \label{sec:pop_method}
In order to evaluate the detection capabilities of the GW detector networks, we construct populations of NSBH systems and use the FIM approach to assess the performance of these networks in detecting GWs from the systems. Sec. \ref{subsec:inj_par} describes the properties of the populations and the rationale behind the assumptions that went into generating them. Sec. \ref{subsec:method} describes the FIM approach and lists the parameters that were used to compute the FIMs. 
\subsection{Injection parameters} \label{subsec:inj_par}
With the limited number of NSBH mergers detected, their population characteristics remain uncertain. While there are studies \cite{Chattopadhyay:2022cnp,Biscoveanu:2022iue,Zhu:2021jbw} that infer the mass and spin distributions from the set of detected NSBH events, the conclusions are susceptible to change with future detections. Due to the uncertainty in the properties of the actual population, we look at two populations to assess the science case of future GW detector networks. 

For the first population, hereinafter referred to as \textit{Pop-1}, we account for the fact that our knowledge of the NSBH population parameters is limited and choose broad distributions to describe the population. The black hole mass distribution is chosen to follow the \texttt{POWER+PEAK} \cite{LIGOScientific:2021psn} distribution between $[3\msun,100\msun]$ and the neutron star mass is sampled from a uniform distribution between $[1\msun,2.9\msun]$, where the upper bound on the NS mass has been set using Ref. \cite{Godzieba:2020tjn}. The spins of both NSs and BHs are assumed to be aligned with the orbital angular momentum of the binary. With $(\boldsymbol{\chi_1}, \boldsymbol{\chi_2})$ denoting the dimensionless spin vectors of the BH and the NS, this implies that $\chi_{1x} = \chi_{1y} = \chi_{2x} = \chi_{2y} = 0$. Neutron stars are assumed to be slowly rotating, with their spins chosen from a uniform distribution between $[-0.05,0.05]$, while BH spins are taken from a uniform distribution between $[-0.75,0.75]$. 

One must note that for Pop-1 the masses and spins of BHs and NSs are sampled independently, ignoring any correlations that may exist between their properties due to physical processes involved in binary formation and evolution. To account for possible correlations, the masses and spins in the second population, hereinafter referred to as \textit{Pop-2}, are derived from the \textit{fiducial} model in Ref. \cite{Broekgaarden:2021iew}. The fiducial model is a binary population synthesis model for NSBH systems that form through the isolated binary formation channel. The model was constructed using \texttt{COMPAS} \cite{COMPASTeam:2021tbl} and involved various physical assumptions summarized in Table 1 of Ref. \cite{Broekgaarden:2021iew}. For BH spins, we use metallicity-dependent fits provided in Eqs. (2) and (3) in Ref. \cite{Chattopadhyay:2022cnp} (restated in Appendix \ref{app:Pop2_params}). It is expected that the angular momentum transport at the time of compact object formation is quite effective in removing most of the rotational energy from the core, making the formed NS/BH nearly non-rotating, if born first. If close enough to its compact object companion, the progenitor of the second-born NS/BH can tidally spin up \cite{Qin:2018vaa,Chattopadhyay:2019xye,Chattopadhyay:2020lff,Bavera:2020inc}. Thus, the fit only applies to systems where the NS progenitor is formed first, allowing the progenitor of the second-born BH to have high spins as it can get tidally spun up by its companion. BHs are assumed to be non-spinning for the rest of the systems where the BH progenitor is formed first. Further, the NS spins are set to be zero. The BH and NS mass and spin profiles for Pop-2 are shown in Fig. \ref{appfig:Pop2_mass_spins} in Appendix \ref{app:Pop2_params}. 

For each population, we generate $250,000$ injections per redshift bin in each of the six redshift bins: $z \in [0.02, 0.05]$, $[0.05, 1]$, $[1, 2]$, $[2, 4]$, $[4, 10]$ and $[10, 50]$. The luminosity distance $D_L$ for each injection is obtained from the redshift $z$ using \texttt{ASTROPY.PLANCK18} \cite{Astropy:2013muo,Astropy:2018wqo}. $\cos(\iota)$ and $\cos(\delta)$, where $\iota$ and $\delta$ are the inclination angle and the declination respectively, are sampled uniformly between $[-1,1]$.  The right ascension $\alpha$ and the polarization angle $\psi$ are sampled from a uniform distribution with bounds $[0, 2\pi]$. $t_c$ and $\phi_c$ are the time and phase of coalescence respectively and can be chosen arbitrarily, we fix them to be $0$. The above-mentioned parameters are summarized in Table \ref{tab:pop_par}.
\begin{table*}[htbp] 
\centering
\caption{\label{tab:pop_par}Parameters that characterize the two populations used in this study to evaluate the detection capabilities of the future detectors.}
\renewcommand{\arraystretch}{1.5}
\begin{tabular}{ l | P{2.2cm}  P{2.2cm} |  P{2.7cm}  P{2.7cm} }
\hhline{=====}
\multirow{2}[2]{*}{Parameter} & \multicolumn{2}{c|}{Pop-1} & \multicolumn{2}{c}{Pop-2} \\
         & Neutron Star & Black Hole & Neutron Star & Black Hole \\
\hhline{-----}
Mass $m$ & [1,2.9] $\msun$ & [3,100] $\msun$ & [1.26,2.50] $\msun$ & [2.6,39.2] $\msun$ \\
\hhline{-----}
Mass Model & Uniform & \texttt{POWER+PEAK} \cite{LIGOScientific:2021psn} & \multicolumn{2}{c}{Derived from the fiducial model \cite{Broekgaarden:2021iew}}\\
\hhline{-----}
Spin $\chi$ & [-0.05,0.05] & [-0.75,0.75] & 0 & [0,1] \\
\hhline{-----}
\multirow{3}[2]{*}{Spin Model} & \multicolumn{2}{c|}{\multirow{2}[2]{*}{Aligned Uniform}} & \multirow{2}[2]{*}{Aligned} & Eqs. (2) and (3) in \\
  & & & & Ref. \cite{Chattopadhyay:2022cnp} (restated\\
  & & & & in Appendix \ref{app:Pop2_params})\\
\hhline{-----}
\multirow{2}[2]{*}{$z$} & \multicolumn{4}{c}{Uniform in six bins: [0.02,0.05],}\\
      &\multicolumn{4}{c}{[0.05,1], [1,2], [2,4], [4,10] and [10,50]} \\
\hhline{-----}
$D_L$ & \multicolumn{4}{c}{$z$ converted using \texttt{ASTROPY.Planck18}}  \\
\hhline{-----}
$\cos(\iota)$ & \multicolumn{4}{c}{Uniform in [-1,1]} \\
\hhline{-----}
$\alpha$ & \multicolumn{4}{c}{Uniform in [0,2$\pi$]} \\
\hhline{-----}
$\cos(\delta)$ & \multicolumn{4}{c}{Uniform in [-1,1]} \\
\hhline{-----}
$\psi$ & \multicolumn{4}{c}{Uniform in [0,2$\pi$]} \\
\hhline{-----}
$t_c$, $\phi_c$ & \multicolumn{4}{c}{0} \\
\hhline{=====}
\end{tabular}
\end{table*}

Once the $250,000$ injections per redshift bin are generated, we logarithmically divide the total redshift range, i.e., $[0.02,20]$, into $50$ bins and randomly pick events from each of the finer bins according to the merger rate corresponding to that bin. This allows us to have a random collection of events in each bin and mitigate selection biases. The calculation of the merger rate as a function of redshift is given in Sec. \ref{subsec:det_rate}. The end product is the population of NSBH mergers assuming an observation time of $10$ years for all the networks.

\subsection{Methodology} \label{subsec:method}
The detector response to a GW signal is given by,
\begin{equation}
\begin{split}
    h^{(A)} (t,\boldsymbol{\mu})&=\,\,F^{(A)}_{+} (\alpha, \delta, \psi, \boldsymbol{\beta})\,h^{(A)}_{+} (t,\boldsymbol{\lambda}) \\
    &+\,\,F^{(A)}_{\times} (\alpha, \delta, \psi, \boldsymbol{\beta})\,h^{(A)}_{\times} (t,\boldsymbol{\lambda}),
\end{split}
\end{equation}
where $h_{+}$ and $h_{\times}$ are the two GW polarizations and $F_{+}$ and $F_{\times}$ are the \textit{antenna pattern functions}. The index $A$ denotes the detector. The antenna pattern functions depend on variables that describe the location of the source of GWs in the sky, i.e., $\alpha$ and $\delta$, the polarization angle $\psi$, and the variables that describe the location of the detector itself, $\boldsymbol{\beta}$ (Tab. III in Ref. \cite{Borhanian:2020ypi} lists the angles that describe the location for several detectors). The strain for each polarization depends on time $t$, the time and phase of coalescence, $t_c$ and $\phi_c$, and on the parameters that describe the source of GWs, $\boldsymbol{\lambda} = \{\mathcal{M}, \eta, \boldsymbol{\chi_1}, \boldsymbol{\chi_2}, \iota, D_L\}$, where $\mathcal{M}$ is the chirp mass, $\eta$ is the symmetric mass ratio, $\iota$ is the inclination angle and $D_L$ is the luminosity distance of the binary. Since we have assumed both the components of the binary to have aligned spins, $(\boldsymbol{\chi_1}, \boldsymbol{\chi_2}) = (\chi_{1z},\chi_{2z})$. For a given detector, $\boldsymbol{\beta}$ is fixed. Thus, the strain $h$ is a time-dependent function of $\boldsymbol{\mu} = \{\alpha, \delta, \psi, \mathcal{M}, \eta, \chi_{1z},\chi_{2z}, \iota, D_L,t_c,\phi_c\}.$

We use the FIM  to estimate the error in the measurement of these parameters using future ground-based GW detectors. The FIM calculation assumes the detector noise to be Gaussian and provides an analytical way to obtain the errors in the form of a covariance matrix $\Sigma$
\begin{equation}
    \Sigma_{ij} = \Gamma^{-1}_{ij} = \left(\frac{\partial h}{\partial \theta_{i}},\frac{\partial h}{\partial \theta_{j}}\right)^{-1},
\end{equation}
where $h$ is the GW waveform, $\theta_i$ is the $i^{th}$ parameter in $\boldsymbol{\mu}$, $(\cdot\,,\cdot)$ is the noise-weighted inner product and $\Gamma$ is the FIM. To obtain the FIM and the measurement errors, we use \texttt{GWBENCH} \cite{Borhanian:2020ypi}, which is a publicly available Python package that calculates the covariance matrix by numerically inverting the FIM. The package can apply numerical differentiation recipes to the GW waveforms that are part of the LIGO Algorithm Library (LAL) \cite{lalsuite}. For our study, we use the \texttt{IMRPhenomXHM} \cite{Garcia-Quiros:2020qpx} waveform model, which is a frequency-domain waveform for non-precessing BBH systems and includes contributions from higher-order modes. The waveform is suitable for BBH systems and cannot account for the tidal effects that can manifest in NSBH mergers. As tidal effects appear at the fifth post-Newtonian (PN) order and only contribute to the strain near the merger, we do not expect their exclusion in the calculation of FIM to significantly alter the results presented in this work. Additionally, we prefer using \texttt{IMRPhenomXHM} instead of a traditional NSBH-suited waveform like \texttt{PhenomNSBH} \cite{Thompson:2020nei} due to the inclusion of higher-modes in the former model which is known to improve the estimation of parameters \cite{Kapadia:2020kss} and allows us to obtain a more realistic estimate of the measurement capabilities of detector networks. To calculate the FIM, the derivatives are taken with respect to parameters $\boldsymbol{\mu}$. \texttt{GWBENCH} also provides the SNR and the $90\%$-credible sky localization error $\Omega_{90}$, which are used to compare the performance of detector networks. 


\section{Network efficiency and detection rate} \label{sec:eff_rate}
With increased detector sensitivities, we not only expect to probe the universe up to higher redshifts but also expect to detect events with unprecedented SNRs. Our expectations can be quantified in terms of the \textit{network efficiency} and \textit{detection rate}.
\subsection{Network efficiency} \label{subsec:eff}
We first introduce the notion of \textit{matched-filtering} SNR and then use it to define the efficiency of a network. The matched-filter SNR $\rho_A$ corresponding to a signal incident on a detector $A$ is defined by the square root of,
\begin{equation}
    \rho^{2}_A = \int_{f_L}^{f_U} \frac{|\Tilde{h}_A(f)|^2}{S^{A}_{n} (f)} df,
\end{equation}
where $\Tilde{h}_A (f)$ is the frequency-domain waveform obtained by taking the Fourier transform of $h_A (t)$, $S^{A}_{n} (f)$ is the power spectral density (PSD) of detector A and $f_L$ and $f_U$ are the lower and upper-frequency cutoffs. Then, for a network with $N$ detectors, the matched-filtering SNR $\rho$ is given by the square root of,
\begin{equation}
    \rho^2 = \sum_{i = 1}^{N} \rho^{2}_{i}.
\end{equation}
At a given redshift, the network efficiency is defined as the fraction of events (at that redshift) that are detected by the network with a matched-filtering SNR greater than the threshold SNR $\rho_{*}$. We calculate the efficiency of networks listed in Table \ref{tab:net} as a function of redshift for two threshold SNRs: $\rho_{*} = 10$ and $\rho_{*} = 100$. The threshold $\rho_{*} = 10$ corresponds to the SNR above which we claim detection and $\rho_{*} = 100$ gives a measure of high-SNR events that can be detected. To calculate the efficiency, we logarithmically divide the total redshift range, from $z = 0.005$ to $z = 20$, into $50$ bins. For each bin  $[z,z+dz]$, the efficiency $\epsilon$ is calculated by,
\begin{equation}
    \epsilon\,(\rho_{*}, z) = \frac{1}{N_z} \sum_{k=1}^{N_z} H\,(\rho_k - \rho_{*}),
\end{equation}
where $N_z$ is the number of events in that redshift bin and $H (x)$ is the Heavyside function. The efficiency for the detector networks as a function of redshift for the two threshold SNRs is given in the left panel of Fig. \ref{fig:eff_rate}. The efficiency curves are well estimated using the three-parameter \textit{sigmoid} functions
\begin{equation} \label{eq:sigmoid}
    f_{\mbox{sigmoid}} = \left(\frac{1+b}{1+b\,e^{a x}}\right)^c.
\end{equation}
The best fit values for $a$, $b$, and $c$ for each efficiency curve are listed in Appendix \ref{app:fit_sigmoid}. Note that Fig. \ref{fig:eff_rate} only shows the efficiency and the detection rate for Pop-1, since they are almost identical for Pop-2. This is also evident in the similarity between the fitting coefficients of the sigmoid functions that approximate the efficiency curves for the two populations (see Table \ref{tab:sigmoid_fit_pars} in Appendix \ref{app:fit_sigmoid}).

To compare the detector networks based on their detection efficiencies, we a measure called the \textit{reach} of the detector. The reach ($z_r$) is defined as the redshift at which the efficiency of the network drops to $0.5$, i.e., it is the redshift at which only half of all the coalescence events will be detected with SNR greater than $\rho_{*}$. As the efficiency is monotonic, this implies that the reach is equal to the redshift at which at least 50\% of the events \textit{within} that redshift will be observed with SNR greater than $\rho_{*}$. We obtain the value of reach using the sigmoid curves in Eq. (\ref{eq:sigmoid}). It must be noted that the definition of reach varies across literature and care must be taken when comparing results from different studies.

The values for the reach of the networks for both populations are given in Table \ref{tab:reach_horizon}. For the detection of NSBH systems in Pop-1, i.e., $\rho_{*} = 10$, the \texttt{HLVKI+} network has a reach of $z = 0.21 \sim 1$ Gpc and is expected to miss events beyond $\sim 6$ Gpc. As most of the high-SNR events are the ones that are situated close-by, the reach for $\rho_{*} = 100$ drops to $z = 0.022$. The Voyager upgrades improve the reach by a factor of $\sim 2$ for both population models. All networks in our study that have at least one of the ET or CE detectors have a reach $z_r > 1$, with the ECS network having the longest detection range of all, with a reach of $z = 6$, probing the star formation rate and metallicity distribution up to high redshifts.

From Table \ref{tab:reach_horizon}, one can also observe the relationship between the reach of the detector and SNR threshold for different detector networks. Specifically, for networks like \texttt{HLVKI+} and \texttt{VK+HLIv} that probe only the nearby universe, the reach can be seen to scale as $\sim 1/\rho$. This is because $\rho \propto 1/D_L$ and, for the local universe, $z$ scales linearly with $D_L$. For the most advanced networks like \texttt{KI+EC} and \texttt{ECS} which are sensitive enough to probe the universe up to higher redshifts, this relationship breaks down as the redshift no longer evolves linearly with $D_L$, with contributions from dark matter and dark energy density.
\begin{table}[htbp] 
  \centering
  \caption{\label{tab:reach_horizon}The reach for the six detector networks for the cases when the threshold SNR $\rho_{*} = 10$ and $\rho_{*} = 100$.}
  \renewcommand{\arraystretch}{1.3} 
    \begin{tabular}{l | P{1.5cm} P{1.5cm} | P{1.5cm} P{1.5cm}}
    \hhline{=====}
    \multirow{2}{*}{Network} & \multicolumn{2}{c|}{\textit{Pop-1}} & \multicolumn{2}{c}{\textit{Pop-2}} \\
    & $\rho_{*} = 10$ & $\rho_{*} = 100$ & $\rho_{*} = 10$ & $\rho_{*} = 100$ \\
    \hhline{-----}
    HLVKI\texttt{+} & 0.21 & 0.022 & 0.23 & 0.023\\
    VK\texttt{+}HLIv & 0.47 & 0.045 & 0.5 & 0.049\\
    HLKI\texttt{+}E & 1.6 & 0.12 & 1.6 & 0.13\\
    VKI\texttt{+}C & 2.7 & 0.19 & 2.9 & 0.2\\
    KI\texttt{+}EC & 3.9 & 0.24 & 4.2 & 0.25\\
    ECS & 6 & 0.32 & 6.6 & 0.34\\
    \hhline{=====}
    \end{tabular}
\end{table}

\subsection{Detection Rate} \label{subsec:det_rate}
\begin{figure*}[htbp] 
\centering
\includegraphics[scale=0.58]{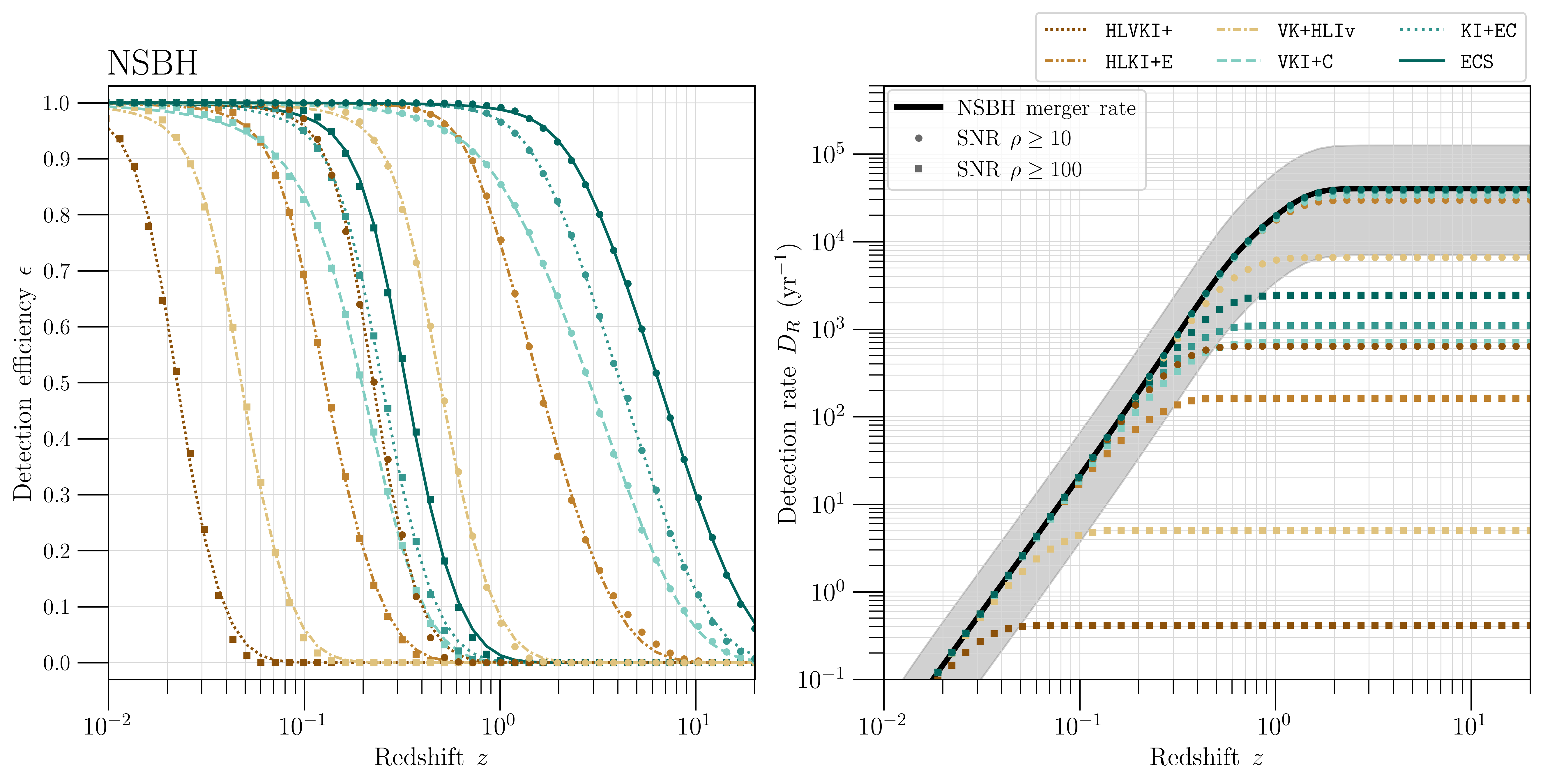}
\caption{\label{fig:eff_rate}\textit{Left panel:} The network efficiency curves for the six next-generation detector networks. The markers represent the efficiency at corresponding redshift values, and the lines are the \textit{best-fit} sigmoid functions which are good approximations of the efficiency curves.  \textit{Right panel:} The detection rate as a function of redshift for the detector networks. The black solid line refers to the total NSBH merger rate. The gray shaded area shows the variation in the total merger rate due to the uncertainty in the value of the local merger rate density.}
\end{figure*}

The detection rate is defined as the number of NSBH mergers up to a given redshift that are detected by a network in a year with a matched-filtering SNR greater than the threshold SNR. It depends on the total merger rate as well as the efficiency of the network. In the detector frame, the total merger rate $R_d$ up to redshift $z$ is given by
\begin{equation} \label{eq:merger_rate_density}
    R_{d} (z) = \int_{0}^{z} \frac{1}{(1+z')}\frac{dR}{dz'} = \int_{0}^{z} \frac{\Dot{n} (z')}{(1+z')} \frac{d V}{d z'}\,dz',
\end{equation}
where $\Dot{n}(z')$ is the merger rate density in the comoving frame, $dV/dz'$ is the comoving volume element (which itself is a function of redshift) and the $(1+z')$ term in the denominator converts the detection rate from the source frame to the detector frame by accounting for the time dilation. Among these mergers, only a fraction are detected by a network, which is determined by the efficiency of the network. Hence, the detection rate $D_R (z)$ is given by
\begin{equation}
D_R (z) = \int_{0}^{z} \epsilon (z',\rho_{*}) \frac{\Dot{n} (z')}{(1+z')} \frac{d V}{d z'}\,dz'.
\end{equation}
Note that the calculation of the detection rate depends on the model used to calculate the merger rate density $\Dot{n}(z)$. Further, $\Dot{n}(z)$ is calibrated by setting the local $(z=0)$ value equal to the observed local merger rate density for NSBH systems. We follow the SFR model described in Ref. \cite{Yuksel:2008cu} which utilizes the gamma-ray burst data to calculate the SFR up to high redshifts. In addition to the SFR, there is a time delay between the formation and the merger of compact binaries, which is described by various time-delay models \cite{Wanderman:2014eza,Virgili:2009ca,deFreitasPacheco:2005ub}. Following Ref. \cite{Zhu:2020ffa}, we choose the log-normal time delay model proposed in Ref. \cite{Wanderman:2014eza} for our study. The redshift distribution based on these assumptions can be expressed in an analytical form \cite{Zhu:2020ffa,Sun:2015bda} and is given in Appendix \ref{app:redshift_dist}. The inferred local merger rate density is reported to lie between 7.8--140 $\mbox{Gpc}^{-3}$ $\mbox{yr}^{-1}$ \cite{LIGOScientific:2021psn}. We calibrate $\Dot{n}(z)$ by fixing the local merger rate density, $\Dot{n}(0) = 45$ $\mbox{Gpc}^{-3}$ $\mbox{yr}^{-1}$, which is the median event-based NSBH merger rate density reported in Ref. \cite{LIGOScientific:2021qlt}, calculated assuming that the observed NSBH systems are a representative of the NSBH population. The obtained curves for detection rate as a function of redshift for the six detector networks are shown in the right panel of Fig. \ref{fig:eff_rate}. The grey region shows the uncertainty in the NSBH merger rate which is due to the uncertainty in the value of the local merger rate density.
Based on the model used, we obtain the cosmic merger rate for NSBH systems, i.e., the number of NSBH mergers up to $z=20$ to be $4.0^{+8.5}_{-3.3} \times 10^{4}$ $\mbox{yr}^{-1}$, where the upper and lower bounds are calculated using the upper and lower bounds of the local merger rate density.

With $\Dot{n}(0) = 45$ $\mbox{Gpc}^{-3}$ $\mbox{yr}^{-1}$, from the right panel of Fig. \ref{fig:eff_rate}, we expect $\mathcal{O}(100)$ detections in \texttt{HLVKI+}, $\mathcal{O}(10^3)$ for \texttt{VK+HLIv} and $\mathcal{O}(10^4)$ for \texttt{HLKI+E}, \texttt{VKI+C}, \texttt{KI+EC} and \texttt{ECS}, with SNR $\rho > 10$ every year. Thus, every network that has at least one XG detector is expected to observe $\mathcal{O}(10^4)$ events every year. Moreover, the \texttt{ECS} network is expected to detect $~ 97\%$ of the cosmic population of NSBH mergers with SNR $\rho > 10$.


\section{Measurement quality and sky localization} \label{sec:measure}
In Sec. \ref{sec:eff_rate}, we noted that the XG networks will not only detect a large number of events, but a significant number of these events will also be detected at high SNRs. Using the methods described in Secs. \ref{subsec:inj_par} and \ref{subsec:det_rate}, we construct the cosmic population of NSBH systems. The expected number of events detected with SNR greater than $10$, $30$, and $100$, are presented in Table \ref{tab:pop_snr}. When comparing the numbers for the Pop-1 and Pop-2 populations, we find them to be of the same order. However, there are more events with higher SNRs in Pop-1 as compared to Pop-2. This is attributed to the different BH mass models in the two populations: about $0.5\%$ of the events in Pop-2 have BH mass greater than $20\msun$, whereas, in Pop-1, $\sim 17\%$ of the events have BH mass greater than $20\msun$. This, along with the fact that higher total mass binaries are expected to be detected at higher SNRs, explains the small differences in the number of systems belonging to the two populations detected with SNRs greater than certain threshold SNRs. The \texttt{HLVKI+} detector is expected to detect $\mathcal{O}(100)$ events with SNR $>10$ and $\mathcal{O}(10)$ with SNR $>30$ every year, but it is unlikely to detect any NSBH mergers with SNR greater than $100$. However, improving the sensitivities of the three LIGO detectors with Voyager upgrades results in the detection of $\mathcal{O}(10^3)$, $\mathcal{O}(100)$ and $\mathcal{O}(1)$ events every year with SNR $>10$, $30$ and $100$ respectively. The inclusion of the Einstein Telescope in the \texttt{HLKI} network with A$+$ sensitivities further improves this number to $\mathcal{O}(10^4)$ detections with SNR $>10$ every year, and $\mathcal{O}(100)$ detections with SNR $>100$ every year. The remaining three networks all of which contain the \texttt{CE-North} detector, i.e., \texttt{VKI+C}, \texttt{KI+EC} and \texttt{ECS} are expected to detect $\mathcal{O}(10^5)$ and $\mathcal{O}(10^4)$ events every year with SNR $>10$ and $30$ respectively. The \texttt{ECS} network is expected to detect the most number of NSBH mergers with SNR $>100$, detecting $\mathcal{O}(10^3)$ events every year, an order of magnitude greater than the expected values for the \texttt{VKI+C} network.
\begin{table}[htbp] 
  \centering
  \caption{\label{tab:pop_snr}The cosmic merger rate per year and the number of events that are detected every year with SNRs greater than $10$, $30$, and $100$ for the six detector networks. The lower and upper bounds in the reported numbers are calculated using the uncertainty in the local merger rate density for NSBH mergers.}
  \renewcommand{\arraystretch}{1.5} 
    \begin{tabular}{l P{2cm}P{2cm}P{2cm}}
    \hhline{====}
    Cosmic Rate & \multicolumn{3}{c}{$4.0^{+8.5}_{-3.3} \times 10^4$ yr$^{-1}$} \\
    \hhline{----}
    SNR $\rho$ & $> 10$ & $> 30$ & $> 100$ \\
    \hhline{----}
    \multicolumn{4}{c}{\textit{Pop-1}}\\
    \hhline{----}
    HLVKI\texttt{+} & $5.1^{+10.8}_{-4.3} \times 10^{2}$ & $1.5^{+3.1}_{-1.2} \times 10$ & $0.0^{+0.5}_{-0.0}$ \\
    VK\texttt{+}HLIv & $5.7^{+12.1}_{-4.7} \times 10^{3}$ & $2.1^{+4.6}_{-1.8} \times 10^{2}$ & $4.1^{+8.8}_{-3.7}$ \\
    HLKI\texttt{+}E & $2.9^{+6.0}_{-2.4} \times 10^{4}$ & $5.5^{+11.6}_{-4.5} \times 10^{3}$ & $1.7^{+3.4}_{-1.4} \times 10^{2}$ \\
    VKI\texttt{+}C & $3.3^{+7.1}_{-2.8} \times 10^{4}$ & $1.3^{+2.7}_{-1.0} \times 10^{4}$ & $6.6^{+14.3}_{-5.5} \times 10^{2}$ \\
    KI\texttt{+}EC & $3.8^{+8.0}_{-3.1} \times 10^{4}$ & $1.8^{+3.8}_{-1.5} \times 10^{4}$ & $1.0^{+2.2}_{-0.8} \times 10^{3}$ \\
    ECS & $3.9^{+8.3}_{-3.3} \times 10^{4}$ & $2.5^{+5.2}_{-2.0} \times 10^{4}$ & $2.3^{+4.8}_{-1.9} \times 10^{3}$ \\
    \hhline{----}
    \multicolumn{4}{c}{\textit{Pop-2}}\\
    \hhline{----}
    HLVKI\texttt{+} & $4.8^{+10.2}_{-4.0} \times 10^{2}$ & $1.1^{+2.9}_{-1.0} \times 10$ & $0.0^{+0.4}_{-0.0}$ \\
    VK\texttt{+}HLIv & $5.5^{+11.6}_{-4.5} \times 10^{3}$ & $1.9^{+4.0}_{-1.6} \times 10^{2}$ & $3.7^{+8.5}_{-3.3}$ \\
    HLKI\texttt{+}E & $2.8^{+6.0}_{-2.3} \times 10^{4}$ & $5.0^{+10.6}_{-4.1} \times 10^{3}$ & $1.2^{+2.6}_{-1.0} \times 10^{2}$ \\
    VKI\texttt{+}C & $3.3^{+7.0}_{-2.7} \times 10^{4}$ & $1.2^{+2.6}_{-1.0} \times 10^{4}$ & $5.4^{+11.3}_{-4.5} \times 10^{2}$ \\
    KI\texttt{+}EC & $3.8^{+8.0}_{-3.1} \times 10^{4}$ & $1.7^{+3.6}_{-1.4} \times 10^{4}$ & $8.5^{+18.0}_{-7.0} \times 10^{2}$ \\
    ECS & $3.9^{+8.3}_{-3.2} \times 10^{4}$ & $2.4^{+5.1}_{-2.0} \times 10^{4}$ & $1.9^{+4.1}_{-1.6} \times 10^{3}$ \\
    \hhline{====}
    \end{tabular}
\end{table}

\begin{figure*}
\centering
\includegraphics[scale=0.33]{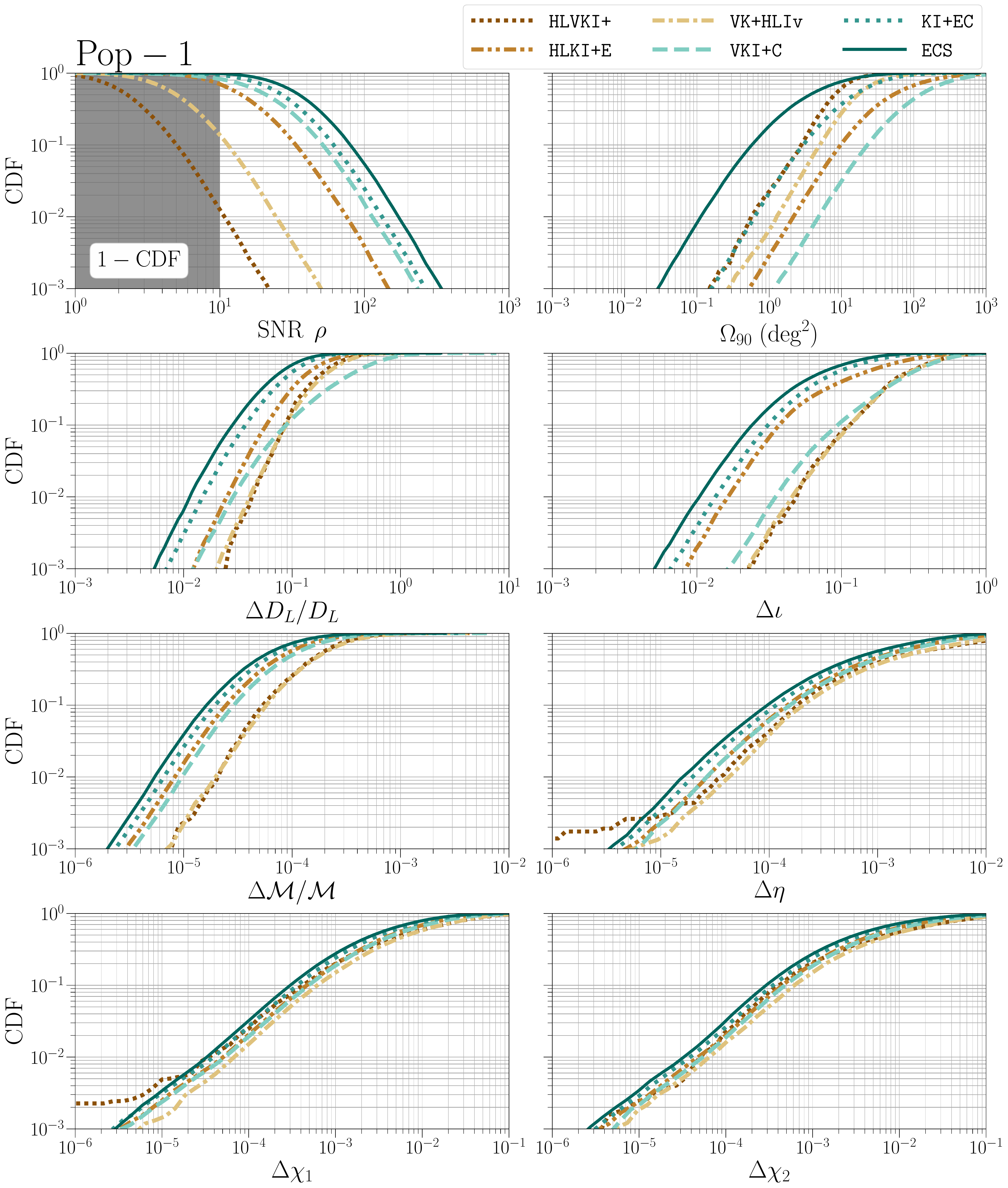}
\caption{\label{fig:cdf_measure}The cumulative density function (CDF) plots showing the trends in SNR $\rho$ and sky-localization $\Omega_{90}$ of the detected events in Pop-1. It also shows the CDFs for fractional errors in chirp mass and luminosity distance, i.e., $\Delta \mathcal{M}/\mathcal{M}$ and $\Delta D_L/D_L$, and absolute errors in inclination angle, symmetric mass ratio and the spins of the BH and the NS, i.e., $\Delta \iota$, $\Delta \eta$, $\Delta \chi_1$ and $\Delta \chi_2$, respectively.}
\end{figure*}

\begin{figure*}
\centering
\includegraphics[scale=0.33]{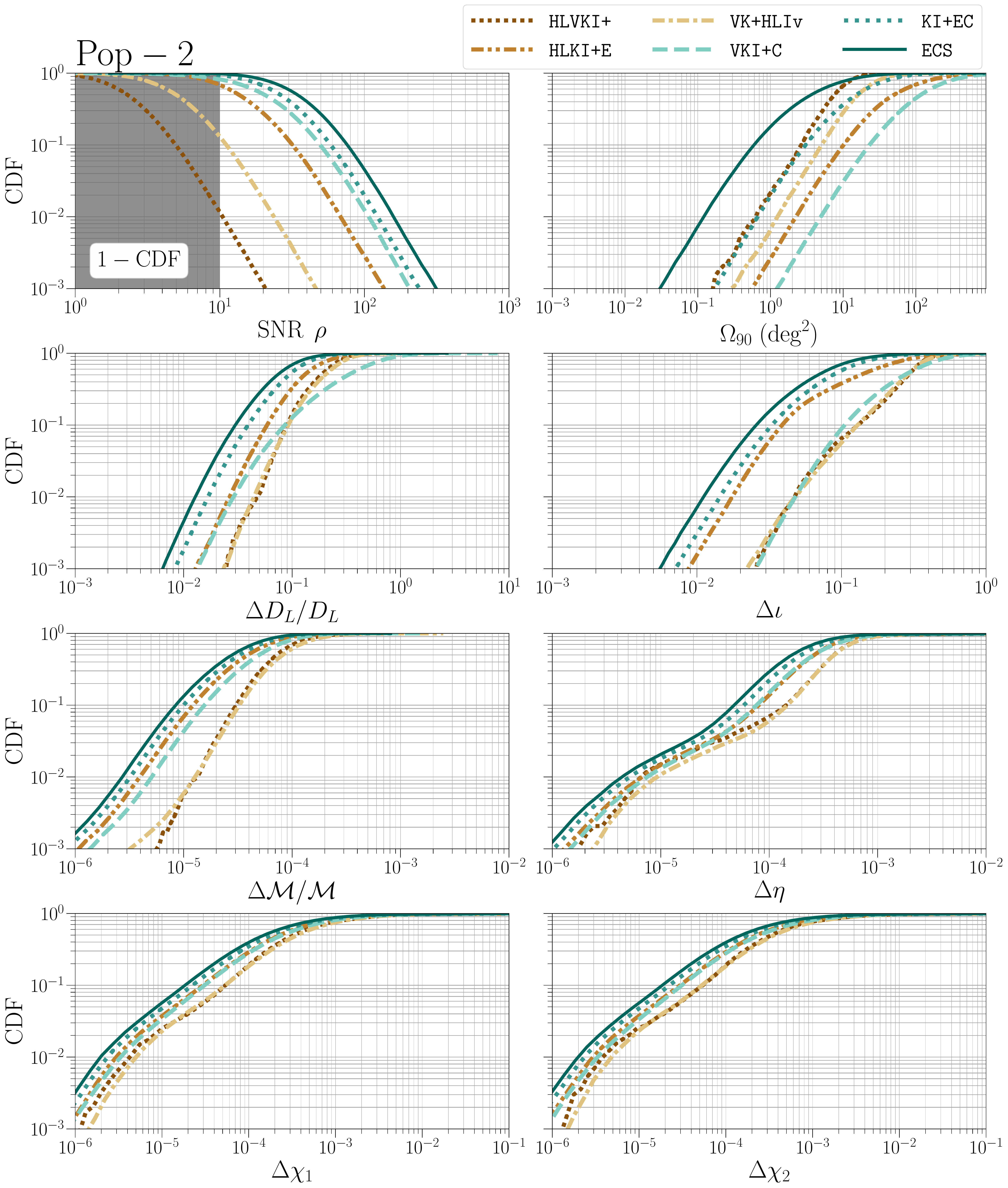}
\caption{\label{fig:cdf_measure2}The cumulative density function (CDF) plots for the SNR, sky-localization, fractional errors in chirp mass and luminosity distance, and absolute errors in the inclination angle, symmetric mass ratio and the spins of the BH and the NS for the detected events in Pop-2.}
\end{figure*}

High fidelity events, i.e., events that are detected with a large SNR, allow for accurate estimation of parameters, like masses and spins, which not only aid in differentiating NSBH from BBH and BNS mergers (based on component masses) but also in testing the predictions of general relativity. Further, high-precision measurements of the masses and spins of the compact objects will unravel the population characteristics of the NSBH systems and help test the predictions of various channels that explain the formation of such systems. Given that a fraction of NSBH mergers is also expected to result in the generation of kilonovae, accurate luminosity distance and sky area measurements will allow an independent measurement of the Hubble constant up to greater distances than BNS systems. In Figs. \ref{fig:cdf_measure} and \ref{fig:cdf_measure2}, we present the cumulative density function (CDF) plots portraying the parameter measurement abilities of the six detector networks for Pop-1 and Pop-2 respectively. In particular, we present the CDFs for SNR $\rho$, $90\%$-credible sky area $\Omega_{90}$, fractional error in the luminosity distance $\Delta D_L / D_L$, absolute error in the inclination angle $\Delta \iota$ (in radians), fractional error in the chirp mass $\Delta \mathcal{M}/\mathcal{M}$, absolute error in the symmetric mass ratio $\Delta \eta$ and the absolute errors in the dimensionless aligned spins of the black hole and the neutron star, i.e., $\Delta \chi_1$ and $\Delta \chi_2$ respectively.

From Figs. \ref{fig:cdf_measure} and \ref{fig:cdf_measure2}, we see that the overall trend in measurement quality when comparing different detector networks is the same for both populations. From the plots for SNR, we see that both \texttt{ECS} and \texttt{KI+EC} detect almost all the NSBH mergers with $\rho > 10$, whereas \texttt{HLKI+E} and \texttt{VKI+C} detect only $\sim 80\%$ of the events with $\rho > 10$.  The fraction falls to $\sim 1\%$ for the \texttt{HLVKI+} network. Voyager upgrades improve the detectability by an order of magnitude, detecting $\sim 15\%$ of the events. For the fractional error in $\mathcal{M}$ and the absolute error in $\eta$, we see that all the networks detect almost all the events with $\Delta \mathcal{M}/ \mathcal{M}$ better than $10^{-4}$ and $\Delta \eta$ better than $10^{-3}$, going up to precision of $\mathcal{O}(10^{-8})$ and $\mathcal{O}(10^{-7})$ respectively for $\mathcal{O}(10)$ events. This points to the estimation of the binary masses with unprecedented precision using future GW detector networks. Further, for all detected events, the spins of both the compact objects can be measured better than an absolute error of $10^{-2}$. This precision in spin measurements will uncover the spin distributions of BHs and NSs involved in NSBH mergers and shed light on the physics involved in the formation of these binaries.  

\begin{table*}
  \centering
  \caption{\label{tab:pop_sa90_logDL}The number of detections per year for the six detector networks with $90\%$-credible sky area less than $10$, $1$, $0.1$ and $0.01$ $\mbox{deg}^2$ and fractional error in luminosity distance less than $0.1$ and $0.01$.}
  \renewcommand{\arraystretch}{1.5} 
    \begin{tabular}{l | P{2.1cm} P{2.1cm} P{2.3cm} P{2cm} | P{2.1cm} P{2.3cm}}
    \hhline{=======}
    Metric & \multicolumn{4}{c|}{$\Omega_{90}\mbox{ (deg)}^2$} & \multicolumn{2}{c}{$\Delta D_L / D_L$} \\
    \hhline{-------}
    Quality & $\leq 10$ & $\leq 1$ & $\leq 0.1$ & $\leq 0.01$ & $\leq 0.1$ & $\leq 0.01$ \\
    \hhline{-------}
    \multicolumn{7}{c}{\textit{Pop-1}} \\
    \hhline{-------}
    HLVKI\texttt{+} & $2.9^{+6.1}_{-2.4} \times 10^{2}$ & $8.7^{+19.5}_{-7.3}$ & $2.0^{+2.0}_{-2.0} \times 10^{-1}$ & $0$ & $6.1^{+12.8}_{-4.8} \times 10$ & $0.0^{+0.1}_{-0.0}$ \\
    VK\texttt{+}HLIv & $1.4^{+2.9}_{-1.1} \times 10^{3}$ & $3.1^{+7.1}_{-2.6} \times 10$ & $6.0^{+13.0}_{-6.0} \times 10^{-1}$ & $0$ & $5.9^{+12.7}_{-4.9} \times 10^{2}$ & $3.0^{+9.0}_{-3.0} \times 10^{-1}$ \\
    HLKI\texttt{+}E & $2.4^{+5.1}_{-2.0} \times 10^{3}$ & $6.0^{+13.4}_{-4.9} \times 10$ & $1.9^{+2.3}_{-1.9}$ & $0$ & $8.2^{+17.5}_{-6.8} \times 10^{3}$ & $9.9^{+22.1}_{-8.4}$ \\
    VKI\texttt{+}C & $8.1^{+17.3}_{-6.6} \times 10^{2}$ & $1.9^{+4.1}_{-1.6} \times 10$ & $2.0^{+6.0}_{-2.0} \times 10^{-1}$ & $0$ & $3.1^{+6.5}_{-2.5} \times 10^{3}$ & $1.1^{+2.2}_{-0.9} \times 10$ \\
    KI\texttt{+}EC & $1.2^{+2.6}_{-1.0} \times 10^{4}$ & $6.1^{+13.0}_{-5.0} \times 10^{2}$ & $1.4^{+3.1}_{-1.2} \times 10$ & $2.0^{+4.0}_{-2.0} \times 10^{-1}$ & $1.9^{+4.0}_{-1.6} \times 10^{4}$ & $7.6^{+15.0}_{-6.0} \times 10$ \\
    ECS & $2.9^{+6.1}_{-2.4} \times 10^{4}$ & $6.1^{+13.0}_{-5.0} \times 10^{3}$ & $2.4^{+5.1}_{-1.9} \times 10^{2}$ & $5.1^{+9.8}_{-4.2}$ & $2.5^{+5.4}_{-2.1} \times 10^{4}$ & $1.8^{+3.7}_{-1.5} \times 10^{2}$ \\
    \hhline{-------}
    \multicolumn{7}{c}{\textit{Pop-2}} \\
    \hhline{-------}
    HLVKI\texttt{+} & $3.0^{+6.4}_{-2.5} \times 10^{2}$ & $9.1^{+20.3}_{-7.7}$ & $0.0^{+0.3}_{-0.0}$ & $0$ & $4.4^{+8.3}_{-3.7} \times 10$ & $0.0^{+0.1}_{-0.0}$ \\
    VK\texttt{+}HLIv & $1.4^{+2.9}_{-1.1} \times 10^{3}$ & $3.2^{+6.8}_{-2.6} \times 10$ & $4.0^{+15.0}_{-4.0} \times 10^{-1}$ & $0$ & $4.7^{+9.7}_{-3.9} \times 10^{2}$ & $0.0^{+0.3}_{-0.0}$ \\
    HLKI\texttt{+}E & $2.3^{+4.9}_{-1.9} \times 10^{3}$ & $5.9^{+12.2}_{-4.8} \times 10$ & $9.0^{+35.0}_{-8.0} \times 10^{-1}$ & $0$ & $8.2^{+17.3}_{-6.7} \times 10^{3}$ & $7.1^{+16.3}_{-6.2}$ \\
    VKI\texttt{+}C & $8.2^{+17.2}_{-6.7} \times 10^{2}$ & $1.8^{+3.8}_{-1.5} \times 10$ & $3.0^{+5.0}_{-3.0} \times 10^{-1}$ & $0$ & $3.3^{+6.9}_{-2.7} \times 10^{3}$ & $4.8^{+8.3}_{-4.0}$ \\
    KI\texttt{+}EC & $1.2^{+2.6}_{-1.0} \times 10^{4}$ & $6.0^{+12.6}_{-4.9} \times 10^{2}$ & $1.2^{+3.1}_{-1.0} \times 10$ & $0.0^{+0.3}_{-0.0}$ & $1.9^{+4.0}_{-1.6} \times 10^{4}$ & $3.7^{+7.5}_{-3.1} \times 10$ \\
    ECS & $2.9^{+6.2}_{-2.4} \times 10^{4}$ & $6.1^{+12.8}_{-5.0} \times 10^{3}$ & $2.3^{+4.9}_{-1.9} \times 10^{2}$ & $4.5^{+11.7}_{-3.8}$ & $2.6^{+5.5}_{-2.1} \times 10^{4}$ & $9.1^{+19.0}_{-7.6} \times 10$ \\
    \hhline{=======}
    \end{tabular}
\end{table*}

\begin{figure*}
  \includegraphics[scale=0.6]{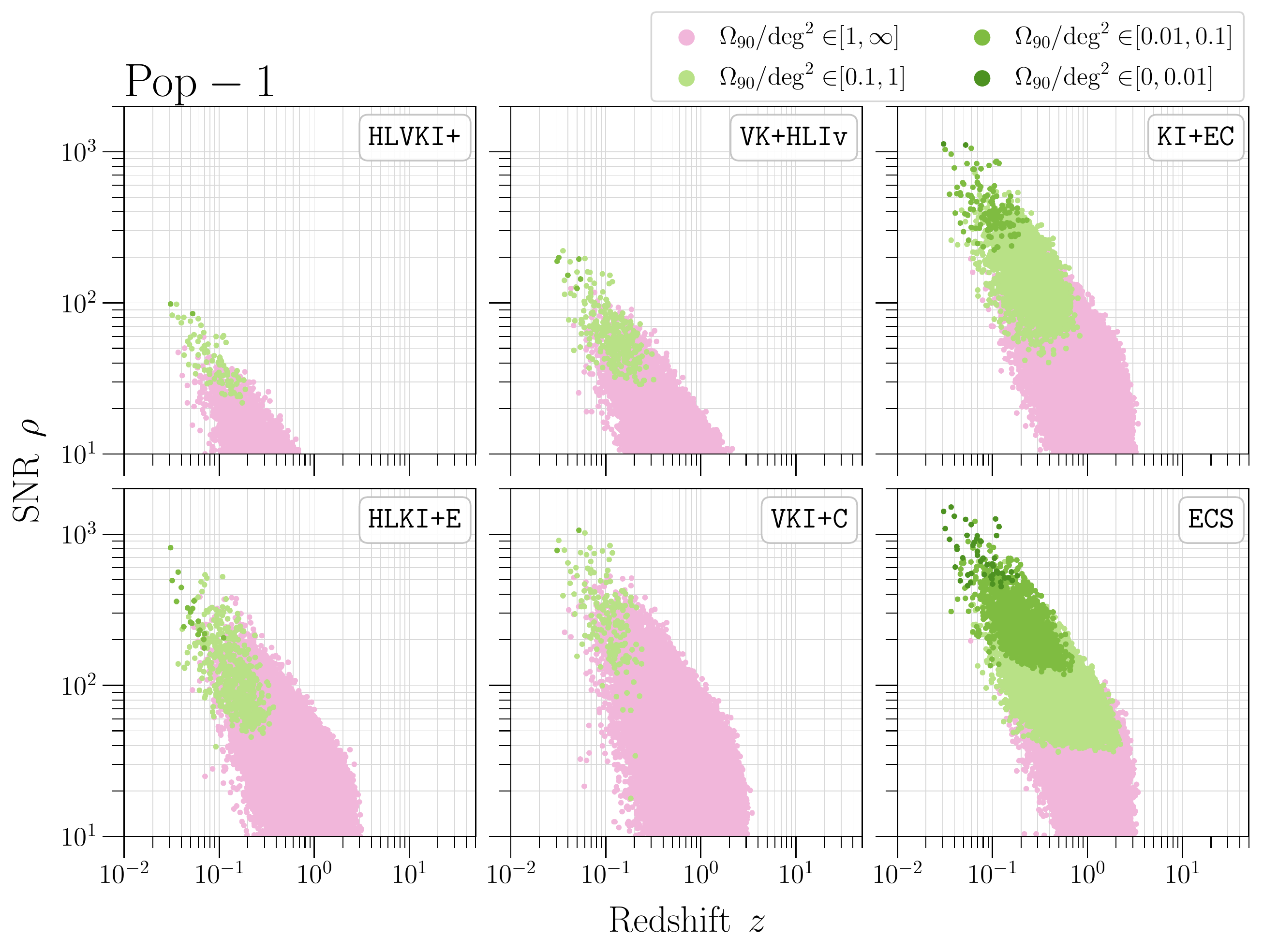}
  \caption{\label{fig:pink_scatter_no_mma_pop1}Plot showing the relationship between SNR $\rho$, sky localization $\Omega_{90}$ and the redshift $z$ for events belonging to the Pop-1 population, corresponding to the six GW detector networks. Each marker is an event detected by the corresponding detector network in an observation time of $10$ years. The color of the marker conveys how well that event can be localized in the sky using GW observation.}
\end{figure*}

\begin{figure*}
  \includegraphics[scale=0.6]{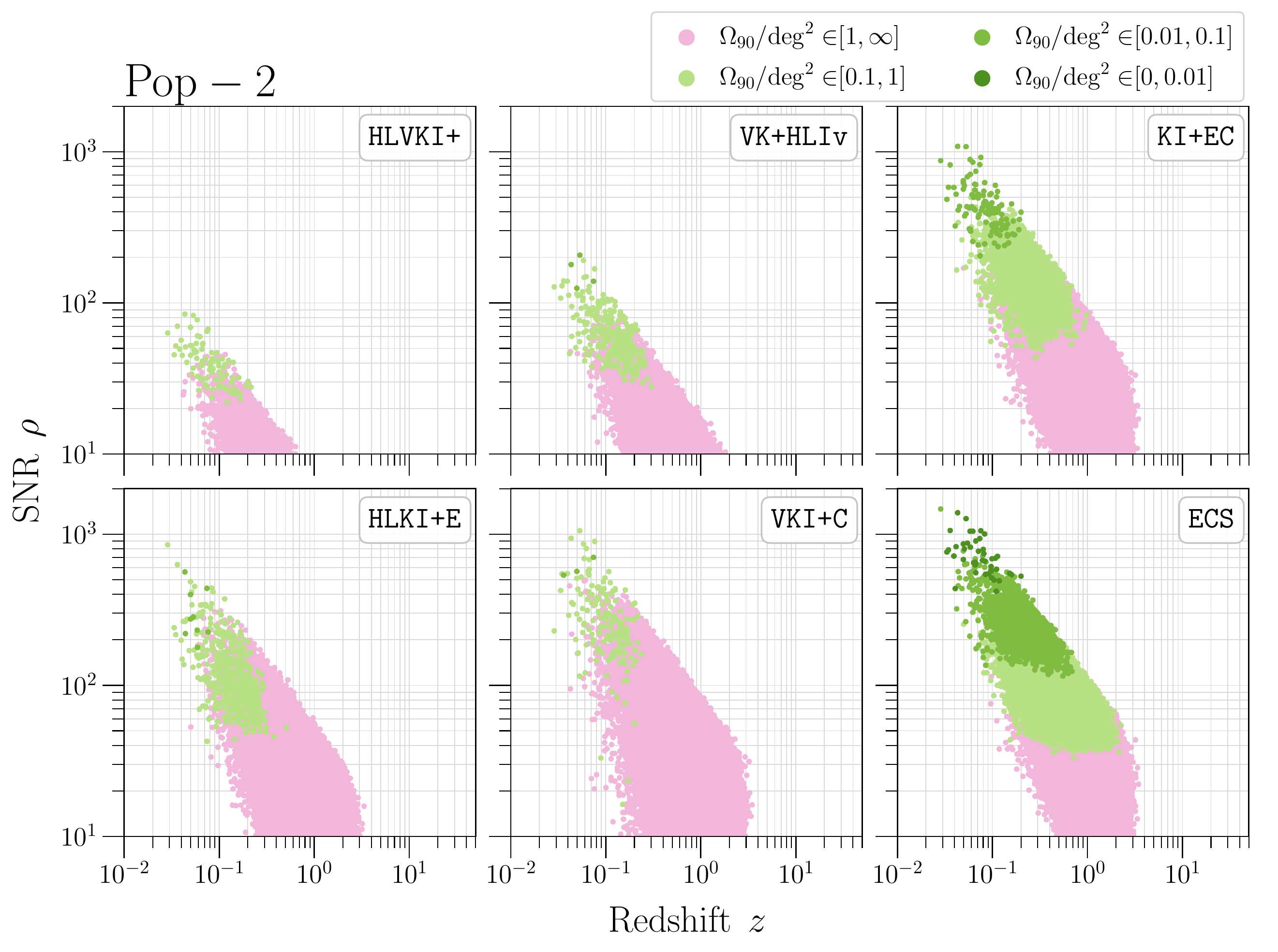}
  \caption{\label{fig:pink_scatter_no_mma_pop2}Plot showing the relationship between $\rho$, $\Omega_{90}$ and $z$ for events belonging to the Pop-2 population, in an observation time of $10$ years.}
\end{figure*}

The visibility of the EM transients that may follow the GW chirp depend on the luminosity distance between the binary and the observer and the orientation of the binary with respect to the line-of-sight of the observer. In addition, precise localization of the event can aid the follow-up efforts of EM telescopes and prove decisive in detecting EM transients. Figures \ref{fig:cdf_measure} and \ref{fig:cdf_measure2} show that the \texttt{ECS} and the \texttt{KI+EC} networks detect almost all their events with fractional errors in the luminosity distance better than $20\%$ and absolute error in inclination angle better than $0.2 \,\mbox{rad}$. While \texttt{VK+HLIv} indeed detects more events than \texttt{HLVKI+}, the overlapping CDF plots for $\Delta \mathcal{M}/\mathcal{M}$, $\Delta \eta$, $\Delta \chi_1$ and $\Delta \chi_2$, $\Delta D_L/D_L$ and $\Delta \iota$ show that the fraction of events detected with certain measurement quality remains the same between the two networks. In fact, the events that \texttt{HLVKI+} does detect, it does a remarkable job at localizing them in the sky, resolving $90\%$ of them to better than $20\,\mbox{deg}^2$, alongside the \texttt{ECS} network. The \texttt{ECS} network performs the best in terms of sky localization as well, detecting $\mathcal{O}(10)$ events every year with a resolution $\sim\,\mathcal{O}(10^{-2}) \,\mbox{deg}^2$. For comparison, the localization of GW170817 using the \texttt{HLV} network was $16\,\mbox{deg}^2$ \cite{LIGOScientific:2018hze}. The number of detections per year for each detector network with $\Omega_{90} \leq\,10$, $1$ and $0.1$, and $\Delta D_L/D_L\leq$ $0.1$ and $0.01$ for both the populations are listed in Table \ref{tab:pop_sa90_logDL}. The corresponding plot is shown in Figs. \ref{fig:pink_scatter_no_mma_pop1} and \ref{fig:pink_scatter_no_mma_pop2} for the two populations, respectively, 
which convey the relationship between the SNR, the $90\%$-credible sky area, and the redshift associated with the binary for events in both the populations. With \texttt{HLVKI+}, we can expect to detect $\mathcal{O}(1)$ event every $10$ years for which the sky position is localized to better than $1\,\mbox{deg}^2$. \texttt{VKI+C} is expected to detect about twice the number of events detected by \texttt{HLVKI+} with the same sky localization, whereas \texttt{VK+HLIv} and \texttt{HLVKI+E} are expected to detect about $4$ and $7$ times as many events, respectively. \texttt{KI+EC} will not only detect $\sim 60$ events every year with the sky localization better than $1\,\mbox{deg}^2$, it is also expected to see $\mathcal{O}(1)$ event \textit{every year} with the localization better than $0.1\,\mbox{deg}^2$. \texttt{ECS} is expected to outperform \texttt{KI+EC} by an order of magnitude, detecting $\mathcal{O}(10)$ events every year with localization better than $0.1\,\mbox{deg}^2$ and $\sim 5$ events in a span of $10$ years localized to an area smaller than $0.01\,\mbox{deg}^2$. As the position of these small number of events is localized to such a small area in the sky, it could even be possible to identify their host galaxies with only the GW signal (subject to the completeness of galaxy catalogues) \cite{Singer:2016eax,Borhanian:2020vyr}.

\section{Multimessenger Astronomy} \label{sec:mma}
During the inspiral phase of an NSBH merger event, the NS can be tidally disrupted by the BH, which can either happen when the NS is outside the innermost stable circular orbit (ISCO), or when the NS is within it. If the tidal disruption occurs when the NS is closer to the BH than the radius of ISCO, $R_{ISCO}$, any tidally-disrupted material is swallowed by the BH and no EM counterpart is generated. When the NS gets disrupted outside $R_{ISCO}$, a fraction of the NS mass can both, be dynamically ejected and form an accretion disk around the remnant BH. This material present outside the remnant powers the EM counterparts which may be detectable by an EM telescope (see Ref. \cite{Kyutoku:2021icp} for a review on NSBH mergers and factors that affect tidal disruption). The possible EM counterparts include short gamma-ray bursts and KN, among others. KN are produced by the radioactive decay of decompressed neutron star material, while the mechanism that produces gamma-ray bursts is still not fully understood, but it is believed to require strong magnetic fields \cite{Kyutoku:2021icp,Metzger:2019zeh}. The possibility of detecting gravitational waves along with the EM counterpart makes NSBH mergers exciting prospects for MMA.
In the following subsections, we will explore the potential of NSBH mergers as candidates for MMA. In Sec. \ref{subsec:subpop_mma}, we discuss the sub-population considered for the MMA study. We also present plots detailing the accuracy in the luminosity distance, inclination angle, and $90\%$-credible sky-area measurement for the events in this sub-population. In Sec. \ref{subsec:ew_mma}, we discuss the possibility of sending \textit{early-warning} alerts to EM telescopes in order to maximize the science output from EM detections. Finally, in Sec. \ref{subsec:kn_mma}, we examine KN as a potential EM counterpart to gravitational waves. We present the expected number of KN detections corresponding to NSBH merger events detected with the six GW detector networks and describe the method used to compute the same.

\subsection{Sub-population for MMA} \label{subsec:subpop_mma}
\begin{figure*}[htbp]
  \includegraphics[scale=0.38]{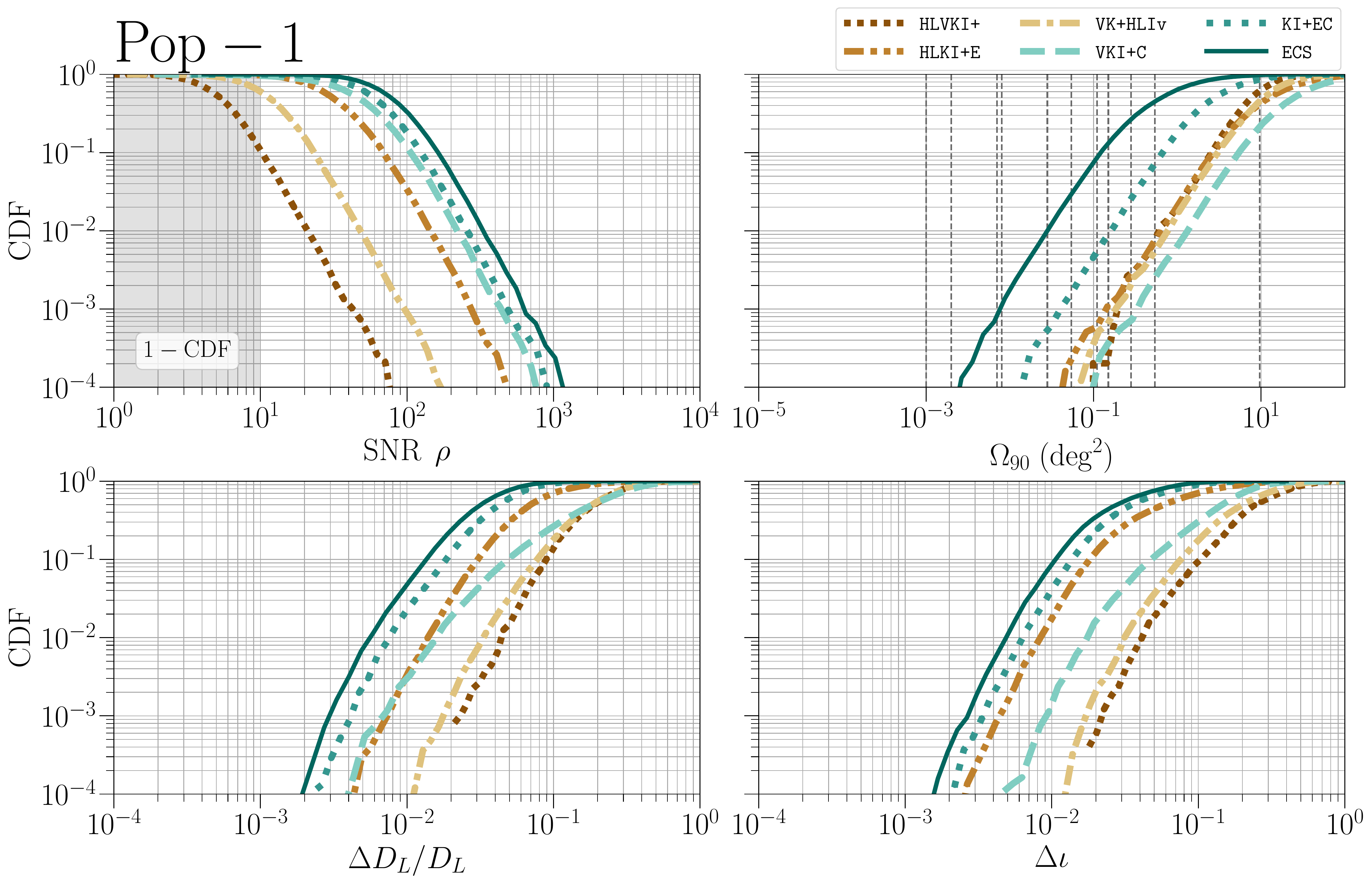}
  \includegraphics[scale=0.38]{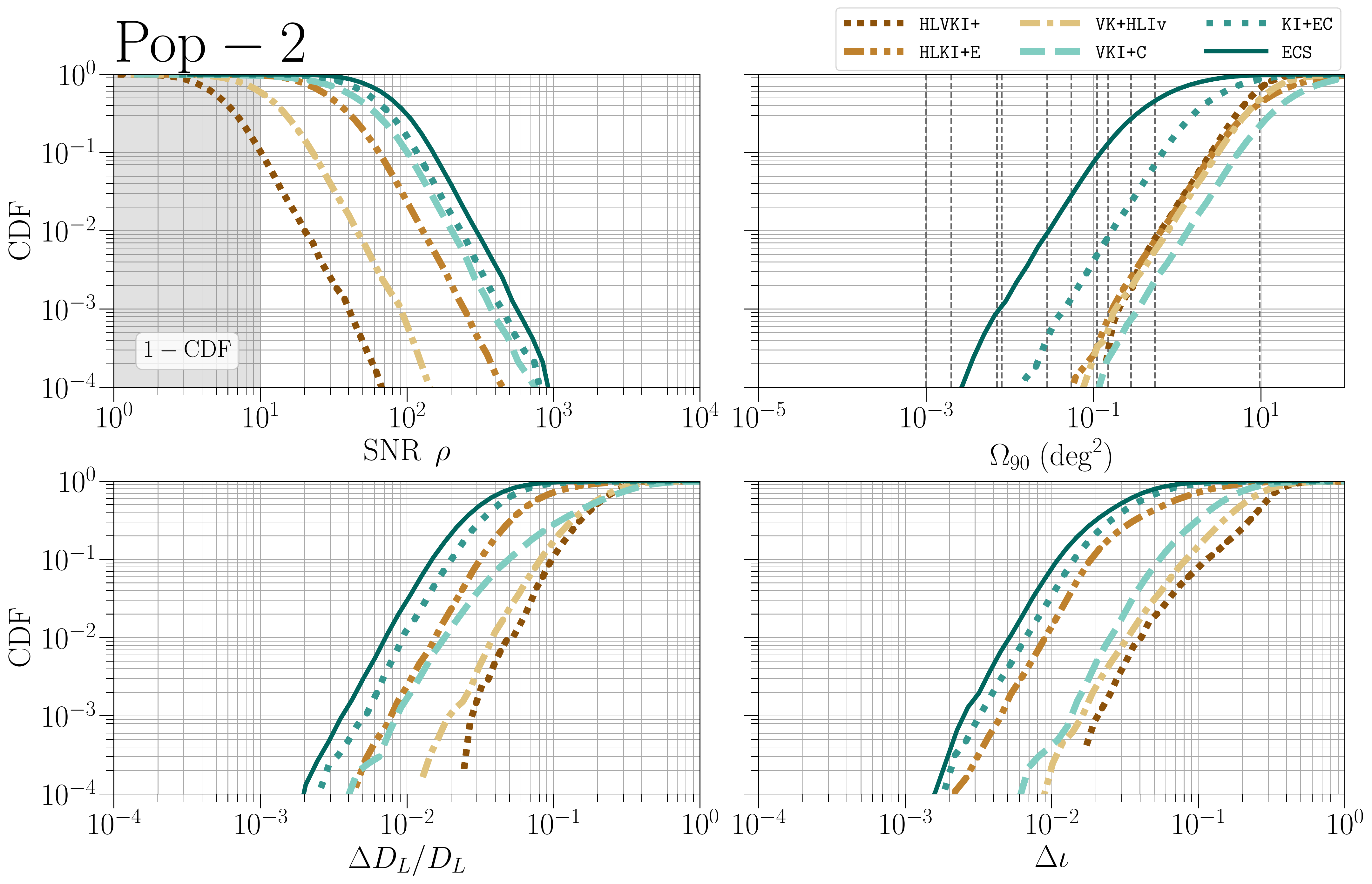}
\caption{\label{fig:mma_cdf} The cumulative density function plots for SNR $\rho$, $90\%$-credible sky area $\Omega_{90}$, fractional error in luminosity distance $\Delta D_L / D_L$ and absolute error in the inclination angle $\Delta \iota$ for the sub-population restricted to $z<0.5$. The vertical black dotted lines in the plot for $\Omega_{90}$ correspond to the FOV of the various EM telescopes listed in Table \ref{tab:telescope_FOVs}.}
\end{figure*}

For the purpose of MMA, we restrict our population to events that lie within a redshift of $z = 0.5$. While short gamma-ray bursts can be detected up to larger redshifts than $z = 0.5$, our study will be focusing on the detection of KN, which are not expected to be visible to the EM telescopes beyond this redshift (as will be seen in Sec. \ref{subsec:kn_mma}). For this sub-population, we look at the measurement accuracy in the luminosity distance, inclination angle, and sky-localization for the six networks. Table \ref{tab:mma_sa90_logDL} shows the number of detections per year for each network with $90\%$-credible sky area $\Omega_{90}$ $< 10$, $1$ and $0.1$, and fractional error in luminosity distance $\Delta D_L/ D_L < 0.1$ and $0.01$. The corresponding CDF plots for SNR $\rho$, $\Omega_{90}$, $\Delta D_L/ D_L$ and $\Delta \iota$ for both the populations are presented in Fig. \ref{fig:mma_cdf}.

\begin{table*}
  \centering
  \caption{\label{tab:mma_sa90_logDL}For the sub-population with events for which $z<0.5$, the table lists the number of detections per year for the six detector networks with $90\%$-credible sky area $\Omega_{90} < 10$, $1$, $0.1$ and $0.01$ $\mbox{deg}^2$ and fractional error in luminosity distance $\Delta D_L / D_L < 0.1$ and $0.01$.}
  \renewcommand{\arraystretch}{1.5} 
    \begin{tabular}{l | P{2.1cm} P{2.1cm} P{2.2cm} P{2cm} | P{2.1cm} P{2.2cm}}
    \hhline{=======}
    Metric & \multicolumn{4}{c|}{$\Omega_{90}\mbox{ (deg)}^2$} & \multicolumn{2}{c}{$\Delta D_L / D_L$} \\
    \hhline{-------}
    Quality & $\leq 10$ & $\leq 1$ & $\leq 0.1$ & $\leq 0.01$ & $\leq 0.1$ & $\leq 0.01$ \\
    \hhline{-------}
    \multicolumn{7}{c}{\textit{Pop-1}}\\
    \hhline{-------}
    HLVKI\texttt{+} & $2.9^{+6.1}_{-2.4} \times 10^{2}$ & $8.7^{+19.5}_{-7.3}$ & $2.0^{+2.0}_{-2.0} \times 10^{-1}$ & $0$ & $6.0^{+12.3}_{-4.7} \times 10$ & $0.0^{+0.1}_{-0.0}$ \\
    VK\texttt{+}HLIv & $1.0^{+2.2}_{-0.9} \times 10^{3}$ & $3.1^{+7.1}_{-2.6} \times 10$ & $6.0^{+13.0}_{-6.0} \times 10^{-1}$ & $0$ & $3.9^{+8.2}_{-3.2} \times 10^{2}$ & $3.0^{+9.0}_{-3.0} \times 10^{-1}$ \\
    HLKI\texttt{+}E & $1.4^{+3.0}_{-1.2} \times 10^{3}$ & $6.0^{+13.4}_{-4.9} \times 10$ & $1.9^{+2.3}_{-1.9}$ & $0$ & $2.5^{+5.3}_{-2.0} \times 10^{3}$ & $9.9^{+22.1}_{-8.4}$ \\
    VKI\texttt{+}C & $6.8^{+14.6}_{-5.6} \times 10^{2}$ & $1.9^{+4.1}_{-1.6} \times 10$ & $2.0^{+6.0}_{-2.0} \times 10^{-1}$ & $0$ & $8.7^{+18.4}_{-7.2} \times 10^{2}$ & $9.4^{+19.6}_{-7.4}$ \\
    KI\texttt{+}EC & $3.2^{+6.7}_{-2.6} \times 10^{3}$ & $5.6^{+12.1}_{-4.7} \times 10^{2}$ & $1.4^{+3.1}_{-1.2} \times 10$ & $2.0^{+4.0}_{-2.0} \times 10^{-1}$ & $3.5^{+7.4}_{-2.9} \times 10^{3}$ & $7.0^{+13.7}_{-5.4} \times 10$ \\
    ECS & $3.7^{+7.8}_{-3.1} \times 10^{3}$ & $2.2^{+4.8}_{-1.9} \times 10^{3}$ & $2.3^{+5.0}_{-1.9} \times 10^{2}$ & $5.1^{+9.8}_{-4.2}$ & $3.6^{+7.7}_{-3.0} \times 10^{3}$ & $1.4^{+2.9}_{-1.1} \times 10^{2}$ \\
    \hhline{-------}
    \multicolumn{7}{c}{\textit{Pop-2}}\\
    \hhline{-------}
    HLVKI\texttt{+} & $3.0^{+6.4}_{-2.5} \times 10^{2}$ & $9.1^{+20.3}_{-7.7}$ & $0.0^{+0.3}_{-0.0}$ & $0$ & $4.4^{+8.3}_{-3.7} \times 10$ & $0.0^{+0.1}_{-0.0}$ \\
    VK\texttt{+}HLIv & $1.1^{+2.3}_{-0.9} \times 10^{3}$ & $3.2^{+6.8}_{-2.6} \times 10$ & $4.0^{+15.0}_{-4.0} \times 10^{-1}$ & $0$ & $3.5^{+7.4}_{-2.9} \times 10^{2}$ & $0.0^{+0.3}_{-0.0}$ \\
    HLKI\texttt{+}E & $1.5^{+3.1}_{-1.2} \times 10^{3}$ & $5.9^{+12.2}_{-4.8} \times 10$ & $9.0^{+35.0}_{-8.0} \times 10^{-1}$ & $0$ & $2.5^{+5.3}_{-2.1} \times 10^{3}$ & $7.1^{+16.3}_{-6.2}$ \\
    VKI\texttt{+}C & $7.1^{+15.1}_{-5.9} \times 10^{2}$ & $1.8^{+3.8}_{-1.5} \times 10$ & $3.0^{+5.0}_{-3.0} \times 10^{-1}$ & $0$ & $9.2^{+19.4}_{-7.6} \times 10^{2}$ & $4.8^{+8.3}_{-4.0}$ \\
    KI\texttt{+}EC & $3.2^{+6.8}_{-2.6} \times 10^{3}$ & $5.7^{+12.0}_{-4.7} \times 10^{2}$ & $1.2^{+3.1}_{-1.0} \times 10$ & $0.0^{+0.3}_{-0.0}$ & $3.5^{+7.4}_{-2.9} \times 10^{3}$ & $3.7^{+7.4}_{-3.1} \times 10$ \\
    ECS & $3.7^{+7.9}_{-3.1} \times 10^{3}$ & $2.3^{+4.9}_{-1.9} \times 10^{3}$ & $2.3^{+4.8}_{-1.9} \times 10^{2}$ & $4.5^{+11.7}_{-3.8}$ & $3.6^{+7.7}_{-3.0} \times 10^{3}$ & $8.6^{+18.1}_{-7.2} \times 10$ \\
    \hhline{=======}
    \end{tabular}
\end{table*}

\begin{table}
  \caption{\label{tab:telescope_FOVs}The field of view (FOV) of some of the electromagnetic (EM) telescopes. Among them, we have used the Rubin Observatory and the \textit{Roman Telescope} to comment on the detectability of kilonovae in Sec. \ref{subsec:kn_mma}. The space telescopes in the list have been \textit{italicized}.}
  \renewcommand{\arraystretch}{1.5} 
    \begin{tabular}{l c}
    \hhline{==}
    Telescope & FOV ($\mbox{deg}^2$)\\
    \hhline{--}
    \textbf{Rubin} \cite{web:Rubin,LSST:2008ijt} & 9.6 \\
    \textit{EUCLID} \cite{Euclid:2021icp} & 0.54 \\
    \textit{Athena} \cite{2013arXiv1308.6785R} & 0.35 \\
    \textit{\textbf{Roman}} \cite{Hounsell:2017ejq,Chase:2021ood} & 0.28 \\
    \textit{Chandra X-ray} \cite{web:Chandra} & 0.15 \\
    \textit{Lynx} \cite{2019JATIS...5b1001G} & 0.13 \\
    \textit{Swift–XRT} \cite{2000SPIE.4140...64B} & 0.12 \\
    Keck \cite{bundy2019fobos} & 0.11 \\
    VLT \cite{web:vlt} & 0.054 \\
    ELT \cite{web:elt} & 0.028 \\
    GMT \cite{2016SPIE.9908E..1UJ} & 0.008 \\
    \textit{HST–WFC3} \cite{web:hst-wfc3} & 0.002 \\
    \hhline{==}
    \end{tabular}
\end{table}

From Fig. \ref{fig:eff_rate}, we see that all the networks with at least one of the XG observatories detect almost all the events up to a redshift of $z=0.5$, with \texttt{ECS} detecting about half of those events with SNRs greater than $100$. The Voyager network detects $\sim 90\%$ of the events whereas \texttt{HLVKI+} detects only $20\%$ of the events. Figure \ref{fig:mma_cdf} shows no significant differences between the CDF plots for the two populations. All the networks measure the luminosity distances for almost all the events better than a fractional error of $30\%$, with \texttt{ECS} and \texttt{KI+EC} constraining the luminosity distance to better than $10\%$ for all the events. 

\begin{figure*}[htbp]
  \includegraphics[scale=0.6]{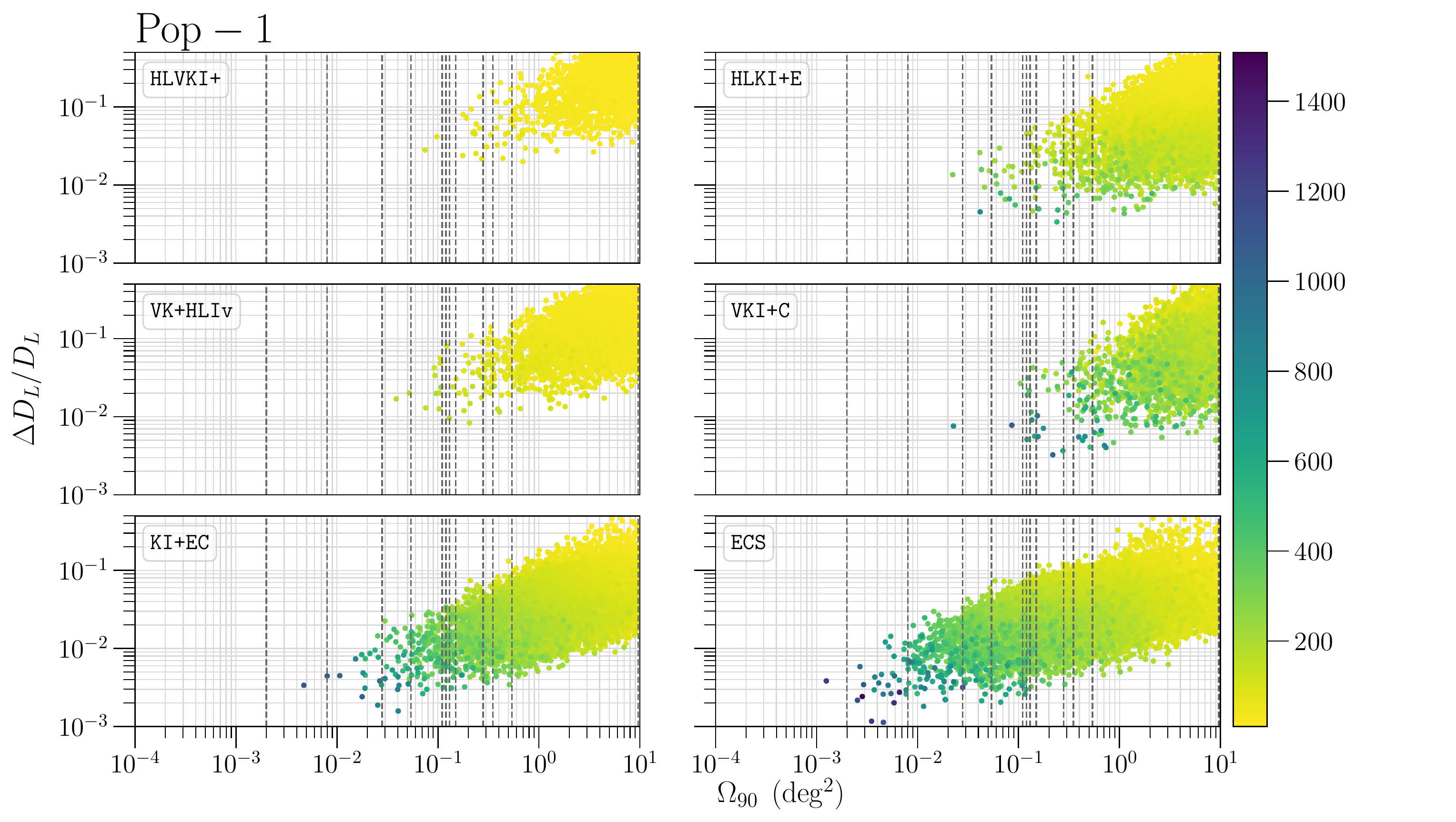}
  \includegraphics[scale=0.6]{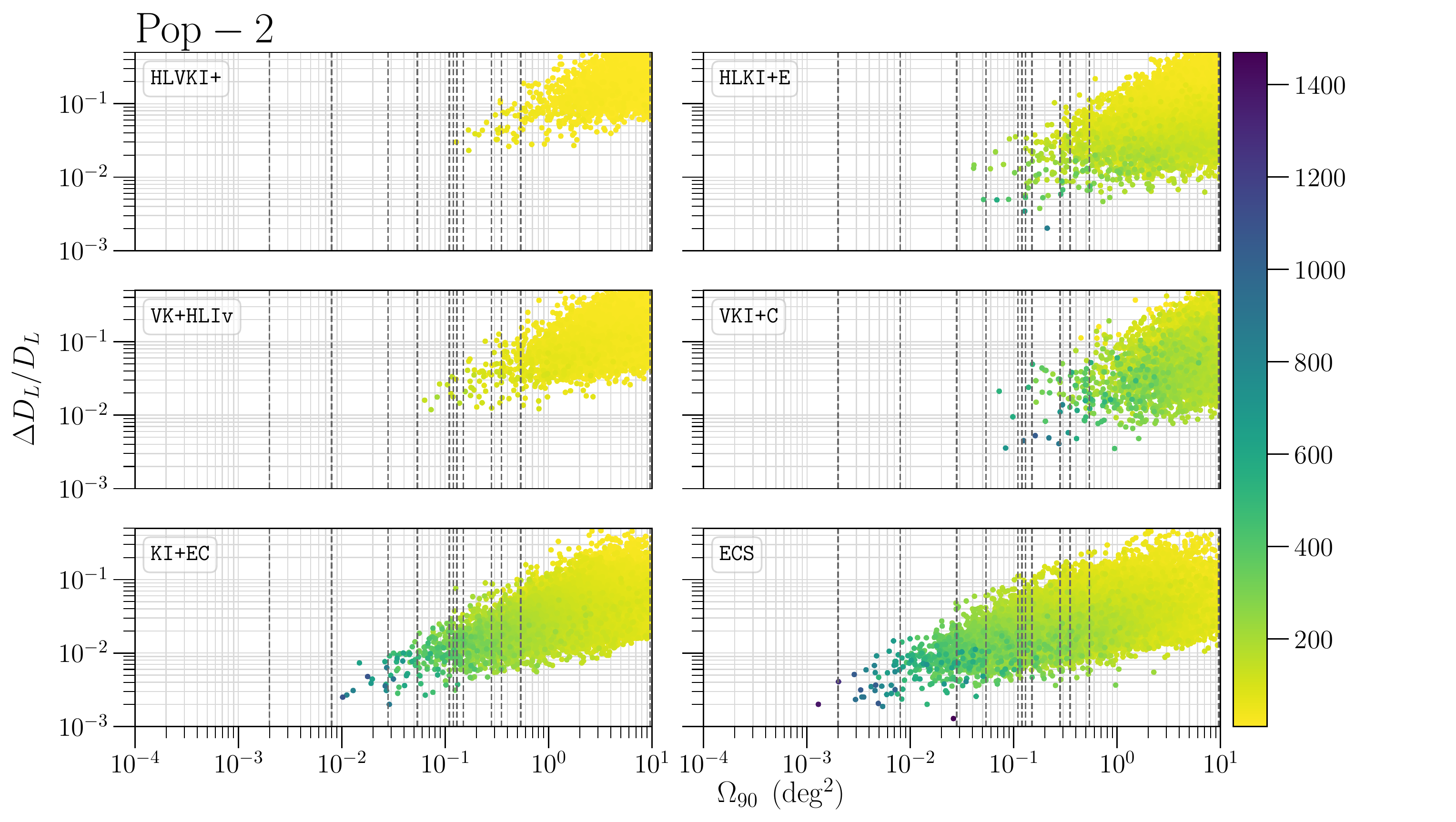}
  \caption{\label{fig:yellow_scatter} The figure shows the relationship between the fractional error in luminosity distance $\Delta D_L/D_L$, $90\%$-credible sky area $\Omega_{90}$ and the SNR (denoted by the color bar) of the events in Pop-1 and Pop-2 for which $z<0.5$. Each of these events, detected in an observation span of $10$ years, appears as a spot placed according to the associated measurement errors in luminosity distance and sky position. The color of the dots represents the SNR with which that particular event was detected in a GW detector network.}
\end{figure*}

To maximize the chances of a telescope detecting the EM counterpart, the estimated sky area from the GW detection should be within the field-of-view (FOV) of the EM telescope. FOVs of some of the EM telescopes are listed in Table \ref{tab:telescope_FOVs} and have been denoted in the plots for $\Omega_{90}$ is Fig. \ref{fig:mma_cdf} and \ref{fig:yellow_scatter}. The FOV of the Rubin observatory is an order of magnitude bigger than any other telescope listed in Table \ref{tab:telescope_FOVs}, allowing it to see many more EM transients compared to any other telescope. As a result, less than $0.5\%$ of the events detected by \texttt{HLVKI+} will be visible to telescopes other than Rubin, whereas only $\sim 10\%$ of the events detected by \texttt{ECS} can be localized in the sky to an area smaller than the FOV of telescopes other than Rubin. In general, EM telescopes can slew and cover multiple patches in the sky, which will increase the number of EM counterparts they will be able to detect. For instance, if the \textit{Roman Space Telescope} with a FOV of $0.28\,\mbox{deg}^2$ can slew and observe five patches in the sky, covering an area of $\sim 1\,\mbox{deg}^2$, then it can detect potential EM counterparts of $\sim3\%$ of the events detected by \texttt{HLVKI+} within $z=0.5$ and $\sim60\%$ of the events detected by the \texttt{ECS} network in the same sub-population. However, the main focus of time-domain survey projects like the Rubin observatory and the \textit{Roman} telescope is to detect supernovae, which are much brighter and evolve much slower than the typical KN. Thus, it is not only difficult for these surveys to detect KN in the first place, but the surveys might also miss the optimal time window to observe a KN without a targeted search. This emphasizes the need for target-of-opportunity (TOO) follow-up to GW events in order to utilize the full potential of MMA \cite{Cowperthwaite:2018gmx,LSST:2018bbx,Zhu:2021jbw}.

\subsection{Early-warning alerts} \label{subsec:ew_mma}
The GW detectors start detecting gravitational radiation from the inspiral phase itself, i.e., much before the actual merger happens. Specifically, if a detector starts detecting the signal at a lower frequency cutoff of $f_L$, then time to coalescence $\tau$ is given by \cite{Sathyaprakash:1991mt}
\begin{equation} \label{eq:tau_ew}
    \tau \approx \left(\frac{0.25}{\eta} \right)\left(\frac{2.8\msun}{M}\right)^{\frac{5}{3}}\left(\frac{5\mbox{Hz}}{f_L}\right)^{\frac{8}{3}}\times 6.4 \times 10^3 \mbox{ s},
\end{equation}
where $M$ is the total redshifted mass of the binary. For the same total mass, the more asymmetric the binary (i.e., the smaller $\eta$ is), the larger the time to merger. However, as the binary gets heavier, the time to merger decreases. Additionally, a smaller $f_L$ means that the detector is able to capture the inspiral phase from an earlier time. For a reference NSBH system with source-frame masses of $m_{NS} = 1.5 \msun$ and $m_{BH} = 8 \msun$ respectively, and $f_L = 20$ Hz with the system at a redshift of $z = 0.1$, the time to coalescence $\tau \approx 30$ seconds, and it increases to about $2.5$ minutes when $f_L$ is lowered to $11$ Hz. Thus, \textit{early-warning} (EW) alerts \cite{Sachdev:2020lfd} with the estimated sky position, based on the data collected by then, can be sent to the EM telescope before the merger, allowing for possible latency in the process, and give time to the telescope to slew in position and still record the EM radiation that is generated during and after the merger. Early observations can allow the EM telescopes to capture prompt emission as well as make early optical and ultraviolet observations that give us information about the r-process nucleosynthesis \cite{Nicholl:2017ahq}.

\begin{figure*}[htbp]
  \includegraphics[scale=0.6]{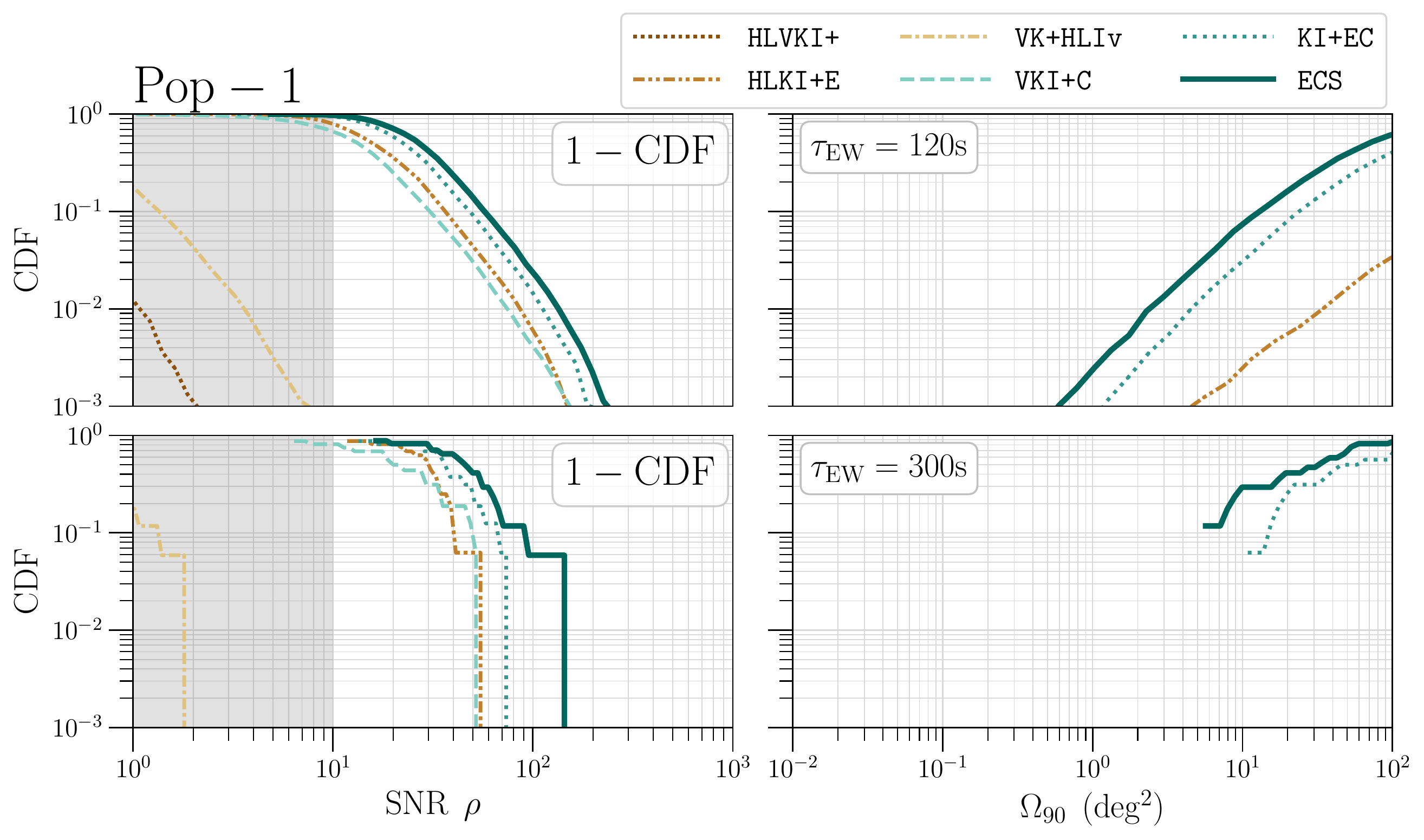}
  \includegraphics[scale=0.6]{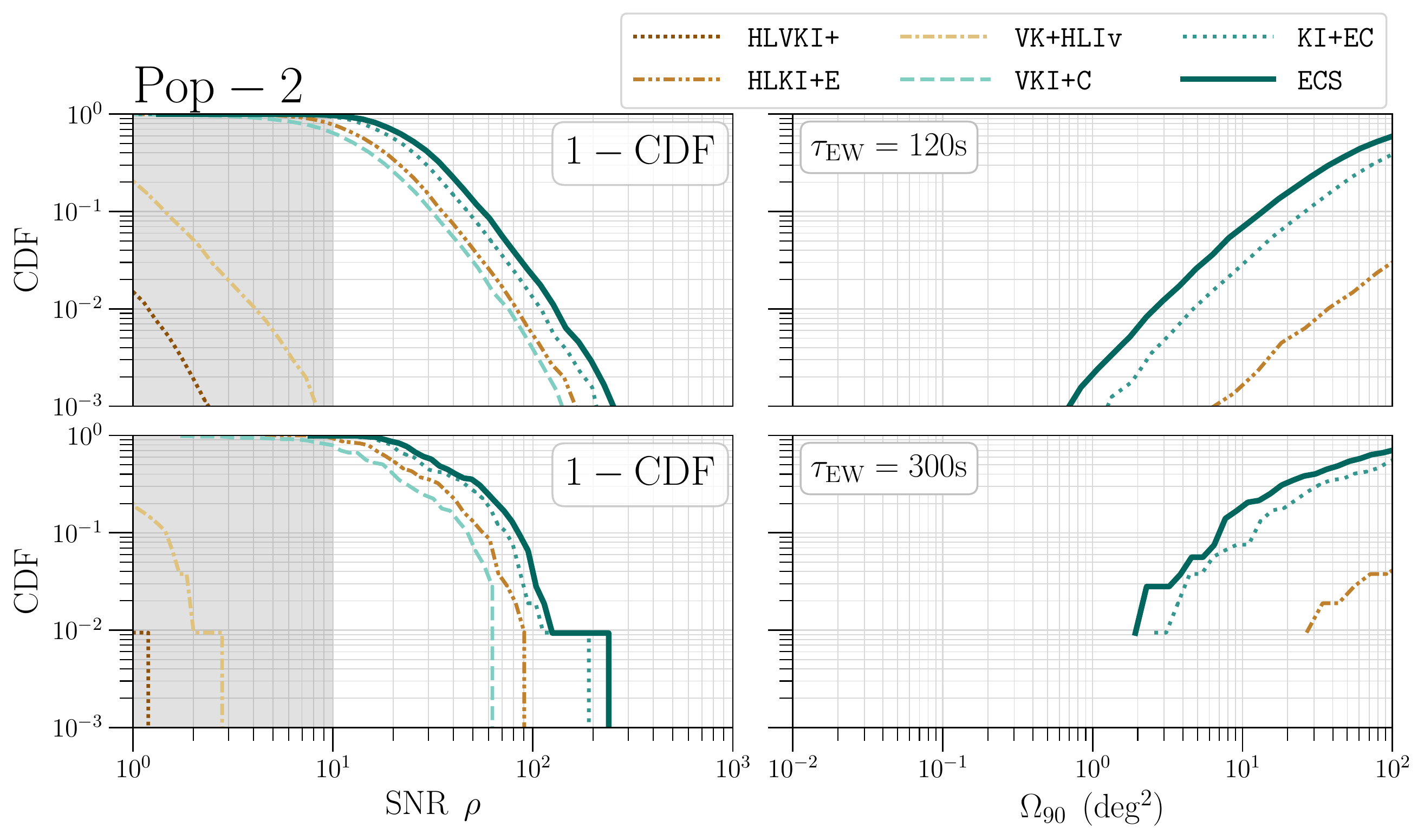}
  \caption{\label{fig:ew_CDFs}The cumulative density function (CDF) plots for events for which early-warning alerts can be sent $2$ minutes and $5$ minutes before their respective mergers.}
\end{figure*}

Following Ref. \cite{Borhanian:2022czq}, we present data for two values of time to coalescence, $\tau_{\mathrm{EW}} = 120$ s and $300$ s. While Ref. \cite{Borhanian:2022czq} also gives data for $\tau_{\mathrm{EW}} = 600$ s, we do not find any NSBH events in either Pop-1 or Pop-2 that qualify that criteria. Equation (\ref{eq:tau_ew}) can be inverted to calculate $f_L$ (referred to as $f_{\mathrm{EW}}$ in this context) for the given values of $\tau_{\mathrm{EW}}$ for every event. $f_{EW}$ is the frequency of the GW signal from which the system has time
$\tau_{\mathrm{EW}}$ left until coalescence. We only consider those events that have accumulated SNR $>10$ in the particular detector network at the time the alert is sent, and for which $f_{EW} > 11$ Hz. However, even if an EW alert is sent to the EM telescope \textit{adequately} early, it is not of much use if the associated sky position reported by the GW detector is too large for the telescope to search for the EM counterpart. As the EM telescopes can look for these counterparts by observing multiple patches of the sky, they can often cover an area that is larger than their FOV. In Table \ref{tab:ew_sa90}, we present the number of observations every year where an EW alert can be sent $120$ s and $300$ s before the merger. We further categorize the events based on how well they are localized, showing numbers for detections that are localized to $\Omega_{90} \leq 100$, $10$ and $1$ $\mbox{deg}^2$ at the time when the alert is sent. We do not list the numbers for \texttt{HLVKI+}, \texttt{VK+HLIv} and \texttt{VKI+C} as no events satisfying the criteria were found, which is also evident from the corresponding CDF plots in Fig. \ref{fig:ew_CDFs}.

From Table \ref{tab:ew_sa90}, we see that EW alerts can be sent to only $\mathcal{O}(10)$ events for \texttt{HLKI+E} such that the sky position of the events is also constrained better than $100\,\mbox{deg}^2$, whereas it can be sent for $\mathcal{O}(100)$ events for the \texttt{KI+EC} and the \texttt{ECS} networks. The corresponding numbers drop by two orders of magnitude if the EW is sent $5$ minutes before coalescence. We also observe that the number of events for which the EW alert can be sent $5$ minutes before the merger is higher in Pop-2 compared to Pop-1. We attribute this difference to the fact that, in general, we expect systems in Pop-1 to have higher total masses compared to systems in Pop-2 (due to the broader distributions of NS and BH masses in Pop-1) which leads to longer signals in Pop-2 [as can be seen from Eq. (\ref{eq:tau_ew})]. In addition, the number of events with $\Omega_{90} \leq 10\,\mbox{deg}^2$ is $\mathcal{O}(10)$ times lower than the number of events for which $\Omega_{90} \leq 100$. For the events for which $10\,\mbox{deg}^2\leq \Omega_{90} \leq 100\,\mbox{deg}^2$, Rubin would need to cover at most $10$ sky patches to follow up a possible EM counterpart, whereas any other telescope listed in Table \ref{tab:telescope_FOVs} would need to slew and cover $\mathcal{O}(100)$ sky patches to detect any possible EM transient.

Figure \ref{fig:ew_CDFs} shows that, while there are events detected by \texttt{VKI+C} with SNR $\rho > 10$ five minutes before their mergers, their sky position cannot be localized to better than $100\,\mbox{deg}^2$. This is consistent with the performance of \texttt{VKI+C} in constraining $\Omega_{90}$ compared to the other networks, as can be seen from Fig. \ref{fig:ew_CDFs}. Moreover, none of the events that are \textit{eventually} detected by \texttt{HLVKI+} and \texttt{VK+HLIv} networks accumulate SNR in excess of $10$ in their respective detector networks $2$ minutes or $5$ minutes before their mergers. 
\begin{table*}[htbp] 
  \centering
  \caption{\label{tab:ew_sa90}The number of detections per year for \texttt{HLKI+E}, \texttt{KI+EC} and \texttt{ECS} for which an EW alert can be sent $120$ s and $300$ s before the merger, with $90\%$-credible sky area measured to be better than $100$, $10$, $1$ $\mbox{deg}^2$ at the time when the alert is sent.}
  \renewcommand{\arraystretch}{1.5} 
    \begin{tabular}{l P{2.1cm}P{2.2cm}P{2.2cm} P{2.2cm}P{2.2cm}P{2cm}}
    \hhline{=======}
    EW Time & \multicolumn{3}{c}{$\tau_{\mathrm{EW}} = 120\mbox{ s}$} & \multicolumn{3}{c}{$\tau_{\mathrm{EW}} = 300\mbox{ s}$} \\
    \hhline{-------}
    $\Omega_{90}\,(\mbox{deg}^2)$ & $\leq 100$ & $\leq 10$ & $\leq 1$ & $\leq 100$ & $\leq 10$ & $\leq 1$ \\
    \hhline{-------}
    \multicolumn{7}{c}{\textit{Pop-1}} \\
    \hhline{-------}
    HLKI\texttt{+}E & $2.9^{+5.9}_{-2.5} \times 10$ & $1.9^{+3.3}_{-1.7}$ & $0.0^{+0.2}_{-0.0}$ & $0.0^{+0.3}_{-0.0}$ & $0$ & $0$ \\
    KI\texttt{+}EC & $3.5^{+7.5}_{-2.9} \times 10^{2}$ & $2.4^{+6.0}_{-2.1} \times 10$ & $5.0^{+24.0}_{-5.0} \times 10^{-1}$ & $9.0^{+30.0}_{-9.0} \times 10^{-1}$ & $0.0^{+0.5}_{-0.0}$ & $0$ \\
    ECS & $5.6^{+11.8}_{-4.6} \times 10^{2}$ & $6.1^{+12.9}_{-5.1} \times 10$ & $1.7^{+5.7}_{-1.5}$ & $1.4^{+3.5}_{-1.3}$ & $5.0^{+12.0}_{-5.0} \times 10^{-1}$ & $0$ \\
    \hhline{-------}
    \multicolumn{7}{c}{\textit{Pop-2}} \\
    \hhline{-------}
    HLKI\texttt{+}E & $2.4^{+5.0}_{-2.1} \times 10$ & $1.2^{+2.4}_{-0.9}$ & $0$ & $4.0^{+9.0}_{-4.0} \times 10^{-1}$ & $0.0^{+0.1}_{-0.0}$ & $0$ \\
    KI\texttt{+}EC & $3.3^{+7.2}_{-2.7} \times 10^{2}$ & $2.2^{+4.7}_{-1.8} \times 10$ & $1.0^{+25.0}_{-0.0} \times 10^{-1}$ & $5.8^{+11.6}_{-4.7}$ & $8.0^{+9.0}_{-8.0} \times 10^{-1}$ & $0$ \\
    ECS & $5.3^{+11.4}_{-4.4} \times 10^{2}$ & $5.7^{+11.9}_{-4.6} \times 10$ & $1.5^{+4.4}_{-0.9}$ & $7.4^{+16.0}_{-5.8}$ & $1.8^{+1.9}_{-1.6}$ & $0.0^{+0.1}_{-0.0}$ \\
    \hhline{=======}
    \end{tabular}
\end{table*}

\subsection{Kilonova Detection} \label{subsec:kn_mma}
During a NSBH merger, the companion BH can cause tidal disruption of the NS resulting in the generation of neutron-rich ejecta. The presence of neutron-rich substances in the expanding ejecta leads to fission and further neutron capture, leading to the production of a wide variety of radioactive elements heavier than iron. These unstable nuclei eventually decay radioactively heating up the ejecta, which leads to approximately isotropic quasithermal emission in the ultraviolet (UV), optical, and infrared (IR) bands. This transient is known as kilonova and serves as one of the most promising EM counterparts to GW radiation from BNS/NSBH systems \cite{Kyutoku:2021icp,Metzger:2019zeh}.

\begin{figure}[htbp]
\centering
\includegraphics[scale=0.57]{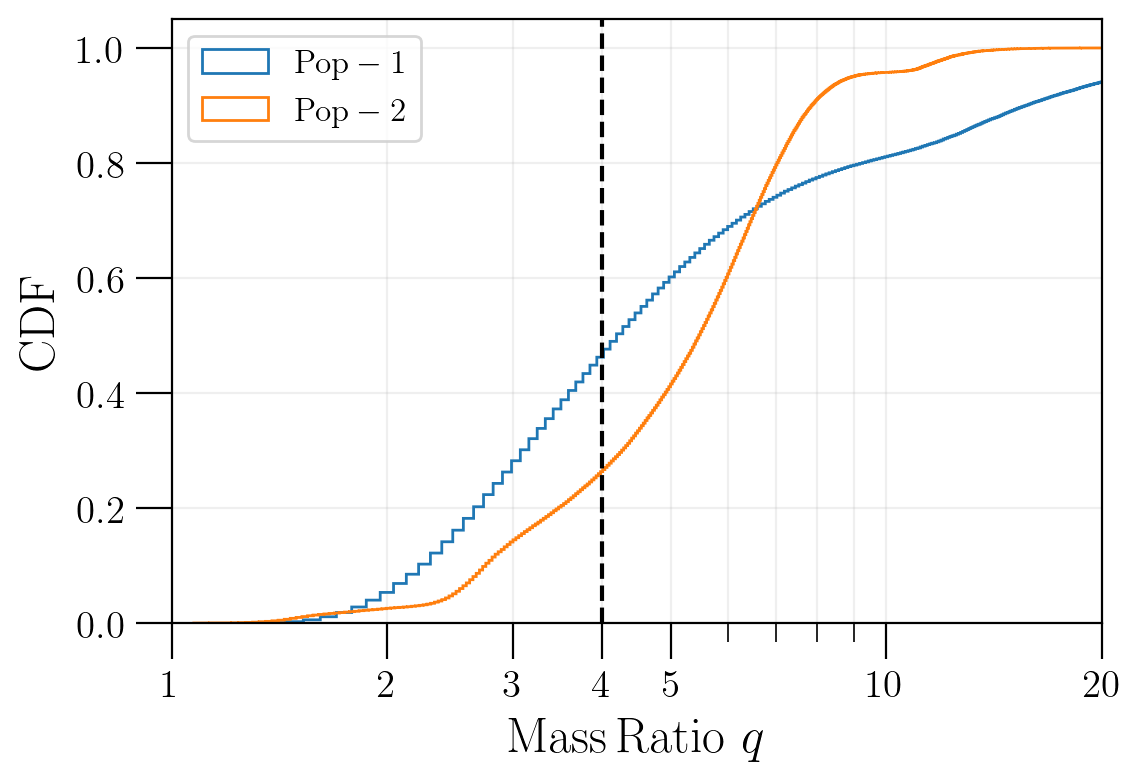}
\caption{\label{fig:CDF_massratio}The cumulative density function (CDF) plots for the mass ratio of the events belonging to the two populations that were detected by the \texttt{ECS} network. The black dashed vertical line separates the $q\leq4$ region from the $q>4$ region. Only events with $q\leq4$ have been considered for the KN study.}
\end{figure}

For a given equation of state (EOS), the tidal disruption of the NS, and consequently the generation of a KN, depends on the mass-ratio $(q = m_{BH}/m_{NS})$ associated with the binary and the dimensionless spin of the BH $\chi_{BH}$. Only BHs with low mass (hence, low $q$) and high prograde $(\chi_{BH}>0)$ spin are expected to tidally disrupt the NS before it passes the $R_{ISCO}$ \cite{Kyutoku:2021icp}. However, given the population parameters used and the bias towards detecting systems with higher masses, we see that higher $q$ systems are preferred over lower $q$ ones (see Fig. \ref{fig:CDF_massratio}). Thus, the GW chirp of only a fraction of NSBH systems is expected to have KN as the EM counterpart \cite{Pannarale:2014rea,Zappa:2019ntl,Fragione:2021cvv,Chatterjee:2019avs}. The detected NSBH events are seen to have $q>4$ and $\chi_{BH} \approx 0$ \cite{Zhu:2021jbw}, which is consistent with the fact that no corresponding KN were detected. Moreover, population-based analysis where the population is based on current observations posits that a tiny fraction, if at all, of NSBH systems are expected to be \textit{EM bright}, i.e., capable of generating an EM counterpart \cite{Biscoveanu:2022iue}.

In this study, we report the number of KN detections per year, for each population model corresponding to the detections made by the six GW detector networks. We select the events from the population with $q\leq4$ and $0\leq\chi_{BH}\leq0.75$, that are detected by each of these detectors with SNR $\rho > 10$, and obtain the bolometric KN light curve (in case of non-zero ejecta mass) for these events. To obtain the light curve, we follow the recipes from Refs. \cite{Kruger:2020gig,Raaijmakers:2021slr} for \textit{BHNS} systems. This involves the calculation of the remnant mass, i.e., the estimated baryon mass outside the BH approximately $10$ ms after the merger, and the mass of the dynamical ejecta. The formula in Ref. \cite{Foucart:2018rjc} gives the \textit{normalized} remnant mass which, when multiplied by the baryonic mass of the initial NS, gives the remnant mass outside the BH. While Ref. \cite{Raaijmakers:2021slr} approximates the baryonic mass of the NS using the formula from Ref. \cite{Lattimer:2000nx}, we calculate it from the TOV data corresponding to each EOS. The mass of the dynamical ejecta is obtained using the fits from Ref. \cite{Kruger:2020gig}. The mass of the disk surrounding the BH is calculated by subtracting the mass of the dynamical ejecta from the remnant mass. A fraction of the disk mass can become gravitationally unbound, which is referred to as disk wind. This fraction is computed by using the formula \cite{Raaijmakers:2021slr}:
\begin{equation}
    \xi = \frac{M_{ej}}{M_{disk}} = \xi_1 + \frac{\xi_2-\xi_1}{1+e^{1.5(q-3)}},
\end{equation}
where $\xi_1 \in (0.04,0.32)$ and $\xi_2 \in (0.04,0.14)$. We set $\xi_1$ and $\xi_2$ to the average values of the upper and lower bounds used in Ref. \cite{Raaijmakers:2021slr}, i.e., $\xi_1 = 0.18$ and $\xi_2 = 0.29$ respectively. The velocity of the dynamical ejecta is approximated using Ref. \cite{Raaijmakers:2021slr} and the velocity of the disk wind is set to $0.1c$ \cite{De:2020jdt,Siegel:2017nub}. The opacities for the dynamical ejecta lie in the range $(1$--$10)$ cm$^2$ g$^{-1}$ due to the Lanthanide rich r-process nucleosynthesis while the disk, after getting irradiated by neutrinos, becomes relatively optically thin with opacity in the range $(0.1$--$1)$ cm$^2$ g$^{-1}$. Because of uncertainties in the nucleosynthetic calculations, we fix dynamical and disk matter opacity to $8$ and $0.5$, respectively. The luminosity curves for both the dynamical ejecta and the unbound disk mass are individually determined by integrating the heating function (which accounts for the heating due to $\beta-$decay), approximated by a power law and implemented by using the numerical fit from Ref. \cite{Korobkin:2012uy}. The luminosity for the dynamical ejecta and the unbound disk mass at each time are added to calculate the total bolometric luminosity curve for the system.
\begin{table}[htbp] 
  \centering
  \caption{The six filters in the Rubin Observatory and the \textit{Roman Telescope}, with the corresponding single-exposure (30s) limiting magnitudes \cite{web:Rubin,Chase:2021ood} and the effective wavelength $(\lambda_{\mathrm{eff}})$ used for each band in order to calculate the photometric band estimates.}
  \renewcommand{\arraystretch}{1.5} 
    \begin{tabular}{P{2cm}P{2.5cm}P{2.2cm}}
    \hhline{===}
    \multicolumn{3}{c}{Vera C. Rubin Observatory}\\
    \hhline{---}
    Band & $m_{\mathrm{lim}}$ (AB mag) & $\lambda_{\mathrm{eff}}$ (\r{A})\\
    \hhline{---}
    $u$ & 23.9 & 3546 \\
    $g$ & 25.0 & 4670 \\
    $r$ & 24.7 & 6156 \\
    $i$ & 24.0 & 7472 \\
    $z$ & 23.3 & 8917 \\
    $y$ & 22.1 & 10305 \\
    \hhline{---}
    \multicolumn{3}{c}{Nancy Grace Roman Space Telescope}\\
    \hhline{---}
    Band & $m_{\mathrm{lim}}$ (AB mag) & $\lambda_{\mathrm{eff}}$ (\r{A})\\
    \hhline{---}
    $R$ & 26.2 & 6160 \\
    $Z$ & 25.7 & 8720 \\
    $Y$ & 25.6 & 10600 \\
    $J$ & 25.5 & 12900 \\
    $H$ & 25.4 & 15800 \\
    $F$ & 24.9 & 18400 \\
    \hhline{===}
    \end{tabular}
  \label{tab:rubin_vals}%
\end{table}

To report the KN detections, we consider the observing EM telescopes to be the Vera Rubin Observatory and the \textit{Nancy Grace Roman Space Telescope}. Table \ref{tab:rubin_vals} lists the photometric bands for the telescopes, the corresponding $5\sigma$ single-exposure limiting magnitudes for point sources $(m_{\mathrm{lim}})$, and the effective wavelength $(\lambda_{\mathrm{eff}})$ for each band. In order to obtain the estimates for the photometric bands from the bolometric luminosity curve, we calculate the blackbody temperature and the radius of the ejecta as a function of time and use them to calculate the associated spectral flux density $f_{\nu}$:
\begin{equation}
    f_{\nu} = \frac{2h\nu^{3}}{c^2} \frac{1}{e^{\frac{h\nu}{kT}}-1}\, \left(\frac{R}{D_L}\right)^2,
\end{equation}
where $h$ is the Planck constant, $\nu$ is the frequency corresponding to $\lambda_{\mathrm{eff}}$ of a particular band, $c$ is the speed of light, $k$ is the Boltzmann constant, $T$ is the blackbody temperature, $R$ is the radius of the ejecta and $D_L$ is the luminosity distance of the system. The spectral flux can then be converted to AB magnitude (AB mag) using
\begin{equation}
    m_{\mathrm{AB}} = -2.5\,\mbox{log}_{10}\,f_{\nu}\,-\,48.6.
\end{equation}

For a given band, if the minimum value of the $m_{\mathrm{AB}}$ time-series is less than the limiting magnitude $m_{\mathrm{lim}}$ for that band (i.e., the peak luminosity of the KN is brighter than the threshold for the band), then we claim that the KN will be \textit{observed} by the corresponding EM telescope. In contrast, the criteria for \textit{detection} of a KN requires more consideration. Specifically, two consecutive exposures with a time lag of $>30$ minutes can be used to rule out fast-moving objects, like asteroids. Note that the model used to generate a KN assumes the emission to be isotropic. Angle dependence in the luminosity function can result in lower peak luminosities than what we obtain, potentially lowering the number of detections. However, as the code used to generate KN light curves is only valid for systems where $\chi_{BH}<0.75$, the number of KN we report is inherently lower than what can be expected for the two populations, as systems with high prograde BH spins are expected to result in KN emission for larger mass ratios. Furthermore, the analysis uses limiting magnitudes for the two telescopes corresponding to single exposure times of $30s$ for Rubin and $67s$ for \textit{Roman} \cite{Chase:2021ood}. Longer exposure times, possibly due to TOO searches, can improve the limiting magnitudes resulting in more KN detections than the ones reported in this study. For a more comprehensive treatment towards detection of KN from NSBH mergers, see Ref. \cite{Zhu:2020ffa}.

\begin{figure}[htbp]
\includegraphics[width=\columnwidth]{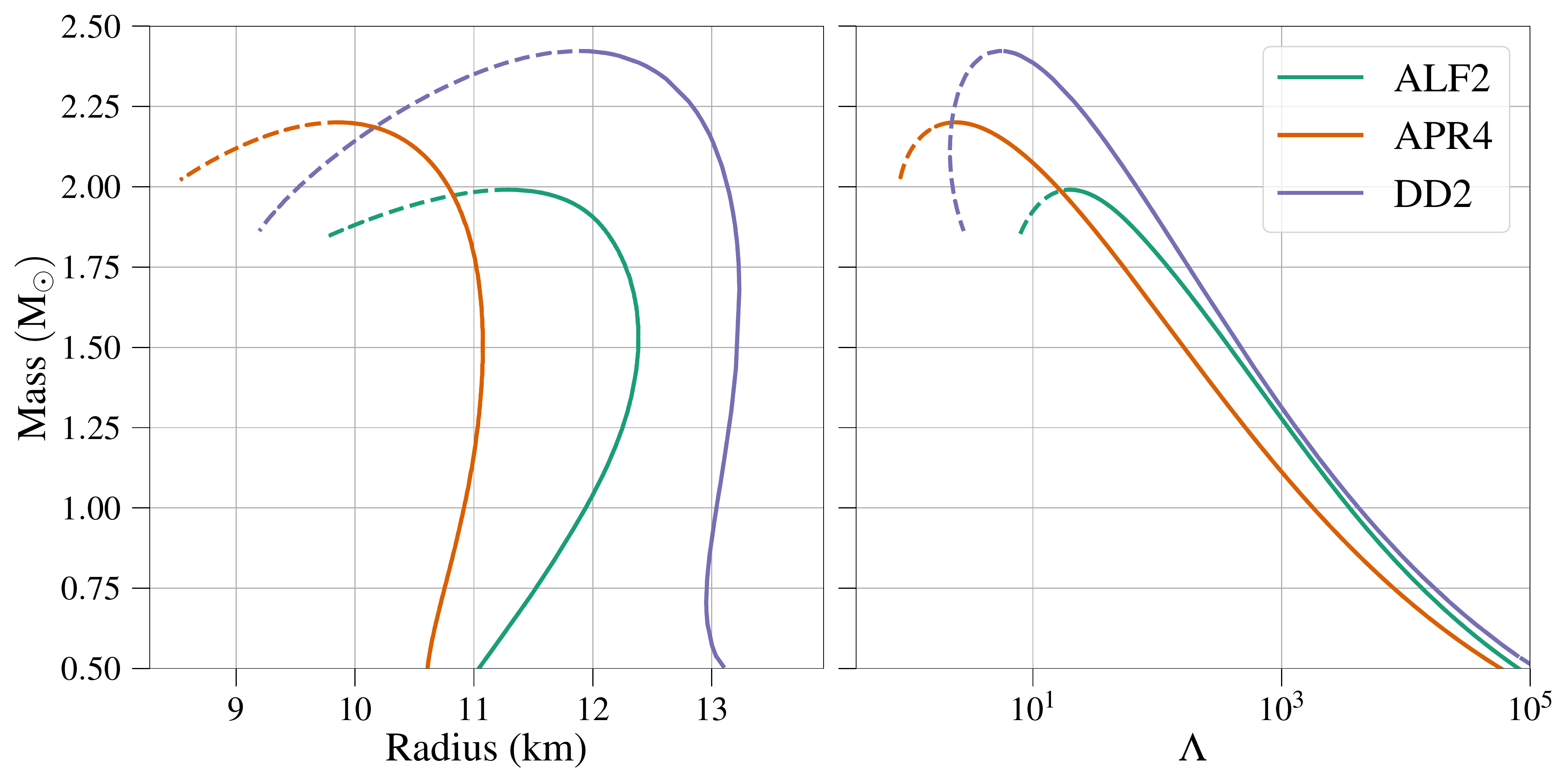}
\caption{\label{fig:EOS_Mass_Radius}The mass-radius relationship for the three equations of state considered in this study, along with the corresponding values of tidal deformability parameter $\Lambda$. The dashed part depicts the unstable branch with $dr/dm>0$.}
\end{figure}

The amount of ejecta in an NSBH merger depends on the unknown NS EOS. To account for this ignorance, for each system we compute the luminosity curves for three EOSs with varying stiffness: \texttt{APR4} \cite{Akmal:1998cf}, \texttt{DD2}  \cite{Typel:2009sy} and \texttt{ALF2} \cite{Alford:2004pf}. The mass-radius curves for the three EOSs and the corresponding curves for the tidal deformability parameter $\Lambda$ are given in Fig. \ref{fig:EOS_Mass_Radius}. We find that the largest number of KN detections are obtained with the $g$ and $r$-filter in the Rubin observatory and the $R$-filter in the \textit{Roman} telescope. We will present a detailed analysis for detections corresponding to the $r$-filter in Rubin and the $R$-filter in the \textit{Roman}. The number of detections for all the filters for the two telescopes can be found in Appendix \ref{appsec:no_of_kn_det} in Tables \ref{apptab:no_of_kn_det_pop1} and \ref{apptab:no_of_kn_det_pop2}.

\begin{figure*}[htbp]
  \includegraphics[width=1.03\textwidth]{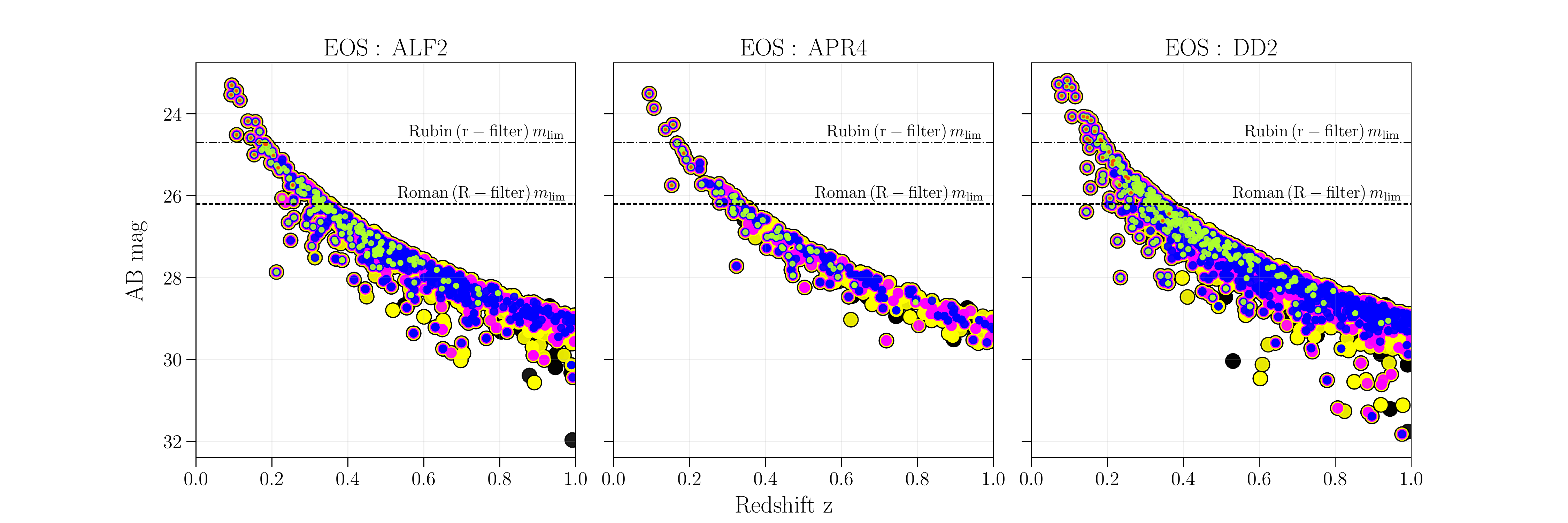}
  \includegraphics[width=1.03\textwidth]{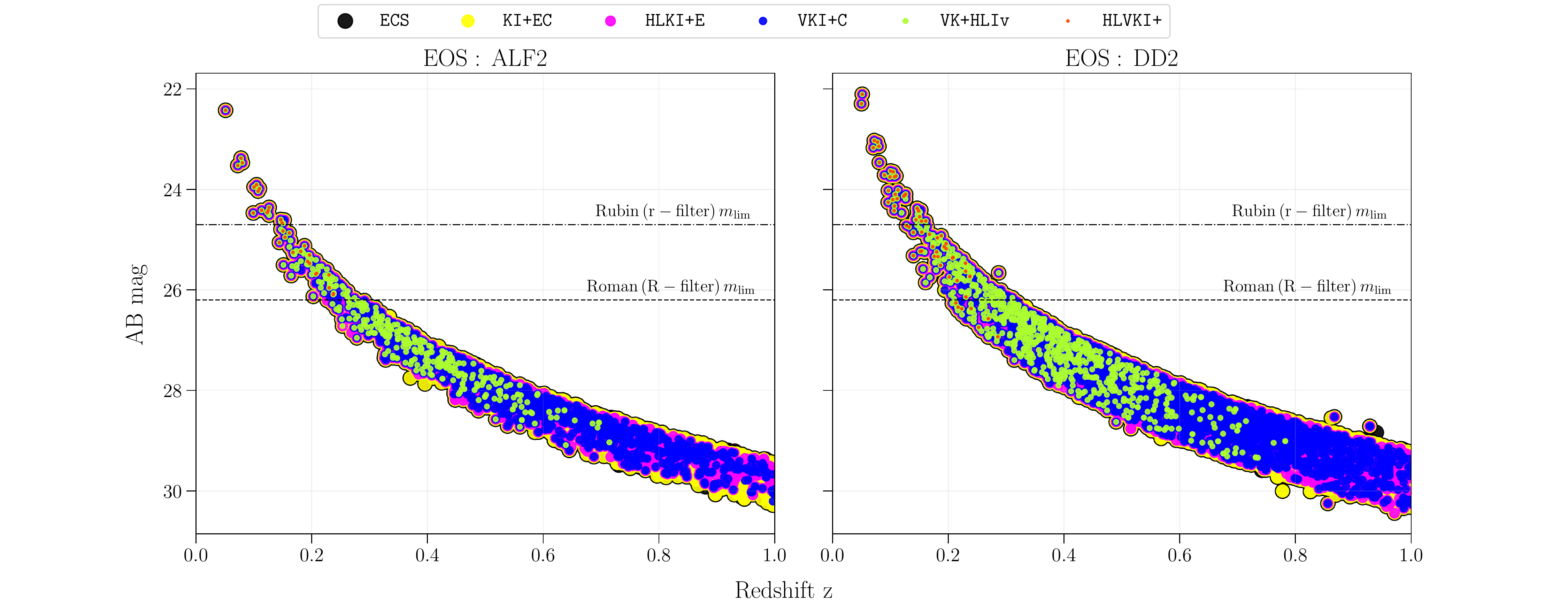}
  \caption{\label{fig:KNs}\textit{Top panel:} The peak luminosity of KN associated with detected NSBH mergers in Pop-1 in a span of $10$ years, as a function of redshift for the three EOSs. The color of the dots denotes the GW detector that detected the corresponding NSBH merger event. The size of the dots denotes the number of merger events detected by the particular GW network, in an observation span of $10$ years, that result in a KN. The dashed and dashed-dotted horizontal lines denote the limiting magnitudes for the $R$-filter in \textit{Roman} and $r-$filter in Rubin respectively. \textit{Bottom panel:} The peak luminosity of KN associated with NSBH mergers in Pop-2 detected in an observation time of $10$ years, as a function of redshift. No KN are obtained for the \texttt{APR4} EOS for Pop-2 events (see Table \ref{tab:r_R_kn_numbers}).}
\end{figure*}

\begin{figure*}[htbp]
\centering
\includegraphics[width=1.05\textwidth]{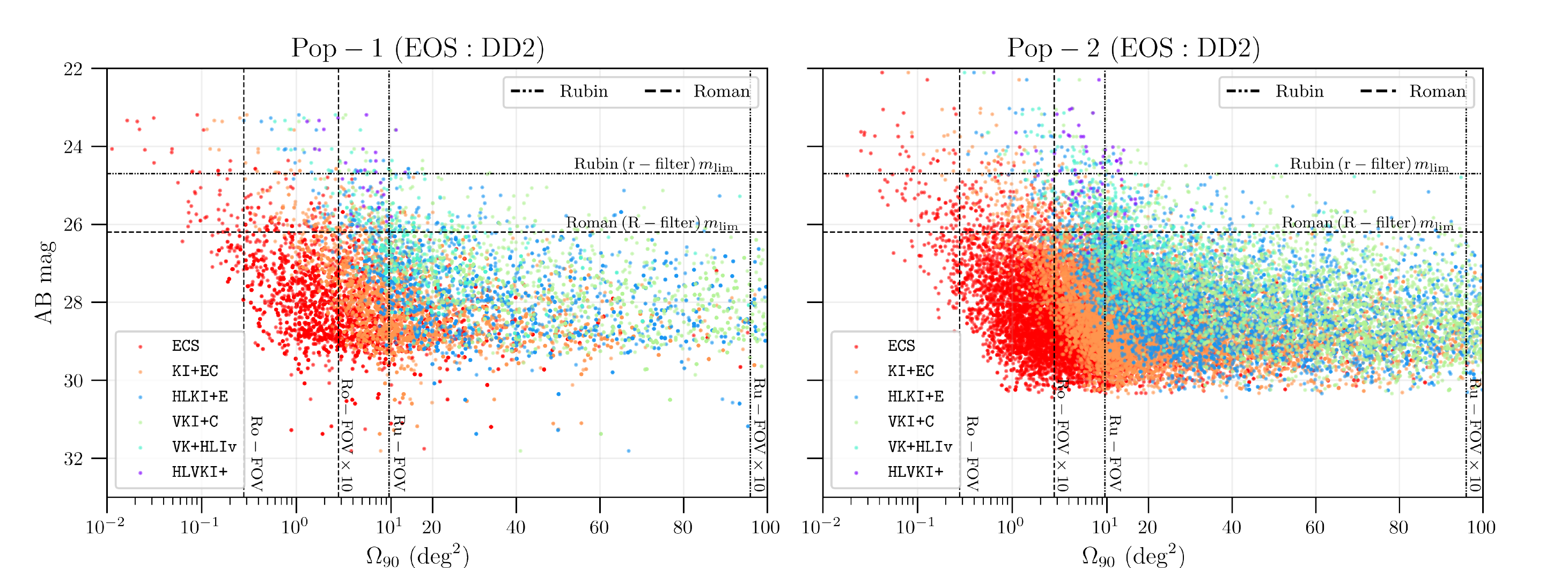}
\caption{\label{fig:KNvsa90}The peak luminosities of KN, when \texttt{DD2} is used as the EOS for the NS, associated with detected NSBH mergers in Pop-1 and Pop-2 in a span of $10$ years, as a function of $\Omega_{90}$ obtained from GW observations. The horizontal dashed and dashed-dotted lines show the limiting magnitudes for \textit{Roman} (Ro) and Rubin (Ru) respectively. The vertical dashed (dashed-dotted) lines show the sky-area corresponding to the FOV and 10 times the FOV for \textit{Roman} (Rubin).}
\end{figure*}

Figure \ref{fig:KNs} shows the peak luminosities of KN for events in Pop-1 and Pop-2 as a function of redshift in an observation time of $10$ years. The largest number of KN are obtained for \texttt{DD2} and the least for \texttt{APR4}. This is consistent with the fact that \texttt{DD2} is the stiffest among the three EOSs considered, whereas \texttt{APR4} is the softest (see Fig. \ref{fig:EOS_Mass_Radius}). The \textit{Roman} telescope has a better limiting magnitude $(m_{\mathrm{lim}}^{R} = 26.2)$ than the Rubin observatory $(m_{\mathrm{lim}}^{r} = 24.7)$ and, consequently, it is expected to observe more KN, which is also seen in Fig. \ref{fig:KNs}. In fact, almost all the events observed by Rubin lie within $z\sim0.2$, whereas \textit{Roman} is able to see KN up to $z\sim0.4$.  However, the number of KN that \textit{Roman} can observe is limited by its small FOV, $\sim 34$ times smaller than the FOV of Rubin. In Table \ref{tab:r_R_kn_numbers}, we present the number of KN observed by each telescope, in an observation time of $10$ years, for the six GW detector networks. The events are categorized based on if they can be localized in the sky using GW observations to better than the FOV of the particular telescope, $10$ times the FOV of the telescope or $100$ $\mbox{deg}^2$. The upper and lower limits with each number in the table show the uncertainty in the number of KN detections from NSBH systems due to the uncertainty in the local merger rate of the NSBH mergers. We see that almost all the events that Rubin will observe will be localized to an area in the sky that is smaller than the FOV of the telescope. This is not the case for \textit{Roman} as only GW events detected by \texttt{KI+EC} and \texttt{ECS} are seen to have $\Omega_{90} <$ FOV(\textit{Roman}). If we assume that \textit{Roman} can slew and cover $10$ sky-patches, it is capable of observing 5--8 times (depending on the EOS) more KN than the Rubin observatory. This is illustrated in Fig. \ref{fig:KNvsa90}, where we have plotted the peak luminosities of KN corresponding to NSBH mergers from the two populations with the corresponding $\Omega_{90}$ obtained from GW observations. Increasing the sky area covered by Rubin to $10$ times its FOV does not significantly increase the number of KN seen by the observatory. For \textit{Roman}, slewing the telescope to cover $10$ times its FOV increases the number of KN detections by $\sim 2$--$10$ times, depending on the GW detector and the EOS, still leaving out $\sim 10$--$50\%$ of the events with $\Omega_{90} \leq 100$ $\mbox{deg}^2$ that it can potentially detect.   

\begin{table*}[htbp] 
  \centering
  \caption{\label{tab:r_R_kn_numbers}The number of KN detections with the $r-$filter of Rubin Observatory and $R-$filter of the \textit{Roman Telescope} for both Pop-1 and Pop-2 for an observation time of 10 years. The events are categorized based on if they can be localized in the sky, using GW observations, better than the FOV of the EM telescope, 10 times the FOV of the EM telescope, or $100\,\,\mbox{deg}^2$. They have been further divided into 3 columns based on the EOS that was used to generate the KN light curves.}
  \renewcommand{\arraystretch}{1.3} 
    \begin{tabular}{l | P{1.3cm} P{1.3cm} P{1.3cm} | P{1.3cm} P{1.3cm} P{1.3cm} | P{1.3cm} P{1.3cm} P{1.3cm}}
    \hhline{==========}
    \multicolumn{10}{c}{\textbf{Pop-1}} \\
    \hhline{----------}
    \multicolumn{10}{c}{Rubin $r-$filter} \\
    \hhline{----------}
    Quality & \multicolumn{3}{c|}{$\Omega_{90}<$ FOV} & \multicolumn{3}{c|}{$\Omega_{90}<$ $10\,\times\,$FOV} & 
    \multicolumn{3}{c}{$\Omega_{90}<$ $100 \,\mbox{deg}^2$}\\
    \hhline{----------}
    EOS & ALF2 & APR4 & DD2 & ALF2 & APR4 & DD2 & ALF2 & APR4 & DD2 \\
    \hhline{----------}
    HLVKI\texttt{+} & $9^{+13}_{-7}$ & $4^{+4}_{-3}$ & $16^{+23}_{-13}$ & $9^{+16}_{-8}$ & $4^{+5}_{-3}$ & $16^{+28}_{-15}$ & $9^{+16}_{-8}$ & $4^{+5}_{-3}$ & $16^{+28}_{-15}$ \\
    VK\texttt{+}HLIv & $14^{+21}_{-12}$ & $4^{+7}_{-3}$ & $23^{+33}_{-19}$ & $14^{+25}_{-13}$ & $4^{+9}_{-3}$ & $23^{+42}_{-22}$ & $14^{+25}_{-13}$ & $4^{+9}_{-3}$ & $23^{+42}_{-22}$ \\
    HLKI\texttt{+}E & $14^{+21}_{-11}$ & $4^{+7}_{-3}$ & $23^{+33}_{-18}$ & $14^{+25}_{-13}$ & $4^{+9}_{-3}$ & $23^{+42}_{-22}$ & $14^{+25}_{-13}$ & $4^{+9}_{-3}$ & $23^{+42}_{-22}$ \\
    VKI\texttt{+}C & $14^{+17}_{-7}$ & $4^{+5}_{-3}$ & $23^{+29}_{-14}$ & $14^{+25}_{-13}$ & $4^{+9}_{-3}$ & $23^{+42}_{-22}$ & $14^{+25}_{-13}$ & $4^{+9}_{-3}$ & $23^{+42}_{-22}$ \\
    KI\texttt{+}EC & $14^{+24}_{-13}$ & $4^{+9}_{-3}$ & $23^{+40}_{-22}$ & $14^{+25}_{-13}$ & $4^{+9}_{-3}$ & $23^{+42}_{-22}$ & $14^{+25}_{-13}$ & $4^{+9}_{-3}$ & $23^{+42}_{-22}$ \\
    ECS & $14^{+25}_{-13}$ & $4^{+9}_{-3}$ & $23^{+42}_{-22}$ & $14^{+25}_{-13}$ & $4^{+9}_{-3}$ & $23^{+42}_{-22}$ & $14^{+25}_{-13}$ & $4^{+9}_{-3}$ & $23^{+42}_{-22}$ \\
    \hhline{----------}
    \multicolumn{10}{c}{Roman $R-$filter} \\
    \hhline{----------}
    Quality & \multicolumn{3}{c|}{$\Omega_{90}<$ FOV} & \multicolumn{3}{c|}{$\Omega_{90}<$ $10\,\times\,$FOV} & 
    \multicolumn{3}{c}{$\Omega_{90}<$ $100 \,\mbox{deg}^2$}\\
    \hhline{----------}
    EOS & ALF2 & APR4 & DD2 & ALF2 & APR4 & DD2 & ALF2 & APR4 & DD2 \\
    \hhline{----------}
    HLVKI\texttt{+} & $0$ & $0$ & $0$ & $3^{+4}_{-3}$ & $1^{+0}_{-1}$ & $6^{+7}_{-6}$ & $16^{+30}_{-15}$ & $7^{+6}_{-6}$ & $29^{+52}_{-27}$ \\
    VK\texttt{+}HLIv & $0$ & $0$ & $0$ & $8^{+16}_{-8}$ & $3^{+5}_{-3}$ & $19^{+42}_{-19}$ & $66^{+131}_{-51}$ & $23^{+50}_{-17}$ & $118^{+242}_{-97}$ \\
    HLKI\texttt{+}E & $0^{+1}_{-0}$ & $0$ & $0^{+1}_{-0}$ & $15^{+28}_{-14}$ & $5^{+7}_{-4}$ & $26^{+61}_{-25}$ & $96^{+168}_{-74}$ & $39^{+68}_{-30}$ & $169^{+313}_{-138}$ \\
    VKI\texttt{+}C & $0$ & $0$ & $0$ & $3^{+5}_{-3}$ & $1^{+1}_{-1}$ & $9^{+18}_{-9}$ & $84^{+151}_{-64}$ & $30^{+56}_{-22}$ & $155^{+293}_{-126}$ \\
    KI\texttt{+}EC & $3^{+5}_{-3}$ & $1^{+1}_{-1}$ & $6^{+8}_{-6}$ & $56^{+100}_{-45}$ & $21^{+36}_{-15}$ & $100^{+193}_{-84}$ & $97^{+170}_{-75}$ & $39^{+68}_{-30}$ & $171^{+318}_{-140}$ \\
    ECS & $26^{+52}_{-20}$ & $12^{+22}_{-9}$ & $50^{+109}_{-41}$ & $88^{+157}_{-68}$ & $33^{+60}_{-25}$ & $156^{+293}_{-128}$ & $97^{+170}_{-75}$ & $39^{+68}_{-30}$ & $171^{+318}_{-140}$ \\
    \hhline{----------}
    \multicolumn{10}{c}{\textbf{Pop-2}} \\
    \hhline{----------}
    \multicolumn{10}{c}{Rubin $r-$filter} \\
    \hhline{----------}
    Quality & \multicolumn{3}{c|}{$\Omega_{90}<$ FOV} & \multicolumn{3}{c|}{$\Omega_{90}<$ $10\,\times\,$FOV} & 
    \multicolumn{3}{c}{$\Omega_{90}<$ $100 \,\mbox{deg}^2$}\\
    \hhline{----------}
    EOS & ALF2 & APR4 & DD2 & ALF2 & APR4 & DD2 & ALF2 & APR4 & DD2 \\
    \hhline{----------}
    HLVKI\texttt{+} & $16^{+16}_{-11}$ & $0$ & $31^{+33}_{-19}$ & $16^{+18}_{-14}$ & $0$ & $31^{+41}_{-26}$ & $16^{+18}_{-14}$ & $0$ & $31^{+41}_{-26}$ \\
    VK\texttt{+}HLIv & $19^{+22}_{-14}$ & $0$ & $37^{+58}_{-28}$ & $19^{+30}_{-16}$ & $0$ & $37^{+82}_{-31}$ & $19^{+30}_{-16}$ & $0$ & $37^{+82}_{-31}$ \\
    HLKI\texttt{+}E & $20^{+24}_{-15}$ & $0$ & $38^{+61}_{-30}$ & $20^{+30}_{-17}$ & $0$ & $38^{+84}_{-32}$ & $20^{+30}_{-17}$ & $0$ & $38^{+84}_{-32}$ \\
    VKI\texttt{+}C & $20^{+20}_{-15}$ & $0$ & $38^{+47}_{-23}$ & $20^{+30}_{-17}$ & $0$ & $38^{+83}_{-32}$ & $20^{+30}_{-17}$ & $0$ & $38^{+83}_{-32}$ \\
    KI\texttt{+}EC & $20^{+30}_{-17}$ & $0$ & $38^{+82}_{-32}$ & $20^{+30}_{-17}$ & $0$ & $38^{+84}_{-32}$ & $20^{+30}_{-17}$ & $0$ & $38^{+84}_{-32}$ \\
    ECS & $20^{+30}_{-17}$ & $0$ & $38^{+84}_{-32}$ & $20^{+30}_{-17}$ & $0$ & $38^{+84}_{-32}$ & $20^{+30}_{-17}$ & $0$ & $38^{+84}_{-32}$ \\
    \hhline{----------}
    \multicolumn{10}{c}{Roman $R-$filter} \\
    \hhline{----------}
    Quality & \multicolumn{3}{c|}{$\Omega_{90}<$ FOV} & \multicolumn{3}{c|}{$\Omega_{90}<$ $10\,\times\,$FOV} & 
    \multicolumn{3}{c}{$\Omega_{90}<$ $100 \,\mbox{deg}^2$}\\
    \hhline{----------}
    EOS & ALF2 & APR4 & DD2 & ALF2 & APR4 & DD2 & ALF2 & APR4 & DD2 \\
    \hhline{----------}
    HLVKI\texttt{+} & $0$ & $0$ & $0$ & $3^{+5}_{-2}$ & $0$ & $6^{+12}_{-4}$ & $32^{+42}_{-27}$ & $0$ & $59^{+92}_{-51}$ \\
    VK\texttt{+}HLIv & $0$ & $0$ & $0$ & $16^{+21}_{-12}$ & $0$ & $29^{+43}_{-24}$ & $93^{+222}_{-76}$ & $0$ & $196^{+435}_{-163}$ \\
    HLKI\texttt{+}E & $0$ & $0$ & $0^{+2}_{-0}$ & $22^{+42}_{-17}$ & $0$ & $44^{+84}_{-36}$ & $122^{+275}_{-99}$ & $0$ & $260^{+542}_{-213}$ \\
    VKI\texttt{+}C & $0$ & $0$ & $0$ & $7^{+10}_{-5}$ & $0$ & $14^{+21}_{-11}$ & $115^{+262}_{-92}$ & $0$ & $243^{+511}_{-197}$ \\
    KI\texttt{+}EC & $7^{+9}_{-5}$ & $0$ & $12^{+18}_{-9}$ & $79^{+173}_{-64}$ & $0$ & $160^{+330}_{-132}$ & $124^{+281}_{-100}$ & $0$ & $267^{+555}_{-218}$ \\
    ECS & $50^{+94}_{-41}$ & $0$ & $100^{+182}_{-83}$ & $115^{+262}_{-92}$ & $0$ & $247^{+512}_{-202}$ & $124^{+281}_{-100}$ & $0$ & $267^{+555}_{-218}$ \\
    \hhline{==========}
    \end{tabular}
\end{table*}
We also note that, in general, the number of expected KN observations corresponding to events in Pop-2 is $\sim 1.5$--$3$ times larger than for events in Pop-1. While a larger fraction of events in Pop-1 have $q<4$ compared to Pop-2 (see Fig. \ref{fig:CDF_massratio}), a significant fraction of systems in Pop-1 contain BHs with retrograde $(\chi_{BH}<0)$ spin, which is a disincentive to tidal disruption before $R_{ISCO}$. However, no KN is detected for events in Pop-2 with \texttt{APR4} as the EOS. This can be explained by noticing that a large fraction of events in Pop-2 contain non-spinning and $5-15$ $\msun$ BHs (see Fig. \ref{appfig:Pop2_mass_spins} in Appendix \ref{app:Pop2_params}) and are unable to tidally disrupt NSs that obey \texttt{APR4}, as \texttt{APR4} leads to the formation of the most compact NSs among the three EOSs (see Fig. \ref{fig:EOS_Mass_Radius}). 

As discussed before, the single-exposure observation criteria might not be suitable in practice, as one needs at least two exposures to differentiate KN emissions from fast-moving objects. For a more realistic picture regarding the number of KN detections, we use a TOO strategy for Rubin which is similar to the approaches discussed in Refs. \cite{Andreoni:2021epw,Branchesi:2023mws}. To claim a KN detection with Rubin, it has to be observed in the $g+i$ filters on two consecutive nights. For each filter, we assume a $600$s single-exposure observation, leading to a limiting magnitude of $26.62$ for the $g-$filter and $25.62$ for the $i-$filter (assuming the most optimistic configurations). To not take a large portion of Rubin's time by making it slew and cover $10$ patches in the sky, we restrict ourselves to KN for which the corresponding GW detections can constrain the sky-area associated with the binary to within $9.6\,\mbox{deg}^2$, i.e., the FOV of Rubin. We also assume a duty cycle of $50\%$ for the Rubin observatory. With these specifications, the number of KN detections with Rubin in an observation span of $10$ years are listed in Table \ref{tab:ToO_Rubin}.
\begin{table}[h] 
  \centering
  \caption{\label{tab:ToO_Rubin}The number of KN detections in an observation span of $10$ years using the $g+i$ target-of-opportunity strategy with the Rubin observatory. The considered events have $\Omega_{90}\leq9.6\,\mbox{deg}^2$, which is equal to the FOV of the Rubin observatory. We also assume a duty cycle of $50\%$.}
  \renewcommand{\arraystretch}{1.4} 
    \begin{tabular}{l P{1.3cm}P{1.3cm}P{1.3cm}}
    \hhline{====}
    \multicolumn{4}{c}{\textbf{Pop-1}} \\
    \hhline{----}
    Network & ALF2 & APR4 & DD2 \\
    \hhline{----}
    HLVKI\texttt{+} & $6^{+12}_{-6}$ & $3^{+2}_{-3}$ & $12^{+18}_{-11}$\\
    VK\texttt{+}HLIv & $12^{+22}_{-10}$ & $4^{+6}_{-4}$ & $20^{+36}_{-16}$\\
    HLKI\texttt{+}E & $12^{+30}_{-10}$ & $5^{+11}_{-5}$ & $22^{+45}_{-19}$\\
    VKI\texttt{+}C & $8^{+16}_{-7}$ & $3^{+3}_{-3}$ & $16^{+28}_{-14}$\\
    KI\texttt{+}EC & $20^{+40}_{-17}$ & $6^{+14}_{-5}$ & $30^{+63}_{-26}$\\
    ECS & $20^{+41}_{-16}$ & $6^{+14}_{-5}$ & $32^{+66}_{-26}$\\
    \hhline{----}
    \multicolumn{4}{c}{\textbf{Pop-2}} \\
    \hhline{----}
    Network & ALF2 & APR4 & DD2 \\
    \hhline{----}
    HLVKI\texttt{+} & $9^{+11}_{-7}$ & $0$ & $17^{+25}_{-14}$\\
    VK\texttt{+}HLIv & $12^{+24}_{-10}$ & $0$ & $29^{+55}_{-25}$\\
    HLKI\texttt{+}E & $14^{+26}_{-12}$ & $0$ & $34^{+67}_{-29}$\\
    VKI\texttt{+}C & $12^{+18}_{-10}$ & $0$ & $23^{+41}_{-19}$\\
    KI\texttt{+}EC & $16^{+39}_{-13}$ & $0$ & $44^{+101}_{-36}$\\
    ECS & $16^{+40}_{-13}$ & $0$ & $45^{+106}_{-37}$\\
    \hhline{====}
    \end{tabular}
\end{table}
We compare the values in Table \ref{tab:ToO_Rubin} with the number of KN observed in a span of $10$ years for all events with $\Omega_{90} < 100$ $\mbox{deg}^2$ specified in Tables \ref{apptab:no_of_kn_det_pop1} and \ref{apptab:no_of_kn_det_pop2}. It is seen that more KN detections are observed following the $g+i$ TOO strategy, compared to the KN observed using a single filter and an exposure time of $30s$, despite the use of a more stringent sky-resolution criteria and $50\%$ duty cycle in the TOO strategy. This is because we use an exposure time of $600s$ for the $g+i$ filters, which significantly improves the limiting magnitude of the $g$ and $i$ filters ($\sim 6\%$) compared to the single 30s-exposure case, leading to a greater number of KN detections. It is safe to assume that a similar strategy for \textit{Roman} will improve the number of KN detections with \textit{Roman} as well. 

Depending on the EOS, one can expect to observe $\mathcal{O}(1)$ to $\mathcal{O}(10)$ KN with Rubin and $0$ to $\mathcal{O}(100)$ KN with \textit{Roman} in an observation span of $10$ years. Even in the best case scenario, less than $10\%$ of the events in the populations result in a KN, out of which less $10\%$ are detected, giving the total number of detected KN to be $< 1\%$ of the cosmic population of NSBH systems, which is consistent with the estimates reported in Refs. \cite{Zhu:2021jbw,Biscoveanu:2022iue}. There is a significant difference between the number of KN expected to be observed based on the EOS used. This points to the possibility of deriving information about the EOS based solely on the number of KN detected in the future. If we assume that the local merger rate is known to be around the median value of $45\,\mbox{Gpc}^{-1}\mbox{yr}^{-1}$, non-detectability of KN from NSBH mergers in the coming years can point in favor of compact NSs governed by softer EOSs. Further, subject to the completeness of galaxy catalogs, detection of KN from NSBH mergers will allow the localization of the host galaxy from which an accurate estimate of the redshift associated with the system can be obtained. Together with the constraints on luminosity distance, NSBH mergers can then be used as an independent tool to measure the Hubble constant \cite{Vitale:2018wlg}.  


\section{Conclusions} \label{sec:concl}
Neutron star-black hole binaries were first discovered in 2020 during the third observing run of the LIGO and Virgo detectors. With two confirmed detections we can be confident that these intriguing systems will be abundantly observed by upgraded detectors and new observatories.  With large mass asymmetries and black hole spins either large or misaligned with the orbital angular momentum, we can expect NSBH signals to reveal relativistic gravity in action with unprecedented detail. Neutron star-black hole binaries will be particularly interesting as they could emit a significant fraction of their energy in higher multipole modes allowing precision tests of general relativity but also enabling accurate measurement of the Hubble parameter. 

In this study, we have evaluated the science potential of NSBH binaries in two networks comprising of upgraded LIGO and Virgo (A\texttt{+} and Voyager upgrades) and four networks comprising of one or more of Cosmic Explorer and Einstein Telescope operating in tandem with upgraded LIGO and Virgo (cf.\, Fig.\,\ref{fig:det_sens} and Table \ref{tab:net}). We consider two different population models for NSBH systems (cf.\, Table \ref{tab:pop_par}) but our main conclusions equally apply to both of the populations. The performance of the networks was evaluated using several metrics as follows:

\paragraph{\bf Detection rate:} 
The cosmological merger rate of NSBH populations, assuming they evolve with redshift in the same way star formation rate does (apart from time delays), is about 40,000 events per year (cf.\,Fig.\,\ref{fig:eff_rate}, right panel and Table \ref{tab:pop_snr}). At the detection signal-to-noise ratio threshold of $\rho_{*}=10,$ the A\texttt{+} and Voyager upgrades will see 1\% and 10\% of the mergers, respectively, while future observatories will observe more than 90\% of this population. There is great utility to large catalogs as they can help discriminate between different astrophysical formation channels of NSBH or facilitate cosmological measurements. For other applications, such as tests of general relativity, the signal quality is of prime importance. While imminent upgrades will not witness high fidelity signals of SNR $>100$, several tens to hundreds of such events will be observed each year by CE and ET.   

\paragraph{\bf Detection efficiency:} 
The merging population of NSBH increases steeply with redshift as the star formation rate grows, but tapers off at around a redshift of $z=2$ (cf.\,Fig.\,\ref{fig:eff_rate}, right panel). While the A\texttt{+} and Voyager upgrades will have a \emph{redshift reach}\footnote{Here, the reach of a network is defined as the redshift at which the network will observe 50\% of all the sources at that redshift.}  of $z\sim 0.2$ and $z\sim 0.5,$ respectively (cf.\,Fig.\,\ref{fig:eff_rate}, left panel), future networks will have a reach of $z\sim 1.5$ to 6 depending on the number of detectors in the network. In particular, the ECS network comprising of one Einstein Telescope and two Cosmic Explorers will observe more than 90\% of all the sources at $z=2.$ This degree of completeness will help mitigate systematics arising from an incomplete catalog. 

\paragraph{\bf Sky localization:} 
A metric of importance is the degree to which a source can be localized in the sky. Precise localization helps in the EM follow-up of GW transients, measurement of cosmological parameters, identification of lensed events, and so on. Imminent upgrades will localize hundreds to thousands of events to within 10 deg$^2$ but it takes a pair of XG observatories to localize 30\% of the events, or tens of thousands, to within the same error region. The number of events that can be localized to within 1 deg$^2$ is typically a factor 30 smaller for all networks except the network with three XG observatories, for which it reduces only by a factor of $5.$ However, not all of these events can be followed up even by the best optical and infrared telescopes, but only mergers within a redshift of 0.5. Within this redshift, the number of available sources for EM follow-up will not change for A\texttt{+} and Voyager networks, but they are ten times smaller, i.e., thousands of mergers, for XG observatories. 

\paragraph{\bf Kilonova detection:}
Kilonova emission in the aftermath of an NSBH merger depends on a number of factors including the ratio of black hole to neutron star mass $q=m_{\rm BH}/m_{\rm NS},$  black hole's spin and the unknown equation of state of neutron stars.  Merger ejecta will be larger, and kilonova brighter, for binaries with smaller mass ratios, 
stiffer equations of state such as \texttt{DD2}, and larger black hole spins. However, kilonovae fail to materialize for large mass ratios and softer equations of state such as \texttt{APR4}. Unfortunately, the number of mergers that could lead to observable kilonova will likely be a few per year even in the most optimistic case of stiffer equations of state.

\paragraph{\bf Early warning alerts:} 
Telescopes can benefit from receiving an alert of an imminent merger minutes earlier as that would help to observe the onset of the EM counterpart and the central engine that triggers the gamma ray burst. A\texttt{+} and Voyager networks will not be able to issue well-localized, early warning alerts two or more minutes before merger. Next generation observatories, on the other hand, should be able to issue alerts to tens or hundreds of NSBH mergers with sky localization of 100 deg$^2$ or less every year. Alerts could be sent 5 minutes before merger for a handful of these and a similar number will be localized to within 10 deg$^2.$

In summary, NSBH mergers will not only be seen in plenty with the next generation observatories, but they will also provide insights into some of the key science questions in astrophysics and cosmology. 

\section*{Acknowledgements} \label{sec:acknowledgements}
We thank Marica Branchesi, Nandini Hazra, Philippe Landry, David Radice and Salvatore Vitale for useful discussions and comments. We would also like to thank Anuradha Gupta for her comments on the work. IG,  AD and BSS were supported by the National Science Foundation, USA, grant numbers PHY-2012083, AST-2006384 and PHY-2207638. SB acknowledges support from the Deutsche Forschungsgemeinschaft, DFG, project MEMI number BE 6301/2-1.  DC is supported by the Science and Technologies Facilities Council, UK, grant ST/V005618/1.

\bibliography{bibliography}
\appendix

\section{Fits for efficiency curves} \label{app:fit_sigmoid}
The detection efficiency curves for the six detector networks can be estimated accurately using three-parameter sigmoid functions [see Eq. (\ref{eq:sigmoid})]. The best fit values for $a$, $b$ and $c$ for the two populations are given in Table \ref{tab:sigmoid_fit_pars}
\begin{table}[htbp] 
  \centering
  \caption{The fitting parameters for sigmoid functions that approximate the efficiency curves for the six detector networks.}
  \renewcommand{\arraystretch}{1.5} 
  \resizebox{8.0cm}{!}{
    \begin{tabular}{ l |  c  c  c | c  c  c}
    \hhline{=======}
    Threshold & \multicolumn{3}{c|}{$\rho_{*} = 10$} & \multicolumn{3}{c}{$\rho_{*} = 100$} \\
    \hhline{-------}
    Parameter & \textit{a} & \textit{b} & \textit{c} & \textit{a} & \textit{b} & \textit{c} \\
    \hhline{-------}
    \multicolumn{7}{c}{\textit{Pop-1}}\\
    \hhline{-------}
    HLVKI\texttt{+} & 32.77 & 0.006026 & 0.2801 & 305 & 0.007523 & 0.3167\\
    VK\texttt{+}HLIv & 14.2 & 0.01263 & 0.2459 & 136.8 & 0.01231 & 0.296\\
    HLKI\texttt{+}E & 8.441 & 0.006133 & 0.081 & 61.09 & 0.0107 & 0.2016\\
    VKI\texttt{+}C & 4.046 & 0.146 & 0.07097 & 21.07 & 0.09252 & 0.3656\\
    KI\texttt{+}EC & 3.31 & 0.02143 & 0.06869 & 25.4 & 0.01808 & 0.2766\\
    ECS &  2.117 & 0.02636 & 0.06671 & 19.87 & 0.01551 & 0.2687\\
    \hhline{-------}
    \multicolumn{7}{c}{\textit{Pop-2}} \\
    \hhline{-------}
    HLVKI\texttt{+} & 32.77 & 0.006037 & 0.2801 & 305.7 & 0.007476 & 0.3156\\
    VK\texttt{+}HLIv & 14.21 & 0.01261 & 0.2455 & 136.7 & 0.01236 & 0.2961\\
    HLKI\texttt{+}E & 8.441 & 0.00613 & 0.081 & 61 & 0.01075 & 0.202\\
    VKI\texttt{+}C & 4.048 & 0.1458 & 0.07094 & 21.08 & 0.09252 & 0.3655\\
    KI\texttt{+}EC & 3.309 & 0.02143 & 0.06873 & 25.41 & 0.01808 & 0.2764\\
    ECS &  2.115 & 0.02644 & 0.06681 & 19.88 & 0.01549 & 0.2684\\
    \hhline{=======}
    \end{tabular}%
    }
  \label{tab:sigmoid_fit_pars}%
\end{table}%

\section{Redshift distributions} \label{app:redshift_dist}
For compact binary mergers, there is a time delay $t_d$ between the formation of the progenitor stars and the merger of the compact binary. The delay for a binary where the progenitors form at redshift $z_f$ and merge at redshift $z_m$ can be calculated using
\begin{equation}
    t_d = \frac{1}{H_0} \int_{z_m}^{z_f} \frac{dz}{(1+z)(\Omega_{\Lambda}+\Omega_{M}(1+z)^3)},
\end{equation}
where $H_0$ is the Hubble constant and $\Omega_{\Lambda}$ and $\Omega_M$ are dark energy and dark matter densities. The delay time depends on the properties of the binary \cite{Broekgaarden:2021iew} and can be expressed as a probability density $P(t_d)$. The merger rate density $\Dot{n}(z)$ is then proportional to the convolution of the SFR $\psi(z)$ with $P(t_d)$ integrated over all possible delay times
\begin{equation}
    \Dot{n}(z) \propto \int_{t_{d,min}}^{t_{d,max}} \psi (z_f(z,t_d))\,P(t_d)\,\,dt_d,
\end{equation}
where $t_{d,min}$ and $t_{d,max}$ are the minimum and maximum delay times. In this study, we use the analytical SFR model proposed in Ref. \cite{Yuksel:2008cu} which is given by \cite{Zhu:2021jbw}
\begin{equation}
    \psi(z) \propto \left[(1+z)^{3.4\eta}+\left(\frac{1+z}{5000}\right)^{-0.3\eta} + \left(\frac{1+z}{9}\right)^{-3.5\eta}\right]^{1/\eta},
\end{equation}
where $\eta = -10$. $P(t_d)$ is chosen to be the log-normal time delay distribution introduced in Ref. \cite{Wanderman:2014eza}, given by
\begin{equation}
    P(t_d) = \frac{1}{\sqrt{2\pi}\,\sigma}\,\,\mbox{exp}\left[\frac{-(\,\mbox{ln}\,t_d - \mbox{ln}\,t_{d,\mu})^2}{2\sigma^2}\right],
\end{equation}
with $t_{d,\mu} = 2.9$ Gyr and $\sigma=$ 0.2. The empirical redshift distribution accounting for the SFR and the time delay model is given by \cite{Zhu:2020ffa,Sun:2015bda},
\begin{multline}
    \Dot{n}(z) \propto \Bigg[\,(1+z)^{4.131\eta}\,+\, \left(\frac{1+z}{22.37}\right)^{-0.5789\eta}\\
    +\,\left(\frac{1+z}{2.978}\right)^{-4.735\eta}\,+\, \left(\frac{1+z}{2.749}\right)^{-10.77\eta}\\ +\,\left(\frac{1+z}{2.867}\right)^{-17.51\eta}\,
    +\,\left(\frac{1+z}{3.04}\right)^{-\frac{0.08148+z^{0.574}}{0.08682}\eta} \Bigg]^{1/\eta},
\end{multline}
where $\eta = -5.51$. This function is normalized by demanding that $\Dot{n}(0)$ is equal to the merger rate density informed by GW observations \cite{LIGOScientific:2021psn} and is plugged in Eq. (\ref{eq:merger_rate_density}) to calculate the merger rate density as a function of redshift.

Another SFR model that is used extensively in literature is the Madau-Dickinson model \cite{Madau:2014bja}
\begin{equation} \label{eq:madau_dickinson}
    \psi(z) = 0.015\frac{(1+z)^{2.7}}{1+[(1+z)/2.9]^{5.6}}.
\end{equation}
Following Ref. \cite{Regimbau:2012ir}, one can take $P(t_d) \propto 1/t_d$ with $t_{d,min} = 20$ Myr and $t_{d,max} = 10$ Gyr. Convolving and integrating over all time delays and fitting the result to a Madau-Dickinson-like form as in Eq. (\ref{eq:madau_dickinson}), we get
\begin{equation}
    \Dot{n}(z) = \phi_0 \frac{(1+z)^a}{1+[(1+z)/c]^b},
\end{equation}
where the fitting coefficients are given by $a=1.803219571,\,b=5.309821767,\, c=2.837264101,\,\phi_0=8.765949529$. 
\begin{figure}[htbp]
\centering
\includegraphics[scale=0.57]{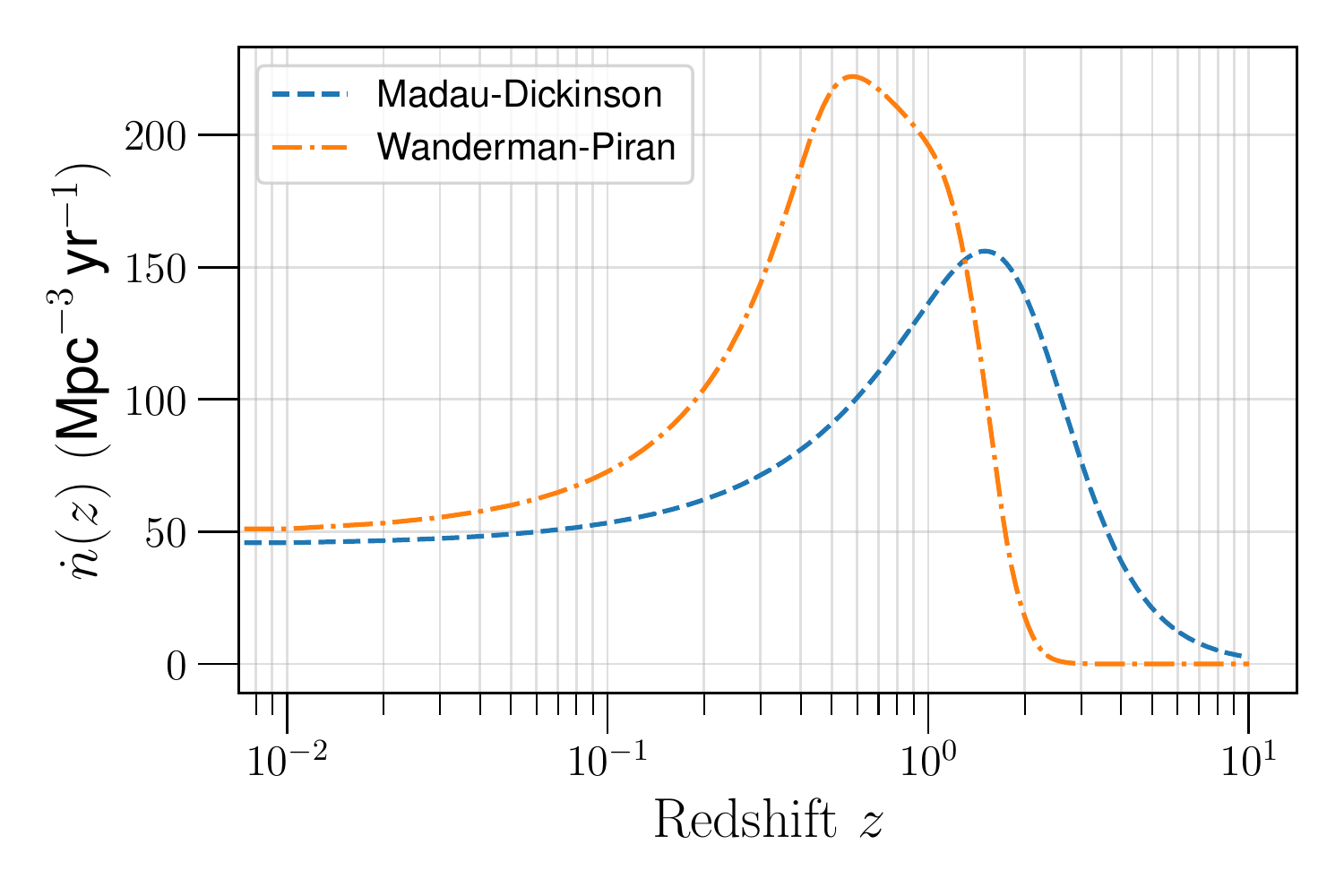}
\caption{\label{appfig:MDvsWP}The merger rate density as a function of redshift for the Madau-Dickinson model with $P(t_d)\propto 1/t_d$ and the Wanderman-Piran model using the log-normal delay time distribution. Both the distributions have been normalized such that the local merger rate $(\Dot{n}(0))$ is equal to $45\,\mbox{Gpc}^{-3}\,\mbox{yr}^{-1}$.}
\end{figure}

Figure \ref{appfig:MDvsWP} compares the merger rate density calculated using the two models discussed here. The merger rate density calculated using the Wanderman-Piran model peaks at a redshift of $z\sim 0.6$, after which it drops steeply, with no NSBH mergers expected after $z\sim 3$. In contrast, the Madau-Dickinson rate peaks at a redshift value close to $z=2$, gradually descending to zero merger rate density near $z = 10$. Using Eq. (\ref{eq:merger_rate_density}), as the differential comoving volume ($dV/dz'$) increases up to a redshift of $z\sim2.3$, the Madau-Dickinson model leads to a higher cosmic merger rate ($\sim 6.6\times10^4\,\mbox{yr}^{-1}$) than the Wanderman-Piran model ($\sim 4.0\times10^4\,\mbox{yr}^{-1}$), with a local NSBH merger rate of $45\,\mbox{Gpc}^{-3}\,\mbox{yr}^{-1}$ for both the distributions.

\section{Mass and spin distributions using the \textit{fiducial} model} \label{app:Pop2_params}
The Pop-2 in our study uses the mass profiles for the NSBH systems from the fiducial model from Ref. \cite{Broekgaarden:2021iew}. In this model, the NSs are assumed to be non-spinning. For BHs, we follow the fits described in Eqs. (2) and (3) of Ref. \cite{Chattopadhyay:2022cnp},
\begin{equation} \label{appeq:chi_BH}
    \chi_{BH}=
    \begin{cases}
      0, & \text{for}\ \mbox{log}_{10}\,P_{orb}>x_1 \\
      1, & \text{for}\  \mbox{log}_{10}\,P_{orb}<x_2 \\
      m\,\mbox{log}_{10}\,P_{orb}+c & \text{for}\ x_2 \leq \mbox{log}_{10}\,P_{orb} \leq x_1
    \end{cases}
\end{equation}
\begin{table}[htbp] 
  \centering
  \caption{Values for $x_1$, $x_2$, $m$ and $c$ for various values of metallicities $Z$ \cite{Chattopadhyay:2022cnp}}
  \renewcommand{\arraystretch}{1.5} 
  \resizebox{8.0cm}{!}{
    \begin{tabular}{ P{1.5cm}  P{1.5cm}  P{1.5cm}  P{1.5cm}  P{1.5cm} }
    \hhline{=====}
    Z & $x_1$ & $x_2$ & $m$ & $c$\\
    \hhline{-----}
    0.0005 & 0.3 & $-0.5$ & $-1.02$ & 0.63 \\
    0.001 & 0.5 & $-0.5$ & $-0.70$ & 0.54 \\
    0.005 & 0.5 & $-0.5$ & $-0.70$ & 0.54 \\
    0.02 & 0.5 & $-0.5$ & $-0.87$ & 0.57\\ 
    \hhline{=====}
    \end{tabular}%
    }
  \label{tab:spin_params}%
\end{table}%

\begin{figure*}[ht]
\centering
\includegraphics[width=0.99\textwidth]{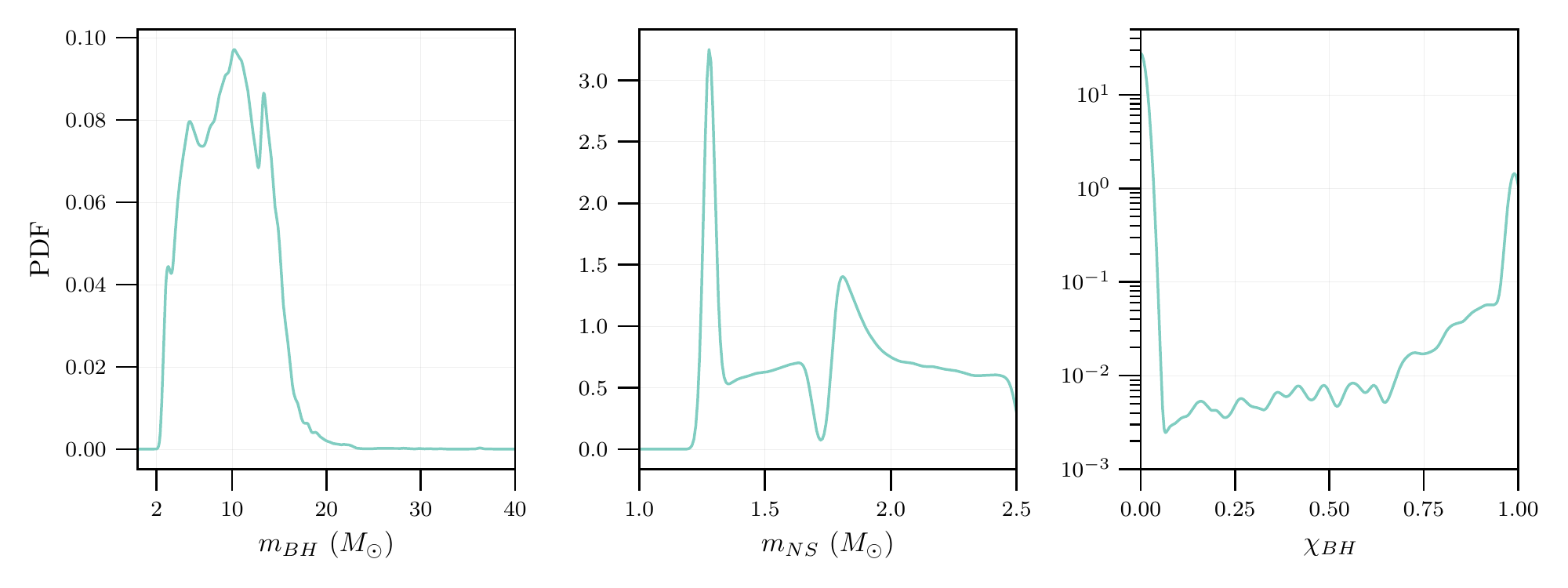}
\caption{\label{appfig:Pop2_mass_spins}The probability density function (PDF) plots for the masses of the BH and the NS in Pop-2 derived from the fiducial model in Ref. \cite{Broekgaarden:2021iew} and the spins of BH derived from the fits in Ref. \cite{Chattopadhyay:2022cnp}.}
\end{figure*}

The values of $x_1$, $x_2$, $m$ and $c$ depend on the metallicity of the environment in which the systems are formed and are given in Table \ref{tab:spin_params}. We perform interpolation to obtain these parameters for systems whose metallicities fall in between the reported values. We also limit the maximum spin of the BH to $1$ and the minimum spin to $0$. The mass profiles for the BH and the NS and the spins of the BH obtained by using these methods are shown in Fig. \ref{appfig:Pop2_mass_spins}.

\section{Number of kilonova detections} \label{appsec:no_of_kn_det}
Our criteria for claiming a KN detection by an EM telescope is that the peak luminosity corresponding to a particular filter must be brighter than the limiting magnitude of the EM telescope corresponding to that filter. In this study, we report the number of detections for two EM telescopes: the Vera C. Rubin Observatory \cite{web:Rubin,LSST:2008ijt} and the \textit{Nancy Grace Roman Space Telescope} \cite{Hounsell:2017ejq,Chase:2021ood}. The field of view (FOV) of the two telescopes can be found in Table \ref{tab:telescope_FOVs}. Both telescopes can observe emissions in six photometric bands listed in Table \ref{tab:rubin_vals}, along with the effective wavelength of each band and the corresponding limiting magnitude. The number of KN detections in 10 years for each band for the two telescopes corresponding to Pop-1 and Pop-2 events are given in Tables \ref{apptab:no_of_kn_det_pop1} and \ref{apptab:no_of_kn_det_pop2}, respectively.
 \begin{table*}[htbp] 
  \centering
  \caption{\label{apptab:no_of_kn_det_pop1}Bandwise number of KN detections for both the EM telescopes, following the GW detections for the six GW detector networks for Pop-1. The numbers are reported for an observation time of 10 years.}
  \renewcommand{\arraystretch}{1.5} 
    \begin{tabular}{l | P{1.3cm} P{1.3cm} P{1.3cm} | P{1.3cm} P{1.3cm} P{1.3cm} | P{1.3cm} P{1.3cm} P{1.3cm}}
    \hhline{==========}
    \multicolumn{10}{c}{\textbf{Vera C. Rubin Telescope}} \\
    \hhline{----------}
    Filter & \multicolumn{3}{c|}{$u$} & \multicolumn{3}{c|}{$g$} & 
    \multicolumn{3}{c}{$r$}\\
    \hhline{----------}
    EOS & ALF2 & APR4 & DD2 & ALF2 & APR4 & DD2 & ALF2 & APR4 & DD2 \\
    \hhline{----------}
    HLVKI\texttt{+} & $0^{+2}_{-0}$ & $0$ & $1^{+4}_{-1}$  & $10^{+17}_{-9}$ & $4^{+5}_{-3}$ & $16^{+30}_{-15}$ & $9^{+16}_{-8}$ & $4^{+5}_{-3}$ & $16^{+28}_{-15}$\\
    VK\texttt{+}HLIv & $0^{+2}_{-0}$ & $0$ & $1^{+4}_{-1}$ & $13^{+25}_{-12}$ & $5^{+9}_{-4}$ & $22^{+46}_{-21}$ & $14^{+25}_{-13}$ & $4^{+9}_{-3}$ & $23^{+42}_{-22}$ \\
    HLKI\texttt{+}E & $0^{+2}_{-0}$ & $0$ & $1^{+4}_{-1}$
  & $13^{+25}_{-12}$ & $5^{+9}_{-4}$ & $22^{+46}_{-21}$ & $14^{+25}_{-13}$ & $4^{+9}_{-3}$ & $23^{+42}_{-22}$ \\
    VKI\texttt{+}C & $0^{+2}_{-0}$ & $0$ & $1^{+4}_{-1}$  & $13^{+25}_{-12}$ & $5^{+9}_{-4}$ & $22^{+46}_{-21}$ & $14^{+25}_{-13}$ & $4^{+9}_{-3}$ & $23^{+42}_{-22}$\\
    KI\texttt{+}EC & $0^{+2}_{-0}$ & $0$ & $1^{+4}_{-1}$ & $13^{+25}_{-12}$ & $5^{+9}_{-4}$ & $22^{+46}_{-21}$ & $14^{+25}_{-13}$ & $4^{+9}_{-3}$ & $23^{+42}_{-22}$\\
    ECS & $0^{+2}_{-0}$ & $0$ & $1^{+4}_{-1}$
 & $13^{+25}_{-12}$ & $5^{+9}_{-4}$ & $22^{+46}_{-21}$ & $14^{+25}_{-13}$ & $4^{+9}_{-3}$ & $23^{+42}_{-22}$\\
    \hhline{----------}
    Filter & \multicolumn{3}{c|}{$i$} & \multicolumn{3}{c|}{$z$} & 
    \multicolumn{3}{c}{$y$}\\
    \hhline{----------}
    EOS & ALF2 & APR4 & DD2 & ALF2 & APR4 & DD2 & ALF2 & APR4 & DD2 \\
    \hhline{-|-|-|-|-|-|-|-|-|-}
    HLVKI\texttt{+} & $5^{+9}_{-4}$ & $2^{+1}_{-2}$ & $10^{+19}_{-9}$ & $2^{+3}_{-2}$ & $0$ & $4^{+7}_{-4}$ & $0$ & $0$ & $0^{+2}_{-0}$\\
    VK\texttt{+}HLIv & $5^{+12}_{-4}$ & $2^{+2}_{-2}$ & $10^{+24}_{-9}$ & $2^{+3}_{-2}$ & $0$ & $4^{+7}_{-4}$  & $0$ & $0$ & $0^{+2}_{-0}$\\
    HLKI\texttt{+}E & $5^{+12}_{-4}$ & $2^{+2}_{-2}$ & $10^{+24}_{-9}$ & $2^{+3}_{-2}$ & $0$ & $4^{+7}_{-4}$  & 0 & 0 & $0^{+2}_{-0}$ \\
    VKI\texttt{+}C & $5^{+12}_{-4}$ & $2^{+2}_{-2}$ & $10^{+24}_{-9}$ & $2^{+3}_{-2}$ & $0$ & $4^{+7}_{-4}$  & $0$ & $0$ & $0^{+2}_{-0}$ \\
    KI\texttt{+}EC & $5^{+12}_{-4}$ & $2^{+2}_{-2}$ & $10^{+24}_{-9}$ & $2^{+3}_{-2}$ & $0$ & $4^{+7}_{-4}$  & $0$ & $0$ & $0^{+2}_{-0}$\\
    ECS & $5^{+12}_{-4}$ & $2^{+2}_{-2}$ & $10^{+24}_{-9}$ & $2^{+3}_{-2}$ & $0$ & $4^{+7}_{-4}$  & $0$ & $0$ & $0^{+2}_{-0}$\\
    \hhline{----------}
    \multicolumn{10}{c}{\textbf{Nancy Grace Roman Observatory}} \\
    \hhline{----------}
    Filter & \multicolumn{3}{c|}{$R$} & \multicolumn{3}{c|}{$Z$} & 
    \multicolumn{3}{c}{$Y$}\\
    \hhline{----------}
    EOS & ALF2 & APR4 & DD2 & ALF2 & APR4 & DD2 & ALF2 & APR4 & DD2 \\
    \hhline{----------}
    HLVKI\texttt{+} & $16^{+30}_{-15}$ & $7^{+6}_{-6}$ & $29^{+52}_{-27}$ & $16^{+29}_{-15}$ & $6^{+5}_{-5}$ & $25^{+47}_{-23}$ & $14^{+29}_{-13}$ & $6^{+5}_{-5}$ & $25^{+45}_{-23}$\\
    VK\texttt{+}HLIv & $66^{+131}_{-51}$ & $23^{+50}_{-17}$ & $118^{+242}_{-97}$ & $43^{+103}_{-34}$ & $11^{+33}_{-10}$ & $69^{+162}_{-53}$ & $38^{+87}_{-30}$ & $11^{+31}_{-10}$ & $62^{+155}_{-49}$\\
    HLKI\texttt{+}E & $96^{+168}_{-74}$ & $39^{+68}_{-30}$ & $169^{+313}_{-138}$ & $61^{+123}_{-49}$ & $24^{+49}_{-21}$ & $96^{+194}_{-76}$ & $54^{+106}_{-44}$ & $17^{+40}_{-15}$ & $84^{+182}_{-68}$\\
    VKI\texttt{+}C & $84^{+151}_{-64}$ & $30^{+56}_{-22}$ & $155^{+293}_{-126}$ & $53^{+112}_{-42}$ & $18^{+39}_{-16}$ & $84^{+178}_{-66}$ & $46^{+95}_{-37}$ & $17^{+36}_{-15}$ & $74^{+168}_{-60}$\\
    KI\texttt{+}EC & $97^{+170}_{-75}$ & $39^{+68}_{-30}$ & $171^{+318}_{-140}$ & $61^{+123}_{-49}$ & $24^{+49}_{-21}$ & $96^{+194}_{-76}$ & $54^{+106}_{-44}$ & $17^{+40}_{-15}$ & $84^{+182}_{-68}$\\
    ECS & $97^{+170}_{-75}$ & $39^{+68}_{-30}$ & $171^{+318}_{-140}$ & $61^{+123}_{-49}$ & $24^{+49}_{-21}$ & $96^{+194}_{-76}$ & $54^{+106}_{-44}$ & $17^{+40}_{-15}$ & $84^{+182}_{-68}$\\
    \hhline{----------}
    Filter & \multicolumn{3}{c|}{$J$} & \multicolumn{3}{c|}{$H$} & 
    \multicolumn{3}{c}{$F$}\\
    \hhline{----------}
    EOS & ALF2 & APR4 & DD2 & ALF2 & APR4 & DD2 & ALF2 & APR4 & DD2 \\
    \hhline{----------}
    HLVKI\texttt{+} & $14^{+29}_{-13}$ & $6^{+5}_{-5}$ & $25^{+42}_{-23}$ & $13^{+24}_{-12}$ & $5^{+5}_{-4}$ & $22^{+42}_{-20}$ & $10^{+17}_{-9}$ & $4^{+5}_{-3}$ & $16^{+28}_{-15}$ \\
    VK\texttt{+}HLIv & $35^{+78}_{-27}$ & $11^{+27}_{-10}$ & $53^{+129}_{-43}$ & $28^{+61}_{-23}$ & $9^{+19}_{-8}$ & $45^{+105}_{-37}$ & $15^{+27}_{-14}$ & $4^{+10}_{-3}$ & $27^{+51}_{-23}$\\
    HLKI\texttt{+}E & $50^{+96}_{-40}$ & $13^{+32}_{-11}$ & $72^{+153}_{-60}$ & $35^{+72}_{-29}$ & $11^{+23}_{-9}$ & $58^{+119}_{-48}$ & $15^{+27}_{-14}$ & $4^{+10}_{-3}$ & $28^{+52}_{-23}$\\
    VKI\texttt{+}C & $44^{+86}_{-35}$ & $13^{+28}_{-11}$ & $65^{+143}_{-54}$ & $35^{+68}_{-29}$ & $11^{+19}_{-9}$ & $53^{+113}_{-44}$ & $15^{+27}_{-14}$ & $4^{+10}_{-3}$ & $28^{+52}_{-23}$\\
    KI\texttt{+}EC & $50^{+96}_{-40}$ & $13^{+32}_{-11}$ & $72^{+153}_{-60}$ & $35^{+72}_{-29}$ & $11^{+23}_{-9}$ & $58^{+119}_{-48}$ & $15^{+27}_{-14}$ & $4^{+10}_{-3}$ & $28^{+52}_{-23}$\\
    ECS & $50^{+96}_{-40}$ & $13^{+32}_{-11}$ & $72^{+153}_{-60}$ & $35^{+72}_{-29}$ & $11^{+23}_{-9}$ & $58^{+119}_{-48}$ & $15^{+27}_{-14}$ & $4^{+10}_{-3}$ & $28^{+52}_{-23}$\\
    \hhline{==========}
    \end{tabular}
\end{table*}
\begin{table*}[htbp] 
  \centering
  \caption{\label{apptab:no_of_kn_det_pop2}Bandwise number of KN detections for both the EM telescopes, following the GW detections for the six GW detector networks for Pop-2. The numbers are reported for an observation time of 10 years.}
  \renewcommand{\arraystretch}{1.5} 
    \begin{tabular}{l | P{1.3cm}P{1.3cm}P{1.3cm} | P{1.3cm}P{1.3cm}P{1.3cm} | P{1.3cm}P{1.3cm}P{1.3cm}}
    \hhline{==========}
    \multicolumn{10}{c}{\textbf{Vera C. Rubin Telescope}} \\
    \hhline{----------}
    Filter & \multicolumn{3}{c|}{$u$} & \multicolumn{3}{c|}{$g$} & 
    \multicolumn{3}{c}{$r$}\\
    \hhline{----------}
    EOS & ALF2 & APR4 & DD2 & ALF2 & APR4 & DD2 & ALF2 & APR4 & DD2 \\
    \hhline{----------}
    HLVKI\texttt{+} & $4^{+5}_{-3}$ & $0$ & $6^{+9}_{-5}$ & $17^{+22}_{-15}$ & $0$ & $33^{+44}_{-28}$ & $16^{+18}_{-14}$ & $0$ & $31^{+41}_{-26}$ \\
    VK\texttt{+}HLIv & $4^{+5}_{-3}$ & $0$ & $6^{+9}_{-5}$ & $21^{+46}_{-18}$ & $0$ & $42^{+93}_{-36}$ & $19^{+30}_{-16}$ & $0$ & $37^{+82}_{-31}$ \\
    HLKI\texttt{+}E & $4^{+5}_{-3}$ & $0$ & $6^{+9}_{-5}$ & $22^{+47}_{-19}$ & $0$ & $43^{+95}_{-37}$ & $20^{+30}_{-17}$ & $0$ & $38^{+84}_{-32}$ \\
    VKI\texttt{+}C & $4^{+5}_{-3}$ & $0$ & $6^{+9}_{-5}$ & $22^{+46}_{-19}$ & $0$ & $43^{+94}_{-37}$ & $20^{+30}_{-17}$ & $0$ & $38^{+83}_{-32}$ \\
    KI\texttt{+}EC & $4^{+5}_{-3}$ & $0$ & $6^{+9}_{-5}$ & $22^{+47}_{-19}$ & $0$ & $43^{+95}_{-37}$ & $20^{+30}_{-17}$ & $0$ & $38^{+84}_{-32}$ \\
    ECS & $4^{+5}_{-3}$ & $0$ & $6^{+9}_{-5}$ & $22^{+47}_{-19}$ & $0$ & $43^{+95}_{-37}$ & $20^{+30}_{-17}$ & $0$ & $38^{+84}_{-32}$ \\
    \hhline{----------}
    Filter & \multicolumn{3}{c|}{$i$} & \multicolumn{3}{c|}{$z$} & 
    \multicolumn{3}{c}{$y$}\\
    \hhline{----------}
    EOS & ALF2 & APR4 & DD2 & ALF2 & APR4 & DD2 & ALF2 & APR4 & DD2 \\
    \hhline{----------}
    HLVKI\texttt{+} & $8^{+13}_{-7}$ & $0$ & $16^{+23}_{-15}$ & $1^{+4}_{-1}$ & $0$ & $6^{+12}_{-5}$ & $0$ & $0$ & $1^{+0}_{-1}$ \\
    VK\texttt{+}HLIv & $8^{+14}_{-7}$ & $0$ & $16^{+29}_{-15}$ & $1^{+4}_{-1}$ & $0$ & $6^{+12}_{-5}$ & $0$ & $0$ & $1^{+0}_{-1}$ \\
    HLKI\texttt{+}E & $8^{+14}_{-7}$ & $0$ & $16^{+29}_{-15}$ & $1^{+4}_{-1}$ & $0$ & $6^{+12}_{-5}$ & $0$ & $0$ & $1^{+0}_{-1}$ \\
    VKI\texttt{+}C & $8^{+14}_{-7}$ & $0$ & $16^{+29}_{-15}$ & $1^{+4}_{-1}$ & $0$ & $6^{+12}_{-5}$ & $0$ & $0$ & $1^{+0}_{-1}$ \\
    KI\texttt{+}EC & $8^{+14}_{-7}$ & $0$ & $16^{+29}_{-15}$ & $1^{+4}_{-1}$ & $0$ & $6^{+12}_{-5}$ & $0$ & $0$ & $1^{+0}_{-1}$ \\
    ECS & $8^{+14}_{-7}$ & $0$ & $16^{+29}_{-15}$ & $1^{+4}_{-1}$ & $0$ & $6^{+12}_{-5}$ & $0$ & $0$ & $1^{+0}_{-1}$ \\
    \hhline{----------}
    \multicolumn{10}{c}{\textbf{Nancy Grace Roman Observatory}} \\
    \hhline{----------}
    Filter & \multicolumn{3}{c|}{$R$} & \multicolumn{3}{c|}{$Z$} & 
    \multicolumn{3}{c}{$Y$}\\
    \hhline{----------}
    EOS & ALF2 & APR4 & DD2 & ALF2 & APR4 & DD2 & ALF2 & APR4 & DD2 \\
    \hhline{----------}
    HLVKI\texttt{+} & $32^{+42}_{-27}$ & $0$ & $59^{+92}_{-51}$ & $29^{+37}_{-24}$ & $0$ & $56^{+80}_{-49}$ & $26^{+35}_{-21}$ & $0$ & $53^{+74}_{-46}$ \\
    VK\texttt{+}HLIv & $93^{+222}_{-76}$ & $0$ & $196^{+435}_{-163}$ & $61^{+144}_{-48}$ & $0$ & $136^{+291}_{-113}$ & $49^{+119}_{-38}$ & $0$ & $119^{+257}_{-98}$ \\
    HLKI\texttt{+}E & $122^{+275}_{-99}$ & $0$ & $260^{+542}_{-213}$ & $71^{+168}_{-56}$ & $0$ & $167^{+341}_{-137}$ & $54^{+131}_{-41}$ & $0$ & $143^{+297}_{-116}$ \\
    VKI\texttt{+}C & $115^{+262}_{-92}$ & $0$ & $243^{+511}_{-197}$ & $67^{+164}_{-52}$ & $0$ & $160^{+329}_{-130}$ & $51^{+129}_{-38}$ & $0$ & $137^{+287}_{-110}$ \\
    KI\texttt{+}EC & $124^{+281}_{-100}$ & $0$ & $267^{+555}_{-218}$ & $71^{+169}_{-56}$ & $0$ & $168^{+347}_{-137}$ & $54^{+132}_{-41}$ & $0$ & $144^{+300}_{-116}$ \\
    ECS & $124^{+281}_{-100}$ & $0$ & $267^{+555}_{-218}$ & $71^{+169}_{-56}$ & $0$ & $168^{+347}_{-137}$ & $54^{+132}_{-41}$ & $0$ & $144^{+300}_{-116}$ \\
    \hhline{----------}
    Filter & \multicolumn{3}{c|}{$J$} & \multicolumn{3}{c|}{$H$} & 
    \multicolumn{3}{c}{$F$}\\
    \hhline{----------}
    EOS & ALF2 & APR4 & DD2 & ALF2 & APR4 & DD2 & ALF2 & APR4 & DD2 \\
    \hhline{----------}
    HLVKI\texttt{+} & $24^{+31}_{-20}$ & $0$ & $51^{+72}_{-44}$ & $20^{+24}_{-17}$ & $0$ & $44^{+66}_{-37}$ & $10^{+16}_{-8}$ & $0$ & $32^{+41}_{-27}$ \\
    VK\texttt{+}HLIv & $40^{+91}_{-31}$ & $0$ & $100^{+223}_{-81}$ & $25^{+64}_{-21}$ & $0$ & $75^{+170}_{-61}$ & $10^{+21}_{-8}$ & $0$ & $38^{+82}_{-32}$ \\
    HLKI\texttt{+}E & $43^{+97}_{-34}$ & $0$ & $113^{+251}_{-92}$ & $26^{+67}_{-22}$ & $0$ & $82^{+184}_{-66}$ & $10^{+21}_{-8}$ & $0$ & $39^{+84}_{-33}$ \\
    VKI\texttt{+}C & $40^{+96}_{-31}$ & $0$ & $108^{+245}_{-87}$ & $26^{+66}_{-22}$ & $0$ & $79^{+181}_{-63}$ & $10^{+21}_{-8}$ & $0$ & $39^{+83}_{-33}$ \\
    KI\texttt{+}EC & $43^{+98}_{-34}$ & $0$ & $113^{+252}_{-92}$ & $26^{+67}_{-22}$ & $0$ & $82^{+185}_{-66}$ & $10^{+21}_{-8}$ & $0$ & $39^{+84}_{-33}$ \\
    ECS & $43^{+98}_{-34}$ & $0$ & $113^{+252}_{-92}$ & $26^{+67}_{-22}$ & $0$ & $82^{+185}_{-66}$ & $10^{+21}_{-8}$ & $0$ & $39^{+84}_{-33}$ \\
    \hhline{==========}
    \end{tabular}
\end{table*}

\end{document}